\documentclass[%
	aps,
	 prm,%
	 amsmath,amssymb,
	preprint,%
	]{revtex4-1}
	
	\usepackage{makecell} %For bold lines on table
	\usepackage{booktabs} %For professional lines of tables

	\usepackage{mdframed}% For frames on definitions
	\usepackage{multirow} % For multiple rows in tables

	\usepackage{graphicx}% Include figure files
	\usepackage{subfigure} %for subfigures
	\usepackage{epstopdf}
	\usepackage{dcolumn}% Align table columns on decimal point
	\usepackage{bm}% bold math
	\usepackage{subfigure}
	\usepackage{rotating}

	\def\0{0}
	\def\US{\textrm{right}} %or cylinder
	\def\DS{\textrm{left}}
	\def\side{\textrm{side}}

	\def\bank{\textrm{bank}}
	\def\snub{\textrm{snub}}
	\def\out{\textrm{out}}
	\def\corr{\textrm{corr}}
	\def\melt{\textrm{melt}}
	\def\freeze{\textrm{freeze}}
	\def\sam{\textrm{sam}}
	\def\addenda{\textrm{addenda}}
	\def\sol{\textrm{sol}}
	
	\def\max{\textrm{max}}
	\def\min{\textrm{min}}
	\def\refR{\textrm{ref}}
    \def\tableerrors{S1}
    \def\tableelectrical{S2}
    \def\figXRDstack{S5} 
    \def\figBigXRD{S6}
	\def\figphotos{S7}
	\def\figSixtyGPaRightSM{S8}
    \def\figNineNineGPaSM{S16}
    \def\figSetup{1}
	\def\figraw{2}
	\def\figsixtyGPaIt{3}
	\def\figsummary{5}
	
	\usepackage[mathlines]{lineno}% Enable numbering of text
    \AtBeginDocument{%
    \setlength{\linenumbersep}{\dimexpr\oddsidemargin+0.7in-\linenumberwidth\relax}%
    }

\begin{document}
% \preprint{Draft}%APS/123-ABC}
%\preprint{Resubmission}
        
	\title{A latent heat method to detect melting and freezing of metals at megabar pressures}
	\author{Zachary M. Geballe$^1$}
	\author{Nicholas Holtgrewe$^{1,2}$}
	\author{Amol Karandikar$^1$}
	\author{Eran Greenberg$^2$}
	\author{Vitali B. Prakapenka$^2$}
	\author{Alexander F. Goncharov$^1$}
	\affiliation{$^1$Earth and Planets Laboratory, Carnegie Institution for Science, Washington, DC 20015, USA}
	\affiliation{$^2$Center for Advanced Radiation Sources, University of Chicago, IL 60637, USA}

	\date{\today}	

	\begin{abstract}
	The high-pressure melting curves of metals provide simple and useful tests for theories of melting, as well as important constraints for the modeling of planetary interiors. Here, we present an experimental technique that reveals the latent heat of fusion of a metal sample compressed inside a diamond anvil cell. The technique combines microsecond-timescale pulsed electrical heating with an internally-heated diamond anvil cell for the first time. Further, we use the technique to measure the melting curve of platinum to the highest pressure measured to date. Melting temperature increases from~$\sim 3000$~K at 34 GPa to~$\sim 4500$~K at 107 GPa, thermodynamic conditions that are between the steep and shallow experimental melting curves reported previously. The melting curve is a linear function of compression over the 0 to 20\% range of compression studied here, allowing a good fit to the Kraut-Kennedy empirical model with fit parameter $C=6.0$. 
\end{abstract}
		
\newcommand{\ud}{\mathrm{d}}
\maketitle

\section{Introduction}
High-pressure melting curves of simple materials provide a fertile testing ground for theories of melting, from simple empirical and semi-empirical models such as the Kraut-Kennedy and Lindemann models  \cite{Kraut1966_PRL,Gilvarry1956}, to atomistic models such as the \textit{ab initio} Z-method \cite{Belonoshko2012}. 
Knowledge of high-pressure melting temperatures is also crucial for understanding the evolution of planetary cores \cite{Hirose2013}.

In order to test simple melting theories, accurate data are needed across a range of densities. In practice, compression up to 10s of percent has been used \cite{Kraut1966_PR}. To achieve this for the relatively incompressible transition metals, pressures of~$\sim 50$~to~$100$~GPa (0.5 to 1 megabar) are required. Recent publications have reported melting curves to pressures above 50 GPa for transition metals including V \cite{Errandonea2019}, Nb \cite{Errandonea2020}, Fe \cite{Sinmyo2019}, Mo \cite{Hrubiak2017}, Ti \cite{Stutzmann2015}, Zr \cite{Parisiades2019}, Pt \cite{Anzellini2019}, and Ta \cite{Karandikar2016}.  Unfortunately, the accuracy of melting data is uncertain for several of the most-studied metals at pressures above 20 GPa, as evidenced by discrepancies among studies of Fe \cite{Sinmyo2019}, Ta \cite{Karandikar2016}, Mo \cite{Hrubiak2017}, and Pt \cite{Anzellini2019}. For platinum, the experimental melting temperatures reported in Refs.~\onlinecite{Errandonea2013,Anzellini2019} are systematically higher than those in Refs.~\onlinecite{Kavner1998,Patel2018,LoNigro2011}, resulting in a discrepancy of at least 1000 K at 70 GPa, the pressure corresponding to 15\% volume compression. 

It may also be possible to test simple analytical models of melting by comparing them to \textit{ab initio} models. For platinum, melting temperature calculations by two different research groups using the recently developed \textit{ab initio} Z-method agree to within 200 K at 10 GPa and within 300 K at 120 GPa \cite{Belonoshko2012,Anzellini2019}. The results imply an approximately linear dependence of melting temperature ($T_m$) with respect to pressure ($P$), but not with respect to volume ($V$), indicating a departure from the Kraut-Kennedy model if the error in calculated melting temperature is less than 400 K. Note that departures from both the Lindemann and Kraut-Kennedy models are common (e.g. \cite{Lazicki2010,Luedemann1968,Wolf1984}), and the Lindemann model has been frequently criticized for its overly simplistic physical basis (e.g. \cite{Stacey1989,Wolf1984}). 
Nevertheless, the accuracy of Z-method calculations is also uncertain, especially in the absence of ``waiting time analyses''  \cite{Alfe2011,Braithwaite2019}. For platinum, the Z-method calculation results match the most recently-published experimental data \cite{Anzellini2019}, but not others \cite{Kavner1998,Patel2018}, underscoring the need for new experimental results, and perhaps new experimental methods that are more reproducible across laboratories than the methods currently used. 

Commonly used melt criteria for diamond cell experiments include visual observation of motion, anomalies in temperature vs. laser power, and the onset of diffuse scattering in X-ray diffraction. These three techniques account for all the experimental data on platinum melting at pressures above 20 GPa \cite{Anzellini2019,Errandonea2013,Kavner1998,Patel2018,Errandonea2013,LoNigro2011}.  The first two criteria are indirectly related to melting. When materials melt, they tend to move and to cause anomalous temperature-power trends, but neither phenomenon is specific to melting, nor do they necessarily occur upon melting \cite{Geballe2012,Parisiades2019,Stutzmann2015}. 

On the other hand, observation of a step-function increase in diffuse X-ray scattering upon increasing temperature would provide strong evidence for melting, because liquids generate much stronger diffuse scattering than crystalline solids. In reality, technical challenges related to large temperature gradients add substantial ambiguity to the identification of the onset of melting by X-ray scattering in laser-heated diamond anvil cells \cite{KarandikarThesis}. For the case of platinum, Anzellini et al. \cite{Anzellini2019} reports precise X-ray based determination of melting temperature up to 30 GPa, but not at higher pressure. The uncertainty in the temperature of ``liquid'' diffraction increases to $\pm 700$~K at 49 GPa, and no diffraction from a liquid is reported at higher pressures. The thesis of Lo Nigro \cite{LoNigro2011} also reports a melting curve from 30 to 90 GPa based on X-ray diffraction, but the platinum diffraction data is noisy in Fig. 3.5 of Lo Nigro \cite{LoNigro2011}, likely due to sample preparation methods designed to study the silicate sample in which a small amount of platinum is embedded. Few details are given about the melt criterion and measurement uncertainties, and the resulting melting curve is~$\sim 200$~K to 1500 K lower than the plateau-based melting data of Anzellini et al. \cite{Anzellini2019}.

To identify melting in a more reproducible way than in previous experiments at pressures above 20 GPa, detecting latent heat could be very useful. All melting transitions have latent heat, and it is typically much larger than the latent heat of solid-solid transitions \cite{Hultgren1963}. In practice, latent heat has been a useful way to identify melting of refractory metals at ambient pressure \cite{Cagran2008}, but it has likely never been identified in static compression experiments at pressure~$>20$ GPa. Albeit, in the case of pulsed-laser heating of hydrogen at 100 to 200 GPa, anomalies in peak temperature versus laser power have been attributed to the latent heat of melting and the latent heat of dissociation of molecular hydrogen \cite{Deemyad2008,Zaghoo2016,Houtput2019}. Nevertheless, the attribution to latent heat is controversial \cite{Montoya2012,GoncharovComment}, and the method of latent heat detection has not been reproduced by any other group, to the best of our knowledge. 

The major experimental challenge in identifying latent heat in high pressure experiments is to deposit heat and measure the sample's temperature (or a proxy for temperature) fast enough and over a large enough sample volume so that little heat is lost to the surroundings. Using finite element models, Geballe and Jeanloz \cite{Geballe2012} showed that the heating timescale should be ns to $\mu$s in order to reveal the latent heat. This timescale results from the inevitably small sample size and inevitably poor thermal insulation in diamond cell experiments. The models also show that latent heat signatures are larger during internal heating than surface heating, suggesting Joule heating of metals is preferable to laser-heating of metals. So far, these extreme requirements have limited the detection of latent heat in static high pressure experiments to the pressure range  below 20 GPa and to devices with larger sample volumes than those in diamond anvil cells \cite{Lazicki2010}.

Here, we report a new technique that records melting by revealing the latent heat of melting of metals in diamond anvil cells at pressures in the range $\sim 7$ GPa to above 100 GPa, and temperatures in the range $\sim 2200$~K to above 4000 K. The technique integrates microsecond-timescale pulsed electrical heating with the internally-heated diamond anvil cell for the first time, thereby creating the short heating timescale and spatial homogeneity needed to reveal latent heat at high pressures. We then use the technique to determine the melting curve of platinum up to 107 GPa.

\section{Experimental Methods}
\subsection{Sample loading}
For each high-pressure run, we use a five-step procedure to prepare a sample of platinum connected to two, three, or four electrical leads and thermally insulated from the diamond anvils by a layer of KCl. The result is an internally heated diamond anvil cell similar to the one used by Zha et al.~\cite{Zha2008} to measure the equation of state of platinum up to 80 GPa and 1900 K.  Details are presented in the Supplementary Methods. Briefly, we first use standard methods to align diamond anvils with 100 to 300~$\mu$m-diameter culets and to make a pre-indented rhenium gasket with an insert made of cubic boron nitride mixed with ND 353 Epotek epoxy (hereafter referred to as ``cBN''). Second, we prepare four outer electrodes that extend from the edge of body of the diamond cell to the edge of the diamond's culet. Third, we prepare the inner electrodes by pressing $\sim 10$~$\mu$m-thick pieces of platinum into the cBN on the culet. 

Fourth, we laser-drill a hole with diameter equal to 40\% of the culet's diameter and fill it with several pieces of KCl and platinum. The pieces of platinum and KCl are stacked so that when the diamond cell is closed, one central piece of platinum of 5 to 30~$\mu$m-width is separated from both anvils by 5 to 10~$\mu$m-thick KCl layers and electrically connected to the four outer electrodes by other pieces of platinum. This central piece is the platinum sample that is eventually melted. Fifth, we dry the KCl by inserting the whole diamond cell in a vacuum oven for at least 45 minutes at 120$^\circ$C followed by an argon-purge. Finally, we close the cell, let it cool, and compress to the target starting pressure. Pressure at room temperature is measured using the shift of the Raman signal from the strained diamond anvil \cite{Akahama2006}. 
After heating, pressure is measured again using the Raman edge or by X-ray diffraction from the~300~K platinum sample \cite{Matsui2009}. For each melting run, the reported pressure at room temperature, $P_0$, is the average of pressures measured before and after heating. 

A simpler version of the above procedure was used for the sample that generated the lowest pressure data presented here. A diamond anvil cell was prepared with 1 mm-diameter culets, without a gasket, and with~$\sim100$~$\mu$m-thick KCl thermal insulation. The relatively large sample was made from a 0.5 mm-long segment of 25~$\mu$m-diameter platinum wire. Strips of gold were cut from 10~$\mu$m-thick foil and used as inner electrodes. The pressure before heating was less than 0.1 GPa.

\subsection{Pulsed heating and electrical measurement}
After compressing each platinum sample to high pressure, we connect it to the home-built electronics that drive current through the sample and measure current and voltage. First, each diamond cell is connected to the electronics, as shown in Fig. \ref{fig:setup}; see Supplementary Methods for details. Second, the capacitor bank is repeatedly discharged by delivery of square waves of 3 to 8~$\mu$s duration to the gate of the transistor (MOSFET). Third, the power of electrical heating pulses is gradually increased by increasing the voltage of the capacitor bank, $V_\bank$, until the platinum sample reaches peak temperatures of 1500 to 2000 K, a temperature range that is high enough for a CCD camera to visualize the thermal emissions from the sample, yet low enough to avoid accidentally melting the sample. The current and voltage of each pulse (or set of pulses) is calculated based on an oscilloscope recording of the outputs of two instrumentation-amplifiers (``in-amps''). One in-amp measures the voltage difference across the reference resistor, while the other measures the voltage difference across the platinum sample. 

\subsection{Thermal emission and X-ray diffraction}
While pulsing electrical power through the high-pressure sample, we measure time-resolved thermal emissions, spatially-resolved thermal emission, and X-ray diffraction. Time-resolved measurements of thermal emissions are the key to detection of melting and freezing temperatures. Spatially-resolved measurements of thermal emission are important for estimating the size of the sample that is melted. X-ray diffraction measurements are important for determining the crystallographic phase of the material that melts and its pressure evolution during heating. 

We use two laboratories to generate the necessary data. The first melting experiment for each sample is performed at the Earth and Planets Laboratory of the Carnegie Institution for Science, where its thermal emissions spectra are recorded with a streak camera, a device that enables measurements with sub-microsecond time-resolution during single-heating-shot experiments. Several samples are subsequently melted at GSECARS, Sector 13 of the Advanced Photon Source at Argonne National Lab. At GSECARS, atomic structure and temperature are monitored by X-ray diffraction and thermal emissions measurements on gated intensified detectors, not streak cameras. The detectors are gated to collect X-ray and optical photons when the sample reaches its highest temperature, the final 1 $\mu$s of the heating pulse. 

\begin{figure}[tbhp]
	\centering
	\includegraphics[width=5.4in]{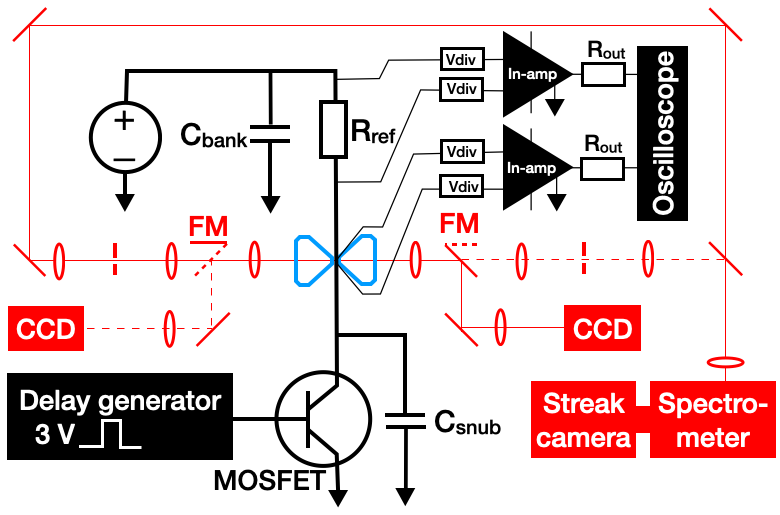} %5.9 in 
	\caption{Schematic of electrical path (black), optical paths (red) and diamond anvils (blue) at the Carnegie Institution for Science. A regulated DC power supply charges a capacitor bank ($C_\bank$: 470~$\mu$F, 70 V electrolytic). When triggered by the Delay generator (SRS DG645), the MOSFET (FQP30N06L) allows current to flow through a reference resistor ($R_\refR = 0.29$~$\Omega$), and the platinum sample that is compressed between diamond anvils. The snubber capacitor ($C_\snub$: 16~$\mu$F, 100 V electrolytic) limits current oscillations. The circuitry for measuring current and four-point-probe voltage are shown in thin black lines. The voltage dividers, Vdiv, reduce input voltage to within the 15 V range of the in-amp (AD842). Each divider is made of two resistors with typical values of 1 k$\Omega$ and 10 k$\Omega$. The in-amp is operated with no gain, referenced to ground, and connected through output resistors ($R_\out$: 105~$\Omega$) to the oscilloscope (Tektronix DPO 3034). A simplified optical path is shown here; see McWilliams et al. \cite{McWilliams2015} for elaboration. During each heating pulse, one flipper mirror (FM) diverts light from the left or right side of the diamond cell to a CCD camera  (Point Grey Grasshopper3 Color) for 2-dimensional imaging of thermal emissions. The other flipper mirror (FM) does not divert the light, allowing it to pass into a confocal filtering system, then into a spectrometer (Princeton Instruments Acton SP2300) and streak camera (Sydor ROSS 1000) for time-resolved measurements of thermal emissions. Solid red lines show the path of light in one configuration; dashed lines show the alternative configuration. Ovals represent lenses, line segments at 45$^\circ$~represent mirrors, and broken line segments represent pinholes.}
	\label{fig:setup}
\end{figure}

In each laboratory, the sample is located at the focal position of the optical system. The Carnegie system is shown schematically in Fig. \ref{fig:setup}, and described in detail in McWilliams et al.~\cite{McWilliams2015}. The GSECARS system is described in Prakapenka et al.~\cite{Prakapenka2008}. At GSECARS, the optical focus is aligned to the X-ray focus. $V_\bank$ is increased until the hottest section of the platinum sample is identified in an imaging camera set to 1 second exposure and maximum gain. Typically, we identify the hotspot by 10 to 100 repetitions of pulsed heating during the 1 second exposure. In all cases, a full cross section of the central platinum strip appears to heat to a nearly uniform temperature (Fig. \figphotos). We then translate the sample so that the hotspot is at the focus of the optical system. 

\begin{figure}[tbhp]
	\centering
	\includegraphics[width=3in]{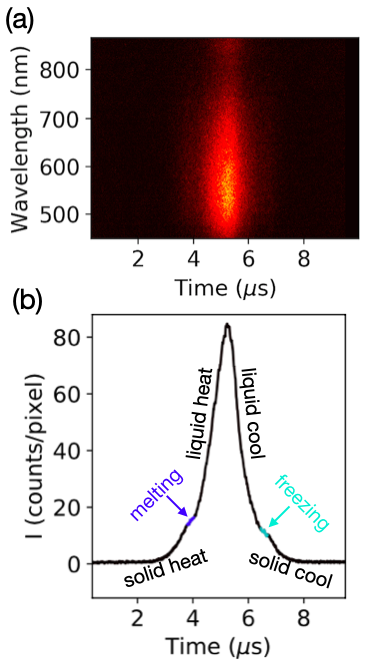}
	\caption{Streak camera image of platinum heated from~$T = 300$~K at~$P = 60\pm3$~GPa to~$T>5000$~K. (a) Raw data. (b) Intensity averaged over the wavelength-dimension. Annotations mark regions interpreted to be melting, freezing, and heating and cooling of solid and liquid platinum.}
	\label{fig:raw}
\end{figure}

At Carnegie, we record thermal emissions on the streak camera (e.g. Fig. \ref{fig:raw}). The measurement's spectral range is 450 to 860 nm in all experiments but one; a higher resolution grating limits the spectral range to 500 to 660 nm for the~$P_0 = 31$~GPa data set. The streak camera is set to 3 or 10~$\mu$s sweep duration for all experiments except for melting the non-gasketed sample ($P_0 = 1$~bar), for which sweep duration is 100~$\mu$s. We record thermal emissions from one side of the sample on the streak camera, and from the other side on a CCD camera. An example of thermal emissions data from one heating pulse to temperatures $>5000$ K at 68 GPa is shown in Fig. \ref{fig:raw}. Anomalies in thermal emission intensity during melting and freezing are easily identified in measurements of intensity versus time.

At GSECARS, temperatures are determined by fitting Planck functions to thermal emissions spectra emitted from a rectangular region of the sample that is 6~$\mu$m~$\times 20$~$\mu$m in area. This fit assumes greybody emission \cite{Benedetti2004}. The X-ray energy is 37 keV and its beam size is 3 x 4~$\mu$m. X-ray patterns are integrated using the Dioptas software \cite{Prescher2015}. The resistive heating pulse duration is 5 to 15~$\mu$s.

For each starting pressure, $P_0$, we collect data at a range of values of $V_\bank$. Then, we change pressure and heat again, if desired. In practice, melting was only documented at different pressures for one sample, first during heating from~$P_0 = 78$~GPa, then during heating from~$P_0 =60$~GPa.

\section{Results}
We report measurements of thermal emissions, voltage, current, and X-ray diffraction of platinum compressed and heated to 107 GPa and~$\sim 5000$~K. We define a ``plateau-like'' region to be one in which a temperature proxy changes anomalously slowly in time, compared to rate of change before and after the plateau-like region. The primary temperature proxy used in this study is the fourth root of thermal emission intensity,~$I^{1/4}$. (The fourth root is motivated by the Stephan-Boltzmann law, $I_\textrm{total} \propto T^4$).

Our main results are (1) plateau-like regions in~$I^{1/4}$~are reproducible and reversible upon cooling, (2) electrical resistance measurements, calorimetric analysis, and X-ray diffraction show that the plateau-like regions are caused by latent heats of melting and freezing, and (3) melting temperatures increase rapidly from 0 to~$\sim 40$~GPa, then more gradually to~$4490 \pm 220$~K at~$107 \pm 9$~GPa (Fig. \ref{fig:summary}). 

\begin{figure}[tbhp]
	\centering
	\includegraphics[width=6in]{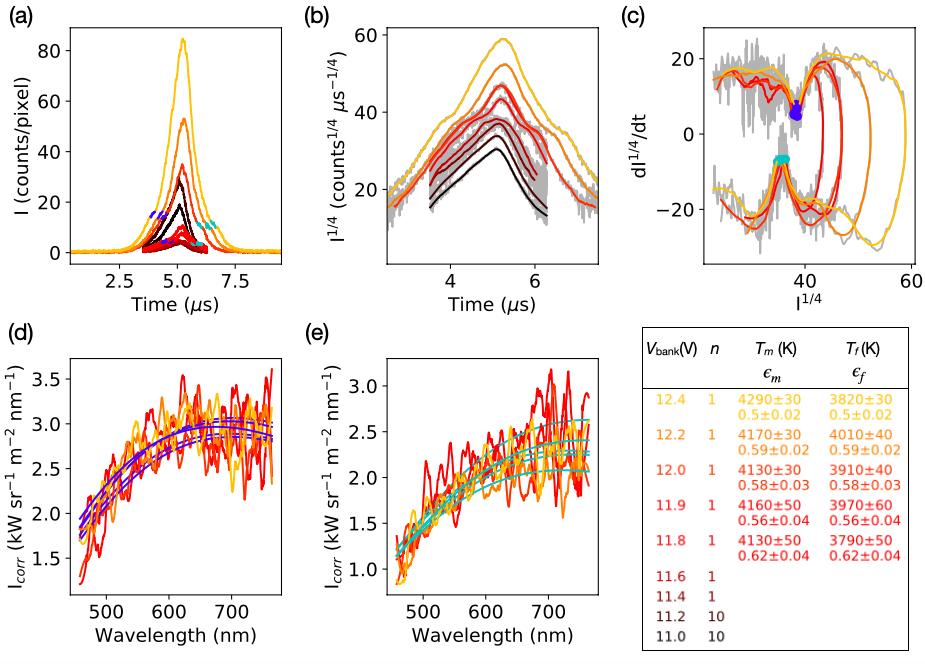}
	\caption{Time-resolved thermal emissions of the left-side of the platinum sample heated from 300 K at~$60 \pm 3$~GPa to past its melting point at~$4060 \pm 140$~K at~$68 \pm 5.0$~GPa. Each warm color (yellow to red to black) represents a set of~$n$~heating pulses driven by the voltage that is listed in the legend ($V_\bank$). Blue and cyan markings indicate melting and freezing. (a) Average counts on the streak camera CCD. (b) Fourth-root of average counts per microsecond, a proxy for temperature. Noisy grey curves show un-smoothed data,~$I^{1/4}$, while colored curves show smoothed data,~$I^{1/4}_s$. (c) Time-derivatives,~$\frac{dI^{1/4}_s}{dt}$~(grey), and smoothed time derivatives, $\frac{dI^{1/4}_s}{dt}_s$ (colors). The smoothing function is a second order Savitzky-Golay filter with timescale~$\tau = 0.4$~$\mu$s for both~$I^{1/4}_s$~and~$\frac{dI^{1/4}_s}{dt}_s$. The minima during heating (blue circles) and maxima during cooling (cyan circles), are interpreted as melting and freezing. The corresponding times,~$t_{\melt}\pm \tau/2$~and~$t_\freeze \pm \tau/2$,~are marked in blue and cyan in (a), and used for the temperature fits in (d) and (e). (d, e) Planck fits (blue and cyan) to thermal emissions spectra during melting and freezing.  Planck fit parameters listed in the legend are melting temperature and emissivity ($T_m$ and $\epsilon_m$), and freezing temperature and emissivity ($T_f$ and $\epsilon_f$). Spectra have been filtered to improve the clarity of the figures using a second order Savitzky-Golay filter with wavelength scale $d\lambda = 20$~nm. Planck fits are performed without filtering the spectra.}
	\label{fig:60GPa_It}
\end{figure}

For each of thirty-three heating runs recorded on the streak camera, the melting region is identified as a plateau-like interval in~$I^{1/4}$; six runs are shown in Fig. \ref{fig:60GPa_It}b and the remainder are shown in Figs. \figSixtyGPaRightSM-\figNineNineGPaSM. The melting temperature measured during an individual melting run is determined by fitting a Planck function to the thermal emissions spectrum collected during the melting interval (Fig. \ref{fig:60GPa_It}d; Supplementary Materials section ``Temperature fits''). The pressure at melting is estimated by adding a heating-induced pressure to the room temperature pressure measurement, ~$P_m = P_0 + \Delta P$. The value of~$\Delta P$~for each melting run is estimated from X-ray diffraction measurements at 30 to 60 GPa, assuming the equation of state of platinum determined by Matsui et al.~\cite{Matsui2009}. Typically,~$\Delta P = 8 \pm 4$~GPa (Supplementary Materials section ``Pressure at melting'').

This process to determine the temperature and pressure of melting, $T_m$ and $P_m$, yields highly reproducible results. Five melting runs are carried out at~$P_m = 68 \pm 5$~GPa while measuring one side of the sample. These data are shown in Fig. \ref{fig:60GPa_It}; the other twenty-five melting runs are shown in Figs. \figSixtyGPaRightSM-\figNineNineGPaSM. For each side of each sample, plateau-like intervals occur at values of~$I^{1/4}$~within 5\% of each other and fitted temperatures are within 160 K ($\pm 80$ K) of each other (Table \tableerrors).  
%Moreover, there are no anomalies in ~$I^{1/4}$~vs. time besides the plateau-like anomalies (Figs. \ref{fig:60GPa_It}, \figSixtyGPaRightSM-\figNineNineGPaSM). 
%These plateau-like regions are apparently much more reproducible than the anomalies in temperature versus laser power documented in Fig. ?? of Shen and Lazor, Fig. ?? of Walker et al., but Sinmyo shows two kind of reproducible,  Fig. ?? of ref ??.

Including data collected from both sides of the sample (left-side and right-side), measured melting temperatures are more scattered (within $\pm 190$~K for all but one sample; within $\pm 250$~K for the sample measured with a narrow spectral range). All measured melting temperatures for each sample and starting pressure are averaged to determine~$T_m$~in a way that weights the two sides of the sample equally (Supplementary Section ``Temperature Fits at Melting''). From sample to sample, the phenomenology of these measurements is reproducible, as shown in the figures of $dI^{1/4}/dt$ vs. $T$ (Figs. \ref{fig:60GPa_It}, \figSixtyGPaRightSM-\figNineNineGPaSM). The reproducibility can also be seen in a more intuitive quantity, the rate of temperature change, $dT/dt$, versus $T$ (Fig. \ref{fig:reproducibility}). In this case, time-resolved temperature is based on a pyrometric measurement described in the Discussion section.

Plateau-like regions are also documented upon cooling in twenty-four of the thirty-three heating runs in which a sample melted (Fig. \ref{fig:60GPa_It}, \figSixtyGPaRightSM-\figNineNineGPaSM). We interpret this as freezing. The value of~$I^{1/4}$~in the plateau-like region is always slightly lower during cooling than during heating, suggesting hysteresis.

The values of melting temperature increase monotonically within uncertainties, from 2170 K at low pressure (our non-gasketed sample) to 4540 K at 107 GPa (Fig. \ref{fig:summary_data}a, Table \ref{table:Tm}). The slope,~$dT_m/dP$, decreases two-fold from~$\sim 40$~K/GPa at ambient pressure to~$\sim 20$~K/GPa at 50 to 100 GPa, but no discontinuities in slope are identified. A fit to the Simon functional form,~$T_m = T_0(P/A + 1)^{1/C}$, yields~$A = 15.1$~and ~$C = 2.60$, assuming the ambient pressure melting temperature, ~$T_0 = 2041$~K. Our measurements of $T_m$ deviate by up to 300 K from the Simon fit, so we summarize them by an error envelope of $\pm 300$~K around the Simon fit~(red shading in Fig. \ref{fig:summary}).

Before describing further experimental results, we summarize the key evidence for our melting interpretation based on the thermal emissions data alone: plateau-like regions are reproducible and reversible, and their temperatures increase monotonically with pressure. Moreover, extrapolation of our measurements to ambient pressure agrees with the known value of melting temperature, 2041 K, to within our measurement uncertainty (Fig. \ref{fig:summary}).

Further evidence that melting and freezing cause the plateau-like regions is provided by combined analysis of thermal emissions measurements with electrical and X-ray measurements. First, electrical resistance typically increases rapidly as a function of temperature during the plateau-like interval, as expected upon melting for a metal (Supplementary Materials  ``Electrical resistance across melting''; Table \tableelectrical).

Second, X-ray diffraction measurements show diminishing intensity of face centered cubic peaks and an increasingly intense diffuse background at temperatures near $T_m$ (Figs. \figXRDstack, \figBigXRD). This rules out the possibility that the latent heat of a crystal-to-crystal phase transition is responsible for the plateau-like regions, at least at the pressures where diffraction was measured near melting (35 to 60 GPa). The X-ray measurements are not used to quantify melting temperature in this study. For details, see the Discussion, the Supplementary Section ``X-ray diffraction near melting'', and Figs. \figXRDstack, \figBigXRD.

Third, the amount of electrical energy deposited during the plateau-like interval is similar to the anticipated value of latent heat plus heat lost to the surroundings. In the Supplementary Materials section ``Latent heat of melting'', we present a quantitative analysis of upper bounds on latent heat, $L_\max$, and entropy change across melting, $\Delta S_\max = L_\max/T_m$. Briefly, we divide the excess Joule heating energy required to overcome the plateau-like region, $E$, by the volume of sample that melts, $V$, times the molar density of crystalline Pt at the melting pressure and temperature, $\rho_m$. Together, $L_\max = E/V\rho_m$. Two uncertainties combine to make this a conservative upper bound on $L$: (1) the quantity $E$ is only partially corrected for heat loss to the surroundings, and (2) we propagate uncertainty in the measurement of $V$ by subtracting the uncertainty $dV$ in order to ensure $L_\max$ is an upper bound. Next, we divide by $T_m$ to calculate an upper bound to the entropy of fusion, $\Delta S_\max$. At $P_m = 34$, 68, and 86 GPa, $\Delta S_\max = 22$~to 37 J/mol/K, which is merely 2 to 3-times the ambient pressure value. This means that a modest entropy change is sufficient to explain plateau-like anomalies.

\section{Discussion}

\subsection{Melting curve of platinum}
\begin{figure}[tbhp]
	\centering
	\includegraphics[width=6.3in]{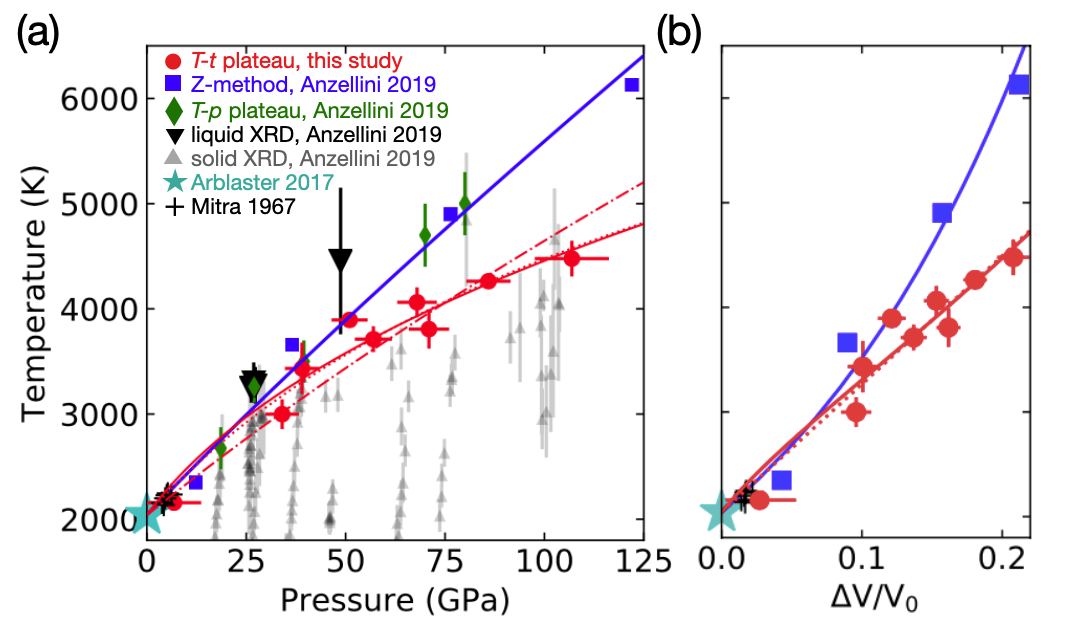}
	\caption{Melting temperature of platinum as a function of (a) pressure and (b) compression.  Experimental data from this study (red circles) are compared to experimental and computational data from Anzellini et al. \cite{Anzellini2019}, Mitra et al. \cite{Mitra1967}, and Arblaster et al. \cite{Arblaster2017}. X-ray diffraction-based identification of solid platinum and of liquid platinum from Anzellini et al. (small grey triangles and large black triangles, respectively) are consistent with our melting data, within the uncertainties. Observations of plateaus in temperature versus laser power (green diamonds) and calculations by the Z-method (blue squares) from Anzellini et al. are consistent with our melting temperatures at pressures up to 40 GPa, but inconsistent at pressures above 60 GPa. Solid curves are Simon fits to the data of this study (red) and to the Z-method calculations of Anzellini et al. (blue). Kraut-Kennedy and Lindemann fits to the data of this study are shown by dotted red and dash-dotted red curves, respectively.}
	\label{fig:summary_data}
\end{figure}

\begin{figure}[tbhp]
	\centering
	\includegraphics[width=4in]{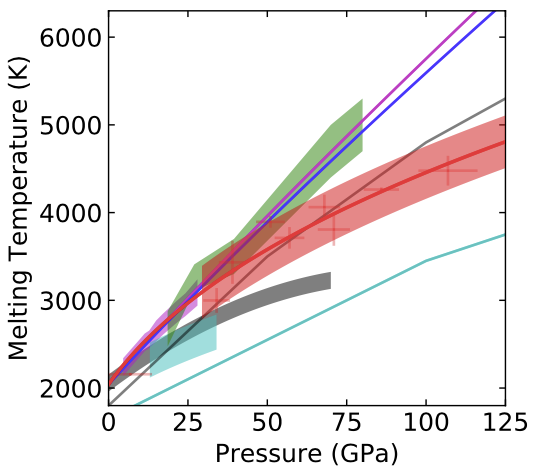}
	\caption{High pressure melting curve of platinum. Melting data of this study (red crosses), the Simon fit to the data (solid red), and an error envelope of $\pm 300$~K at pressures above 30 GPa (red shading). Past experimental studies are summarized by error envelopes: Anzellini et al. \cite{Anzellini2019} (green), Errandonea \cite{Errandonea2013} (magenta), Kavner and Jeanloz \cite{Kavner1998} (grey), Patel and Sunder \cite{Patel2018} (cyan). Theoretical results are shown in solid curves: Belonoshko and Rosengren \cite{Belonoshko2012} (magenta), Anzellini et al. \cite{Anzellini2019} (blue), Jeong and Chang \cite{Jeong1999} (grey), and Liu et al. \cite{Liu2010} (cyan).}
	\label{fig:summary}
\end{figure}

The melting temperature of platinum increases from 2041 K at ambient pressure to 3300 K at 40 GPa, in line with the steep slopes documented in Refs.~\onlinecite{Mitra1967,Errandonea2013,Anzellini2019,Belonoshko2012} (Fig. \ref{fig:summary}). Above 50 GPa, however, the slope is much shallower than reported by Anzellini et al. \cite{Anzellini2019} and Belonoshko and Rosengren \cite{Belonoshko2012}. We find~$dT_m/dP < 25$~K/GPa at all pressures from 50 to 110 GPa. This decreasing slope is expected according to the Kraut-Kennedy empirical model, which predicts that~$T_m$~depends linearly on volume, not pressure \cite{Kraut1966_PRL}. Indeed, the volume dependence of latent-heat based measurements of~$T_m$~clearly approximates a line that includes the ambient pressure melting point,~$T_0 = 2041$~K (Fig. \ref{fig:summary_data}b). The Z-method calculations could also be fitted to a line that includes ambient pressure melting, but the deviation would be $\sim$ 400 K to 500 K at 12 GPa and 122 GPa. %Instead, Anzellini et al. \cite{Anzellini2019} uses a Simon fit, which results in a maximum misfit of 300 K (Fig. \ref{fig:summary_data}b). %In other words, our data show the validity of the Kraut-Kennedy empirical model for Pt melting from 0 to 110 GPa, while the Z-method results suggests a departure from this behavior. 
It is possible that Z-method calculations which use a ``waiting time analysis'' would generate lower values of melting temperature \cite{Alfe2011,Braithwaite2019}.

Both the Lindemann and Kraut-Kennedy functions can be used to fit our melting data with one free parameter, and the Kraut-Kennedy fit has a lower root mean square deviation. Note that the Lindemann model is sometimes used with zero free parameters, using known or assumed values of the Gruneisen parameter, $\gamma_0$, and its pressure dependence, $q$, as well as an assumed value for the Lindemann parameter. Here, we use the formulation of the Lindemann model in Anderson and Isaak \cite{Anderson2000}, in which the melting temperature at ambient pressure is fixed to its known value. We fix the value of $\gamma_0$ to 2.7 and allow $q$ to be a fitting parameter, motivated by the fact that three experimental studies find similar values of $\gamma_0$ but very different values of $q$. Matsui et al. \cite{Matsui2009}, Fei et al. \cite{Fei2007},  and Zha et al. \cite{Zha2008} find $(\gamma_0, q) = (2.70, 1.1)$, (2.72, 0.5), and (2.75, 0.25 to 0.01), respectively. The Lindemann model is \cite{Anderson2000},
\begin{equation}
    T_m = T_0 \left( \frac{V}{V_0} \right) ^{2/3} \exp\left( \frac{2\gamma_0}{q}(1-(V/V_0)^q) \right)
\end{equation}
The Kraut-Kennedy model \cite{Kraut1966_PRL} has one free parameter, $C$.
\begin{equation}
    % T_m = T_0 (1+C \Delta V/V_0)
    T_m = T_0 \left( 1 + C (1- V/V_0) \right)
\end{equation}
Here, $V$ is volume, $V_0$ is the volume at ambient pressure. In both cases, we assume $T_0 = 2041$ K \cite{Arblaster2017}, and the room temperature equation of state determined by Matsui et al.~\cite{Matsui2009}. Note that here $V$ refers to values along the melting curve, as in Refs.~\onlinecite{Gilvarry1956,Anderson2000}, unlike in Ref.~\onlinecite{Kraut1966_PRL}. The best fit parameter is $q = 1.04$ for the Lindemann model and $C = 6.0$ for the Kraut-Kennedy model (Fig. \ref{fig:summary}). Note that the value $q=1.04$ is very close to 1.10, the value found in the equation of state study of Matsui et al. \cite{Matsui2009}.  
%Note that the fitted value, $q=1.04$, is so close enough to the value determined by Matsui et al. \cite{Matsui2009}, $q=1.10$, that a single iteration is sufficient to reach self-consistency between the equation of state and the Lindemann melting models. 
Nevertheless, the root mean square deviation of Kraut-Kennedy fit to data is smaller than that of the Lindemann fit (190 K compared to 270 K), so we prefer the Kraut-Kennedy fit. Conveniently, the Kraut-Kennedy and Simon fits are nearly identical over the pressure range 0 to 120 GPa (Fig. \ref{fig:summary_data}). We highlight the Kraut-Kennedy fit in this manuscript rather than the Simon fit because it uses one free parameter rather than two. 

Despite the agreement of our data to the melting curves of Refs. \cite{Mitra1967,Errandonea2013,Anzellini2019,Belonoshko2012} at pressures below 40 GPa, our melting data are discrepant with previous experimental and computational results in several ways (Figs. \ref{fig:summary_data}-\ref{fig:summary}). In the pressure range from 40 to 80 GPa, the range of slopes of our melting curve, 25 to 18 K/GPa, is inconsistent with the 40 K/GPa slope reported in Anzellini et al. \cite{Anzellini2019}. We associate the discrepancy to a difference in melt detection method. The only experimental constraint with~$< 1000$~K uncertainty for the melting curve of Anzellini et al. at pressures above 40 GPa is the saturation in temperature as the power of a continuous-wave laser is steadily increased, a phenomenon that is not specific to melting. Rather, it can be caused by surface reflectivity changes or movement of material within a solid or liquid phase \cite{Geballe2012}. In the pressure range 50 to 80 GPa, our melting temperatures are 300 to 1500 K higher than those reported in Lo Nigro \cite{LoNigro2011} and in Kavner and Jeanloz \cite{Kavner1998}, in which melting was determined by X-ray diffraction and visual observation, respectively. In the pressure range 80 to 120 GPa, our melting temperatures are 600 to 1000 K lower than those calculated by the Z-method \cite{Anzellini2019,Belonoshko2012}.

Our melting curve is consistent the X-ray diffraction data of Anzellini et al. \cite{Anzellini2019} (Fig. \ref{fig:summary_data}a), and with our own X-ray diffraction data (Supplementary Section ``X-ray diffraction near melting''), albeit within~$\sim1000$~K uncertainties in determination of $T_m$ from most of the X-ray diffraction measurements. At 30 GPa, Anzellini et al. reports a narrowly constrained melting temperature based on X-ray diffraction, and it agrees with the latent heat melting temperatures documented here (Fig. \ref{fig:summary_data}). At 50 GPa, Anzellini et al. reports the transition from solid to liquid diffraction in the range 3040 K to 5130 K, the low-temperature-end of the error bar for solid diffraction to the high-temperature-end of the error bar for liquid diffraction. This~$\pm 1000$~K range spans our latent-heat melting data at~$50 \pm 10$~GPa  (Fig. \ref{fig:summary_data}a). At 60 to 100 GPa, Anzellini et al. reports solid X-ray diffraction only, with error bars that overlap our melting data in all cases. The X-ray diffraction data from the present study are described in detail in Supplementary Section ``X-ray diffraction near melting''. Briefly, we measured temperature and X-ray diffraction during the pulsed electrical heating of four samples during five heating runs to peak temperatures above the quantity $(T_m - 1000$ K), where $T_m$ is the melting temperature based on our latent heat criterion. One heating run shows no kink in the plot of diffuse scattering intensity versus temperature (Fig. \ref{fig:XRD_stack}o), while the other four all show kinks within~$\pm 1000$~K of~$T_m$~(Fig. \ref{fig:XRD_stack}c,f,i,l). Only two of the runs showed kinks within $\pm 300$ K of the latent heat melting temperature (Fig. \ref{fig:XRD_stack}c,f). In summary, there is agreement to within~$\pm 1000$~K between the latent heat melting temperatures and the X-ray diffraction data from this study and from Anzellini et al. \cite{Anzellini2019}. To reduce the uncertainty in X-ray determination of melting, it may be important to invent new ways to contain a molten sample at pressures above 40 GPa and temperatures above 3000 K for longer times, and/or to use more intense X-ray sources.

\subsection{Reproducibility of electrical heating and latent heat detection}

The shape of the latent heat anomaly in~$I^{1/4}$~versus~$t$~is reproducible at all pressures from~$6.8 \pm 6.8$~GPa to~$106.9 \pm 9.3$~GPa. Figs. \ref{fig:60GPa_It} and~\figSixtyGPaRightSM-\figNineNineGPaSM~show thirty plateau-like regions in which the quantity~$dI^{1/4}/dt$~consistently decreases temporarily before increasing again. But rather than rely on ten figures to document the reproducibility of the new melt-identification method, we can further process the data and generate a single, easy-to-read figure. 

We convert intensity, $I$, to temperature, $T$, using a two-step process that assumes constant emissivity during each heating run.First, we use spectroradiometry, as in the determination of $T_m$ described above. Planck functions are fit to thermal emissions spectra averaged over a single time-interval, using two free parameters, temperature and emissivity. The time interval is the plateau-like melting interval if exists, and the most intense~$\sim 1$~$\mu$s otherwise. Second, fixing the fitted value of emissivity, $\epsilon$, we use pyrometry to determine temperature. We numerically solve for the following equation for temperature, $T$, at each time, $t$:
\begin{equation}
    \int_{\lambda_1}^{\lambda_2}{\epsilon \times \textrm{Planck}( T, \lambda) d\lambda} = \int_{\lambda_1}^{\lambda_2}{I_\sam(\lambda,t) d \lambda} 
\end{equation}
Here, ``Planck'' is the Planck function for blackbody radiation, $\lambda_1 = 450$~nm, and $\lambda_2=860$~nm for all data sets except the data set with~$P_m = 39$~GPa, for which~$\lambda_1 =500$~and~$\lambda_2 = 660$~nm. The measured intensity,~$I_\sam$, is corrected for optics and camera efficiency by the usual calibration with a standard tungsten lamp. 

\begin{figure}[tbhp]
	\centering
	\includegraphics[width=6.3in]{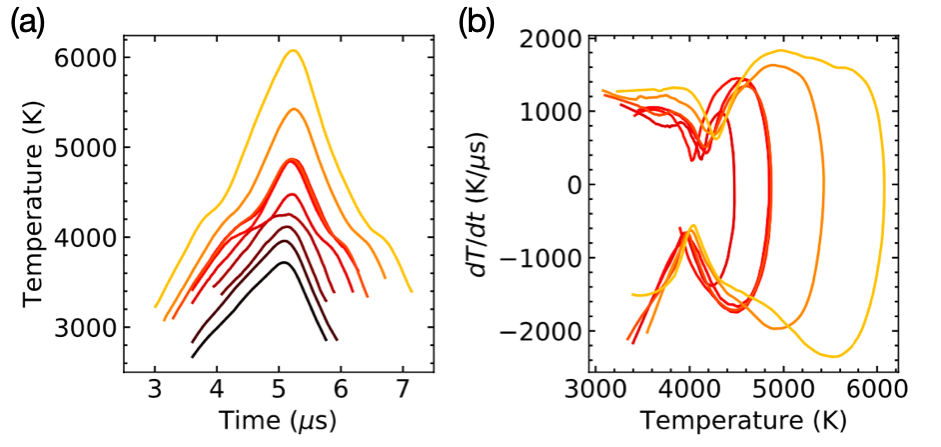}
	\caption{Temperature evolution of platinum heated from room temperature at~$60 \pm 3$~GPa during the same nine sets of heating runs shown in Fig. \figsixtyGPaIt. Emissivity is fitted to emissions spectra from a narrow region of each curve. (a) Temperature, $T$, versus time, $t$. (b) Heating and cooling rates, $dT/dt$, versus $T$.
	}
	\label{fig:60GPa_Tt}
\end{figure}

\begin{figure}[tbhp]
	\centering
	\includegraphics[width=3in]{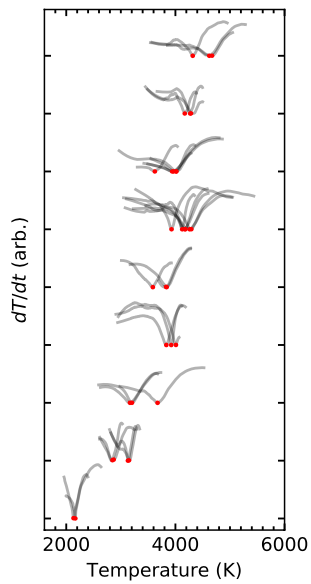}
	\caption{Signature of latent heat absorption during all melting runs documented in this study. The rate of temperature change, $dT/dt$, is plotted against temperature, $T$ (grey curves). The dip in each curve is caused by the latent heat of melting. Values of $dT/dt$ are scaled and offset so that each cluster of curves reflects all the melting data generated by heating from a single starting pressure. 
	}
	\label{fig:reproducibility}
\end{figure}

The temperature evolution is shown in Fig. \ref{fig:60GPa_Tt}a for nine heating runs starting at~$P_0 = 60$~GPa. The temperature-time function has been filtered through to a second-order Savitzky-Golay filter with the same timescale, $\tau$ used in plots of $I^{1/4}$ vs. t. Then temperature is differentiated with respect to time, and a second, identical Savitzky-Golay filter is used to reduce the noise in $dT/dt$. The resulting values of $dT/dt$ versus $T$ are plotted in Fig. \ref{fig:60GPa_Tt}b for the melting data at $P_0 = 60$~GPa ($P_m = 68$~GPa), and truncated to show only the melting region in Fig. \ref{fig:reproducibility} for all thirty-three melting runs at~$P_m = 7$~to~107 GPa. 

\begin{figure}[tbhp]
	\centering
	\includegraphics[width=6.3in]{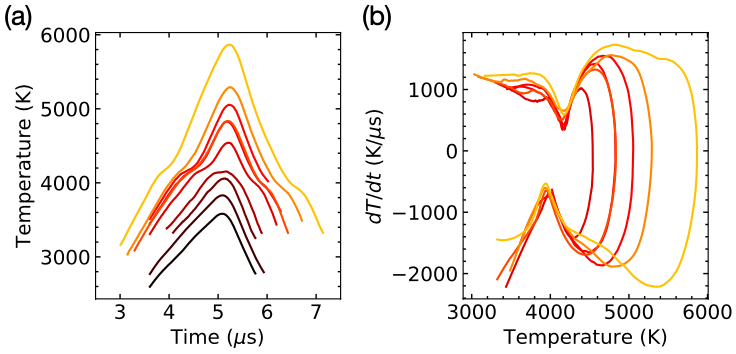}
	\caption{Temperature evolution of platinum heated from room temperature at~$60 \pm 3$~GPa, assuming a fixed emissivity of 0.58. The data are from the same nine sets of heating runs shown in Figs. \figsixtyGPaIt, \ref{fig:60GPa_Tt}. (a) Temperature, $T$, versus time, $t$. (b) Heating and cooling rates, $dT/dt$, versus $T$.
	}
	\label{fig:60GPa_Tt_fixed_b1}
\end{figure}

Fig. \ref{fig:reproducibility} shows the signature of melting in all data used to generate the melting curve of platinum to 107 GPa. Latent heat absorption manifests as clear dips in the plots of $dT/dt$ versus $T$. Moreover, the dips in $dT/dt$ are transient in all cases; temperature increases again after latent heat is absorbed. The variation in temperature of~$dT/dt$~minima in Figs. \ref{fig:60GPa_Tt} and \ref{fig:reproducibility} seems to be caused by the uncertainty in Planck fits. If instead of using a two parameter Planck fit, we fix the value of emissivity for several streak camera images collected from one side of one sample, we find much less variation. An example is shown in Fig. \ref{fig:60GPa_Tt_fixed_b1}. By fixing emissivity to the 0.58, the mean of emissivities fitted using two-parameter Planck fits,~$dT/dt$~minima range from 4140 to 4180 K, which is seven times less variation than the range of~$dT/dt$~minima found when emissivity is allowed to vary from image to image (4020 to 4290 K). In other words, the precision of our measurement of plateaus in~$I^{1/4}$~propagates to~$\pm 20$~K uncertainty in temperature, but the precision of the temperature measurement itself is only~$\pm 140$~K since it is affected by uncertainties in $I^{1/4}$ and emissivity. The reproducibility of measurement of plateau temperature from side to side and sample to sample is~$\pm 300$, suggesting this is the accuracy of the melting curve. 

The latent heat plateaus documented in this study are different than plateaus documented in studies that use continuous laser heating. First, the observation interpreted as a melting ``plateau'' in temperature versus laser power rarely show temperatures increasing again after the plateau region \cite{Shen1995,Lord2009,Lord2010,Lord2014,Dewale2007,Dewaele2010,Kimura2017,Sinmyo2019,Parisiades2019}. Second, some studies show that the shape of the temperature-laser power anomaly is not reproducible, with sample temperature increasing after a plateau during some heating runs and decreasing after a plateau in other heating runs \cite{Lord2009}. This variability can be caused by changes in the sample surface, which causes changes in the efficiency of laser-absorption \cite{Geballe2012}. Whereas the properties of a metal's surface can change at temperatures below or above the melting temperatures and can result in more or less absorption, the latent heat of melting is only absorbed upon melting and only released upon freezing. This may crucial be to the reproduciblity of the plateau-like anomalies in the data presented here.

The relatively high reproducibility of heating platinum to a liquid state may be useful for future studies, since containing a liquid in a diamond cell is a major technical challenge. In some cases, pulsed resistively heated samples can be repeatedly heated to well above their melting points. The two most outstanding heating runs were performed on one sample at $P_m = 51$ GPa, and one sample at $P_m = 71$ and 86 GPa. The former was melted several hundred times while monitoring X-ray diffraction and electrical resistance. The latter was reproducibly melted nine times, reaching more than 1000 K above the melting temperature during one pulse. In both cases, the stress state inside the gasket hole was relatively isotropic, as evidenced by the lack of increasing hole diameter upon compression at room temperature prior to the melting experiment. By contrast, in cases where the gasket hole visibly expanded during compression, which suggests significant axial stress, the melted segment of the sample seemed to narrow. This narrowing caused the peak temperature to increase when repeatedly heating with a constant driving voltage, $V_\bank$. 

For several samples, resistivity increases during melting provide a second indication of melting, and can be identified at every melting repetition using an oscilloscope. This melt identification technique could be used in an automated feedback loop to reproducibly heat a sample to slightly above its melting temperature. In fact, a manual feedback-loop was employed during some of the X-ray diffraction measurements. We manually adjusted~$V_\bank$ during sequences of 1000 melting shots so that the onset of melting, as observed by a kink in 4 point probe voltage, occurred $\sim 2$~$\mu$s before the end of the heating pulse.

\subsection{Latent heat versus other sources of anomalous temperature change}

This is likely the first time that latent heats have been detected in static compression experiments at pressures~$> 20$~GPa, despite several claims of latent heat detection in diamond cells. Most previous studies have suffered from slow heating timescales ($\gg \mu$s for diamond-cell-sized samples), which causes thermal conduction out of the sample to dominate the temperature evolution.

Five alternative explanations for the plateau-like regions are possible, but unlikely. First, the plateau-like regions could be caused by a solid-solid phase transition to a high-temperature solid with entropy nearly as high as that of liquid platinum. In this scenario, the latent heat of melting would be dwarfed by the latent heat of the solid-solid transition, obscuring the melting plateau while highlighting the solid-solid plateau. Two pieces of evidence make this unlikely. First, such a solid is not predicted for platinum at high pressure, and not observed for any elemental metal at ambient pressure. Even solid Fe and Ti, whose entropies increase substantially upon solid-solid transitions above 1000 K, still maintain entropies that are significantly smaller than their liquids \cite{Hultgren1963}. Second, the X-ray diffraction data at 35 to 55 GPa reveal no crystalline peaks besides fcc platinum, even when the temperature of the heated region of the sample exceeds the temperature of the plateau-like region.

A second alternative explanation is that the latent heat of fusion of KCl causes the plateau-like regions. This scenario would require very large values for thermal conductivity of KCl so that the sample's surface temperature evolution is significantly affected by heat absorption in KCl. In reality, we expect the sample's surface temperature to be much more strongly affected by the highly conductive platinum than the low thermal conductivity KCl in part because of the contrast in thermal conductivities and in part because Joule heat is deposited in the platinum only. Still, thermal modeling would be required to quantify possible effects of the latent heat of KCl on the temperature evolution of the platinum surface.

Third, an approximately 10-fold increase in thermal conductivity of the KCl medium would decrease the slope of temperature versus time, as modeled in Fig. 7 of Geballe and Jeanloz \cite{Geballe2012}. However, the decrease would be maintained at all temperatures above the transition temperature. To reproduce the plateau-like observations, a sequence of transitions would be required in which the thermal conductivity of KCl increased~$\sim10$-fold and then decreased~$\sim 10$-fold. This sequence would be unprecedented for an alkali halide at any pressure, to the best of our knowledge. 

A fourth alternative explanation is that platinum transitions to a low resistivity phase at high temperature, causing a plateau in Joule heating power. This would lead to a plateau-like region in the same way that reflectivity increases have been shown to cause plateau-like regions in models of pulsed laser heating \cite{Geballe2012,GoncharovComment,Montoya2012}. However, we infer the opposite from our electrical data: resistance increases with temperature by~3~to~10\%~in the plateau-like region for several of the samples (Table \ref{table:electrical}), and no decrease in resistance with increasing temperature is detected for any sample.

A fifth possibility is that a near-melting phenomenon, such as fast recrystallization, surface premelting, or bulk premelting, causes the plateau-like regions. Fast recrystallization of several metals has been detected at temperatures that are 100s of K below melting in diamond cells, using sequences of $\sim 1$ second X-ray diffraction images (e.g. Refs. \cite{Anzellini2013,Parisiades2019,Stutzmann2015}). However, recrystallization at the $\sim$~1 second timescale would not affect our microsecond-timescale melting experiments. Premelting would introduce anomalously high specific heat at temperatures below melting, biasing the temperature measurement of the plateau-like region to lower values. However, we do not know of any prediction of bulk pre-melting for platinum. If premelting were restricted to surface (i.e. less than a few nanometers), we would expect a very small downward shift in the temperature of the plateau-like region since the heat capacity of the sample's interior has a much larger effect than the surface heat capacity on temperature evolution at our heating timescale; a 1~$\mu$s timescale yields a thermal diffusion lengthscale of $\sqrt{D\tau} = 7$~$\mu$m at $\sim 50$~GPa and 2000 K, assuming thermal conductivity from McWilliams et al. \cite{McWilliams2015}, the equation of state from Matsui et al. \cite{Matsui2009}, and a heat capacity of three times the gas constant. 

\section{Conclusions}
Using the new method, detection of melting and freezing by latent heat is reproducible and reversible. Plateaus-like regions in thermal emission intensity versus time are reproducible to~$\pm 5\%$ intensity, which is equivalent to~$\pm 20$~K. Planck fits to determine temperature are reproducible to~$\pm 140$~K. Reproducibility is~$\pm 190$~K among both surfaces of all samples, excluding the one sample measured with a narrow spectral range. Moreover, the shape of plateau-like anomalies in $I^{1/4}$ versus time is reproducible for both surfaces of all samples at all pressures. These successes suggest that the new technique is an excellent candidate for further studies of melting and freezing experiments on a wide range of metals at megabar pressures and temperatures to at least 5000 K.

The melting curve of platinum measured by the latent heat method is steeply sloped from ambient pressure to~$\sim 40$~GPa. At higher pressure the slope, $dT_m/dP$, decreases smoothly to $\sim 15$~K/GPa at 100 GPa, departing from the results of \textit{ab initio} Z-method calculations published so far. As a function of compression, on the other hand, melting temperature increases linearly over the 0 to 20\% range of compression studied here, allowing a good fit to the Kraut-Kennedy empirical model with fit parameter C = 6.0.

\clearpage

\begin{table}[ht]
\begin{tabular}{		c			c		}	
								
\hline\hspace{5mm} 	$P_m$ (GPa)	\hspace{5mm} &\hspace{5mm} 	$T_m$ (K)\hspace{5mm}  \\
	\hline
$	6.8	\pm	6.8	$&$	2160	\pm	20	$ \\
$	34	\pm	4.2	$&$	3000	\pm	140	$ \\
$	39	\pm	4.3	$&$	3430	\pm	250	$ \\
$	51	\pm	4.5	$&$	3890	\pm	70	$ \\
$	57	\pm	4.7	$&$	3710	\pm	120	$ \\
$	68	\pm	5	$&$	4060	\pm	140	$ \\
$	71	\pm	5.1	$&$	3810	\pm	190	$ \\
$	85.9	\pm	5.6	$&$	4260	\pm	30	$ \\
$	106.9	\pm	9.3	$&$	4480	\pm	170	$ \\
\hline								
\end{tabular}																
\caption{Melting points}
\label{table:Tm}
\end{table}

\clearpage
\begin{acknowledgements}
We thank Suzy Vitale for milling the platinum samples with the FIB. We thank Maddury Somayazulu, Paul Goldey, Mike Walter, and Viktor Struzhkin for productive discussions. Portions of this work were performed at GeoSoilEnviroCARS (The University of Chicago, Sector 13), Advanced Photon Source (APS), Argonne National Laboratory. GeoSoilEnviroCARS is supported by the National Science Foundation – Earth Sciences (EAR – 1634415) and Department of Energy- GeoSciences (DE-FG02-94ER14466). This research used resources of the Advanced Photon Source, a U.S. Department of Energy (DOE) Office of Science User Facility operated for the DOE Office of Science by Argonne National Laboratory under Contract No. DE-AC02-06CH11357.

\end{acknowledgements}
% \nocite{*}
%\bibliography{Pt_melting_PRM}% Produces the bibliography via BibTeX.

\clearpage
\resetlinenumber
\begin{center}
\textbf{\large Supplemental Materials for ``A latent heat method to detect melting and freezing of metals at megabar pressures''}
\end{center}
%%%%%%%%%% Prefix a "S" to all equations, figures, tables and reset the counter %%%%%%%%%%
\setcounter{equation}{0}
\setcounter{figure}{0}
\setcounter{table}{0}
\setcounter{section}{0}
\setcounter{page}{1}
 
\makeatletter
\renewcommand{\theequation}{S\arabic{equation}}
\renewcommand{\thefigure}{S\arabic{figure}}
\renewcommand{\thetable}{S\arabic{table}}
\renewcommand{\thesection}{S\arabic{section}}

\begin{figure}[tbhp]
	\centering
	\includegraphics[width=5in]{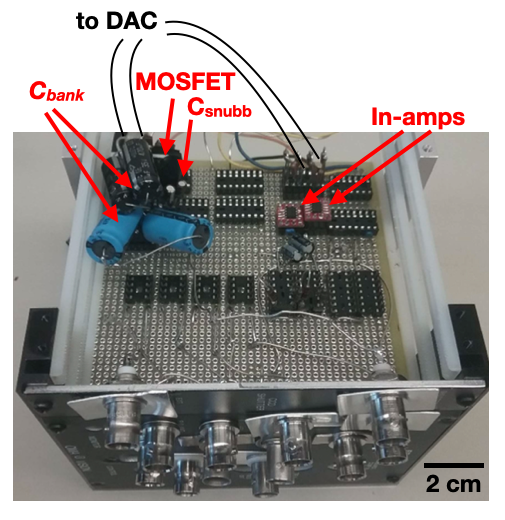}
	\caption{Photo of the electrical pulser. The viewing angle is slightly rotated from a bird's-eye view. The box, vector board, and dual in-line pin connectors are mostly empty of components since they were constructed for four pulsers to be used in parallel and in series. The components plugged into the dual-in-line pin connectors in the photo are those needed for the single pulser experiments used in this study and shown schematically in Fig. \figSetup.}
	\label{fig:pulser_photo}
\end{figure}

\begin{figure}[tbhp]
	\centering
	\includegraphics[width=6in]{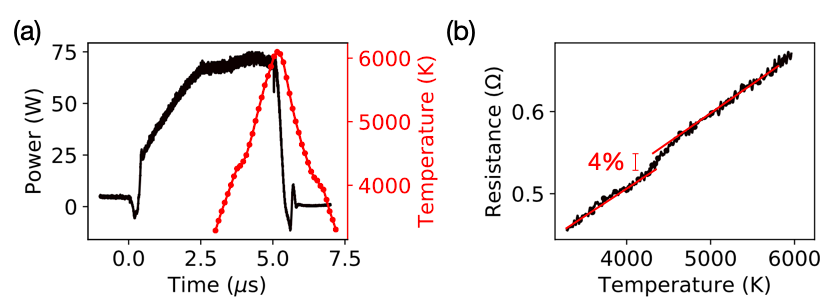}
	\caption{Power and resistance of platinum sample \#6 heated from room temperature at~$60 \pm 3$~GPa during the heating run that achieved the highest temperature. (a) Power (left axis) and temperature (right axis) versus time. (b) Resistance versus temperature during heating (black), and linear guides-to-the-eye (red) used to estimate the resistance increase across melting. A second-order Savitzky-Golay filter with timescale $\tau = 0.2$~$\mu$S is used to filter the raw measurements of voltage and current before computing power and resistance.}
	\label{fig:60GPa_electrical}
\end{figure}

\begin{figure}[tbhp]
	\centering
	\includegraphics[width=6.3in]{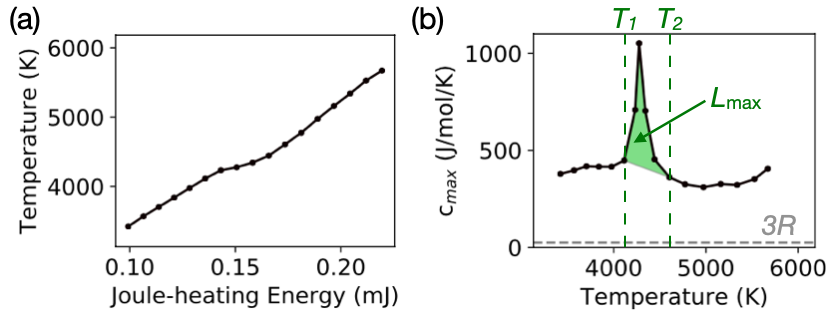}
	\caption{Calorimetry of platinum sample \#6 heated from room temperature at~$60 \pm 3$~GPa during the heating run that achieved the highest temperature. (a) Temperature versus Joule heating energy, $E_J$. (b) Upper bounds of specific heat capacity of the sample, $c_\max$ (black), and of latent heat, $L_\max$ (green shaded area) (Eqs. \ref{eq:cmax} and \ref{eq:Lmax}). The width of the plateau like region is marked as $\Delta T_p$. The grey dashed line shows the  Dulong-Petit limit, $c = 3R$. A second-order Savitzky-Golay filter with 50 K temperature scale is used to smooth the derivative $dE_J/dT$.}
	\label{fig:60GPa_latent}
\end{figure}

The text of this supplement elaborates on experimental and analytical methods. Tables \ref{table:errors} and \ref{table:electrical} detail results for each melting and freezing run. Figs. \ref{fig:pulser_photo}-\ref{fig:photos} document details that are mentioned in the main text, including pressure measurements at high temperature, spatial distributions of temperature, electrical measurements, and latent heat analysis. Figs. \ref{fig:60GPa_SM}-\ref{fig:99GPa_SM}, along with Fig.~\figsixtyGPaIt~of the main text, show that the new melt criterion is reproducible for all platinum samples.

%\normalsize
% \small
% \footnotesize
% \scriptsize
\begin{flushleft}
\linespread{1.0}\footnotesize

Table \ref{table:errors}: Pressure and temperature of melting and freezing for each individual melting run.

Table \ref{table:electrical}: Sample dimensions, resistance measurements, and calorimetric properties.

Fig. \ref{fig:pulser_photo}: Photo of the electrical pulser.

Fig. \ref{fig:60GPa_electrical}: Power and resistance of platinum heated from room temperature at~$60 \pm 3$~GPa.

Fig. \ref{fig:60GPa_latent}: Calorimetry of platinum heated from room temperature at~$60 \pm 3$~GPa.

Fig. \ref{fig:60GPa_Tx}: Spatial distribution of thermal emissions and of temperature of platinum heated from room temperature at~$60 \pm 3$~GPa.

Fig. \ref{fig:XRD_stack}: Stack of X-ray diffraction patterns from sample \#4 at 39 to 47 GPa.

Fig. \ref{fig:Big_XRD}: X-ray diffraction from 30 to 60 GPa.

Fig. \ref{fig:photos}: Photos of six samples at room temperature and during heating.

Fig. \ref{fig:60GPa_SM}: Time-resolved thermal emissions of platinum heated from room temperature at~$60 \pm 3$~GPa. %sample \#6

Fig. \ref{fig:0GPa_SM}: Time-resolved thermal emissions of platinum heated from room temperature at~$<0.1$~GPa. %sample \#1

Fig. \ref{fig:26GPa_SM}: Time-resolved thermal emissions of platinum heated from room temperature at~$26 \pm 1.3$~GPa. %sample \#2

Fig. \ref{fig:31GPa_SM}: Time-resolved thermal emissions of platinum heated from room temperature at~$31 \pm 1.6$~GPa.  %sample \#3

Fig. \ref{fig:43GPa_SM}: Time-resolved thermal emissions of platinum heated from room temperature at~$43 \pm 2.2$~GPa.  %sample \#4

Fig. \ref{fig:49GPa_SM}: Time-resolved thermal emissions of platinum heated from room temperature at~$49 \pm 2.4$~GPa.  %sample \#5

Fig. \ref{fig:63GPa_SM}: Time-resolved thermal emissions of platinum heated from room temperature at~$63 \pm 3.2$~GPa.   %sample \#7

Fig. \ref{fig:78GPa_SM}: Time-resolved thermal emissions of platinum heated from room temperature at~$78 \pm 3.9$~GPa.   %sample \#6

Fig. \ref{fig:99GPa_SM}: Time-resolved thermal emissions of platinum heated from room temperature at~$99 \pm 5$~GPa.   %sample \#8

\end{flushleft}
\clearpage

\section{Methods}

\subsection{Sample preparation}
We use five steps to load the platinum samples in a diamond cell. First, we use standard methods to prepare a Zha-style symmetric diamond anvil cell \cite{Zha2017} with tungsten carbide seats, standard-cut anvils with 100 to 300~$\mu$m-diameter culets, and a gasket with an electrically insulating insert. The gasket is made from a piece of 250~$\mu$m-thick rhenium and a~$\sim 0.5$~mm chunk of cBN-epoxy mixture. The rhenium is pre-indented to 20 to 30 GPa, and the whole culet area is removed by laser drilling. The cBN-epoxy (hereafter referred to as ``cBN'') is a hard tack made from mixing~$0.25$~$\mu$m-grain size cubic boron nitride with ND 353 Epotek epoxy, and curing it at room temperature. The chunk of cBN is placed on the culet inside the indented rhenium, and pre-indented to 20 to 30 GPa. A ring of epoxy is added around the top of the indent in the cBN to protect it from chipping off and allowing hence to avoid short-circuits. The ring dimensions are~$\sim 30$~$\mu$m-thick,~$\sim 600$~$\mu$m inner diameter, and~$\sim 900$~$\mu$m outer diameter.

Second, we prepare a set of four outer electrodes held on an aluminum sleeve that slides into a 15 mm-diameter bore surrounding the tungsten carbide seat on the piston side of the diamond anvil cell (see the bore in Fig. S1 of Zha et al. \cite{Zha2017}). Each electrode is cut from either 18~$\mu$m-thick platinum foil or a spool of 125~$\mu$m-diameter platinum wire, and soldered to a 1/64 to 1/32 inch-thick copper clad board that is glued to the aluminium sleeve with epoxy. Each electrode is shaped to a narrow point~$\sim 50$~$\mu$m-wide, and bent so that its tip is within~$\sim 50$~$\mu$m of the edge of the culet. 

Third, we prepare four inner electrodes by pressing short pieces of 25~$\mu$m-diameter platinum wire into the cBN on the culet and nearby facets so that they are fully embedded and make electrical contact with the outer electrodes. 

Fourth, we drill a hole of diameter~$\sim0.4$-times the culet diameter and fill it with pieces of KCl and platinum. The KCl pieces are chipped off of slabs of KCl that have been pressed to~$10\pm5$~$\mu$m thickness. The platinum pieces are cut from pieces of wire (99.95\% purity, 25~$\mu$m-diameter, Alfa Aesar 7440-06-4) that has been pressed to~$10 \pm 5$~$\mu$m-thickness. At least five pieces of platinum and several pieces of KCl are stacked so that when the diamond cell is closed, a central piece of platinum of 5 to 30~$\mu$m-width is separated from both anvils by KCl layers and electrically connected to the four outer electrodes. This central piece is the platinum sample that is eventually melted. Platinum cutting is performed by hand with a razor blade for~$P_0 =$~31, 43, and 63 GPa, and into barbell-shaped pieces with a FIB for~$P_0 =$~26, 49, 60, 78, 99 GPa. The barbell shape facilitates electrical connection by providing a wider target than in the case of a razor-cut piece of platinum. Once a sample is loaded to high pressure, however, there is no apparent difference in the FIB-ed samples and razor-cut samples.

Fifth, we place the whole diamond cell in a vacuum oven for at least 45 minutes at 120$^\circ$C, with anvils spaced 0.2 to 0.5 mm apart to allow evaporation of H$_2$O from the KCl. We purge the oven with argon. Then, within 10 seconds, we remove the cell and compress the sample to at least 2 GPa. After the cell cools to room temperature, we compress to the target pressure and wait for at least 30 minutes for stresses to relax. Finally, we measure pressure,~$P_0$, using the Raman shift of the strained diamond anvil \cite{Akahama2006}.

\subsection{Electrical connections}
To connect the electrical leads from inside the diamond cell to the pulser and measurement electronics, we use a sequence of strain-relieved wires of increasing diameter. Briefly, we use solder and two sets of barrel-connectors (CUI PJ-202BH and PP3-002B) to connect eight segments of 3 to 8 cm-long, 0.2 mm-diameter copper wire and eight segments of 5 to 10 cm-long, $\geq 1$~mm-diameter stranded copper to the circuitry of Fig. \figSetup. Strain relief for the 0.2 mm-diameter wires are provided by pieces of copper clad board that are mechanically attached to the diamond cell. The 10 cm-long segments of stranded copper wire are used for the voltage measurement across the sample. The shorter, 5 cm-long segments are used to deliver the heating current.

\subsection{Electrical Pulser}
The electrical pulser is centered around a power MOSFET, Fairchild Semiconductor part number FQP30N06L (Fig. \figSetup). The electrical architecture is a simplified version of the primary side of the transformer described in Giesselmann et al. \cite{Giesselmann2005}. It is also inspired by the vacuum-tube pulsers used for the microsecond pulsed-calorimetry experiments at one atmosphere \cite{Gallob1986,Cezairliyan1987}. Our homemade capacitor bank consists of four aluminum electrolyte capacitors with 470~$\mu$F capacitance and 35 V maximum rating, which are connected half in parallel and half in series to make a 470~$\mu$F, 70 V capicator bank. Three 1~$\Omega$~resistors are connected in parallel to make the reference resistor, whose resistance is measured to be ~$R_\refR =0.29$~$\Omega$ using a four point probe measurement. The RC timescale for capacitor discharge is therefore at least $R_\refR C_\bank = 136$~$\mu$s, which allows steady pulses for all experiments presented here. No flyback diode is used, unlike the circuit in Giesselmann et al. \cite{Giesselmann2005}. A snubber capacitor is used to limit unwanted oscillations. The capicitance value of 16~$\mu$F was found to be optimal. Surprisingly, aluminum electrolyte capacitors, which have a polarity, were much more effective in limiting oscillations than non-polar ceramic capacitors with similar capacitance. We hypothesize that the aluminum electrolyte capacitor acts as a diode in parallel with a capacitor, and thereby provides some of the functionality of a flyback diode. 

The electrical measurement probes are centered around a pair of in-amps (AD8429). Each of their four inputs is protected by an identical voltage divider made from two resistors (1 to 10 k$\Omega$, 1\% accuracy) that reduce the voltage at the input by 3.5-fold or 11-fold so that it falls within the~$\sim 15$~V range of the in-amp. This protection is crucial; one in-amp was apparently damaged prior to the 107 GPa melting run, causing massive distortion of the four-point probe voltage measurement. 

All the components described above are attached to the vector board shown in Fig. \ref{fig:pulser_photo} by dual in-line pin connectors, so that testing different combinations of components can be done without soldering, including the replacement of in-amps that may be damaged. 

Delay generators drive the gate of the MOSFET directly with 3 to 5 V pulses. For comparison, a MOSFET driver, an ignitron tube, and a thyrotron tube are used in Giesselmann et al. \cite{Giesselmann2005}, Cezairliyan and McClure \cite{Cezairliyan1987}, and Gallob et al. \cite{Gallob1986}, respectively. It is possible that the delay generators used here provide marginal current to the gate of the MOSFET, generating unwanted fluctuations in pulse power. 

\section{Analysis}

First, this Supplementary section details the analytical procedure used to (A) identify the melting interval in each streak camera data set, and (B) determine its temperature. Next, section (C) describes two sources of potential uncertainty that have not been included in the analysis of melting temperature. Section (D) describes determination of pressure at the melting temperature. Finally, sections (E)-(H) describe four analytical procedures that constrain properties beyond the melting curve: (E) the time-dependence of temperature, (F) electrical resistance across melting, (G) latent heat of melting, and (H) atomic structure near melting.

\subsection{Melting identification}

To determine the melting temperature of platinum at high pressure, the key data are the streak camera images of time-resolved, wavelength-resolved thermal emissions (e.g. Fig. \figraw). In each image, the intensity of thermal emissions increases as a function of time until the MOSFET is switched off, at which point the intensity decreases. At high heating powers, two pronounced anomalies are evident in the plot of average intensity, one upon increasing intensity and one upon decreasing intensity (Fig. \figraw b). These anomalies are present in all curves of intensity vs. time above a  threshold in driving voltage for each sample (e.g.~$V_\bank \geq 11.8$~V in Fig. \figsixtyGPaIt). For ease of viewing, we plot~$I^{1/4}$, the fourth root of thermal emission intensity per cumulative microsecond of streak camera exposure, a proxy for temperature, in Fig. \figsixtyGPaIt b. We also plot its derivative,~$\frac{dI^{1/4}}{dt}$, versus~$I^{1/4}$~in Fig. \figsixtyGPaIt c. 

Note that before computing the derivative, we filter the raw~$I^{1/4}$~data with a second order Savitzky-Golay lowpass filter and truncate it to avoid propagating noise from the region of negligible signal. After computing the derivative, we apply the same Savitzky-Golay filter. Un-filtered data is shown in grey curves overlaid by the filtered data in colored curves. Table \ref{table:errors} lists the filter timescale:~$\tau =0.3$~to~1.2~$\mu$s for the high-pressure samples, and~$\tau = 8$~$\mu$s for the non-gasketted sample.

\subsection{Temperature fits at melting}

To determine the thermodynamic melting temperature at each pressure studied, we assume that (1) the sample radiates and transmits out of the diamond cell as a greybody, (2) the optical path transmits all colors to the detector with an efficiency that is calibrated by the standard procedure (i.e. measurement of a standard tungsten lamp), (3) the temperatures during the plateau-like regions (i.e. the centers of the anomalies) are the melting and freezing temperatures of the platinum sample, and (4) the melting temperature, not the freezing temperature, is the equilibrium temperature of the solid-liquid transition. Assumption (1), the greybody assumption, is typical in diamond-cell experiments and has been discussed extensively (e.g. Benedetti and Loubeyre \cite{Benedetti2004}). Similarly, assumption (2) is typically assumed. It is true for sufficiently achromatic optical systems measuring sufficiently broad hot spots, though $\sim 100$s of K errors may be possible in diamond cell heating experiments \cite{Giampaoli2018}. Here, we use apochromatic objective lenses (Mitutoyo NIR 20x Long Working Distance; see McWilliams et al.\cite{McWilliams2015} for details), and measure hotspots that are~$\sim 8$~to 17~$\mu$m wide (FWHM of temperature; e.g. Fig. \ref{fig:60GPa_Tx}). Assumption (4) is almost always true at ambient conditions, even during~$\mu$s-timescale pulsed heating experiments \cite{Ubbelohde1978}. Assumption (3) is likely correct within the~$\sim 100$~K uncertainties estimated below because of two factors that limit temperature gradients in our experiments. First, a positive temperature coefficient of resisitivity causes the hot cross section of the platinum sample to increase in temperature in a relatively uniform way. Even though the surfaces cool rapidly due to conduction through the thin KCl to the diamond anvils, a negative feedback loop causes the surface to be Joule-heated more than the interior since electrons seek the path of least resistance. Second, the high thermal conductivity of platinum relative to the KCl medium causes the majority of the temperature decrease from the interior of the sample to~$\sim 300$~K at the anvil surface to occur in the KCl medium \cite{McWilliams2015, Geballe2020}. Together, these two phenomena limit temperature gradients, making the platinum's surface temperature very similar to the temperature of the platinum material near the surface, which in turn, has a dominant effect on the temperature evolution of the surface. 

For each heating run, we determine the melting and freezing temperatures,~$T_m$~and~$T_f$, by fitting a greybody curve to the thermal emissions spectra during the plateau-like region of each anomaly (Fig. \figsixtyGPaIt d). The center of each plateau-like region is identified by an extremum in~$\frac{dI^{1/4}}{dt}$. The width of the plateau-like region is set equal to half the filtering timescale,~$\tau /2$. We fit to the Planck function times emissivity, $\epsilon$, 
\begin{equation}
	I_\corr = \epsilon \frac{2 h c^2}{\lambda^5} \frac{1}{e^{\frac{hc}{\lambda k_B T}}-1}
	\label{eqn:greybody}
\end{equation}
and require~$\epsilon \leq 1$. Here,~$c$~is the speed of light,~$h$~is Planck's constant,~$k_B$~is Boltzmann's constant. The fitting parameters are temperature,~$T$, and emissivity,~$\epsilon$. Fitted temperatures are reproducible for both melting and freezing. For each side of the sample, the temperatures are reproducible to within their~$\pm 50$~to~$\pm 200$~K precision regardless of how many times the sample has been melted and regardless of how much power is used melt the sample (Fig. \figsixtyGPaIt, Table \ref{table:errors}).

Note that the precision of $T_m$ and $T_f$ is the quadrature sum of two sources: (1) the standard error calculated with Python's function ``scipy.optimize.curve\_fit'' when fitting our data to the Planck function (Eq. \ref{eqn:greybody}), 
and (2) the error due to fluctuations in the streak camera, which we estimate by the function
\begin{equation}
	\sigma_T = 0.05 \times T \times \sqrt{2.5/N} \times (1+ 0.00035 (T-2685)) 
	\label{eqn:streak_noise}
\end{equation}
  Here,~$N$~is the number of counts per pixel in the wavelength dimension and $I_\corr$ is the intensity of thermal emissions corrected for the system response by measuring a standard tungsten lamp.  In typical spectroradiometry experiments in diamond-cells, the standard error of the fit is dominant because of the high signal-to-noise ratio achieved with CCD cameras for exposures of milliseconds to seconds when temperatures are sufficiently high ($> 1500$~K). By contrast, here we use a less stable streak camera, which we choose for its outstanding speed and gain. We determine Eq. \ref{eqn:streak_noise} by measuring emission from a standard tunsten lamp at 2685 K several times at each of 10-, 20-, 50-, and 100-sweeps per streak image.    

 For each sample and pressure, we average individual measurements of $T_m$ and $T_f$ by a two-step process that weights both sides of the sample equally. First, we average the temperatures measured on each side independently,
 \[
 T_{m,\side} = \sum_{i \in \side}{w_{m,i}T_{m,i}}
 \]
 for individual measurements,~$T_{m,i}$, and weights~$w_i = dT_{m,i}^{-2}$,~ where~$dT_{m,i}$~is the uncertainty on an individual measurement. The error in~$T_{m,\side}$~is the variance-weighted standard error:
 \[
 dT_{m,\side} = \sqrt{ \sigma^2 \chi^2_\nu}
 \]
 Here,~$\sigma$ is the the standard error of the weighted mean,
 \[
 \sigma = \frac{1}{\sum{w_i}}
 \]
 and~$\chi_\nu^2$ is the reduced chi-squared,
 \[
 \chi^2_\nu= \sum{w_i(T_{m,i} - T_{m,\side})^2}/(N-1)
 \]
where~$N$~is the number of measurements of melting. The analogous weighted averages are computed for~$T_{f,\side}$~and~$dT_{f,\side}$.

Second, we average~$T_{m,\DS}$~and~$T_{m,\US}$~with equal weights,~$T_m = (T_{m,\DS} + T_{m,\US})/2$, and propagate uncertainty by the formula,
\[
dT_m = \sqrt{ \left( \frac{T_{m,\DS} - T_{m,\US}}{2}\right) ^2+\left( \frac{1}{dT_{m,\DS}\ ^2}+\frac{1}{dT_{m,\US}^2}\right) ^{-1}}
\]
We calculate~$T_f$~and~$dT_f$~in the analogous way.
  
Table \ref{table:errors}  shows that melting temperatures from left and right side are within their mutual uncertainty for several samples and starting pressures, but not for all of them. Two likely causes of the deviations are (1) slight deviations from greybody behavior (i.e., wavelength-dependence of emissivity) of the platinum-KCl interface that differ depending on the surface's history and (2) difference in thermal losses to the KCl and cold regions of the platinum itself. Nevertheless, the total uncertainties, including this deviation, are relatively small (Table \ref{table:errors}). 
Melting temperatures are reproducible from left-side to right-side within~$\pm 20$~to~$\pm 190$~K at all pressures from 34 to 107 GPa, with one exception: the measurement $P_m =39$~GPa, $T_m = 3430 \pm 250$~K, for which the spectral range of thermal emissions measurement was limited to 500 to 660 nm. 

We document the identification of melting and freezing, and the corresponding temperature fits at all pressures in Figs. \ref{fig:0GPa_SM}-\ref{fig:99GPa_SM}. For each sample and pressure, plateau-like regions are documented upon melting in all measurements during which the pulsed-heating power is sufficient to superceed the melting temperature~by~$\sim 10\%$. Albeit, three other samples not reported here did not exhibit plateau-like regions during heating. In these cases, the hotspots were extremely short lengths of platinum ($l_\sam < 5$~$\mu$m), suggesting that insufficient volumes of sample were melted to generate plateau-like regions with signal greater than noise. A fourth sample that did not exhibit plateau-like regions was sample \#8 after it was compressed from 99 to 120 GPa. Note that shiny platinum surface darkened after repeated melting runs from~$P_0 =99$~GPa. The darkening may be due to a reaction at the Pt-KCl surface, though no new peaks appeared in X-ray diffraction measurements.

\subsection{Potential sources of added uncertainty in~$T_m$}
Two sources of uncertainty may bias the values of melting temperature reported here. First, the color temperature of thermal emissions might not equal the true temperature of the platinum surface. Although ambient pressure platinum emits at all wavelengths with approximately constant emissivity, this might not be true at high pressure near melting. Although KCl is transparent to all wavelengths of light at ambient pressure, this might not be true at high pressure and temperature (e.g. Arveson et al. \cite{Arveson2018}). Both uncertainties exist in nearly all melting experiments of platinum at~$> 10$~GPa. Albeit, one study \cite{Kavner1998} varied the pressure medium, reducing the possibility that absorption of pressure-media biases color temperatures.

Second, as discussion in the main text, it is possible, though unlikely, that the surface temperature of platinum during the plateau-like region is significantly lower than the temperature of the material inside the platinum sample. The plausibility of this uncertainty could be tested by finite element models that couple the heat equation and Ohm's law.

\subsection{Pressure at melting}
To determine the pressure at which melting occurred, we measure pressure at room temperature,~$P_0$, and add an estimate of heating-induced pressure. First, we average the pressure at room temperature before and after heating to determine~$P_0$. Pressure typically drops by 0 to 3 GPa during a set of one to ten melting repetitions. Pressures at ambient temperature are determined by the Raman-edge of the diamond anvils \cite{Akahama2006}. The values are listed in Table \ref{table:errors}. Raman-derived pressures from the two diamond anvils typically match to within 2 GPa, even at 99 GPa. For simplicity, we assume a 5\% uncertainty in~$P_0$. 

Second, for all pressures from 10 to 90 GPa, we add the following estimate of heating-induced pressure:
\begin{equation}
\Delta P = \frac{1}{2}  P_{th} (T') \pm \frac{1}{4} P_{th} (T')
\label{eqn:deltaP}
\end{equation}
where~$T' =$~min($T$,2500 K), and~$P_{th}$~is the Mie-Gruneisen-Debye thermal pressure at constant volume. The equation of state is from Matsui et al. \cite{Matsui2009}, where the Vinet formalism is used, and the parameters are~$V_0 = 60.38$~\AA$^3$/unit cell,~$K_0 = 273$~GPa,~$K' = 5.2$,~$\gamma_0 = 2.7$,~$\theta_0 = 230$~K,~$q = 1.1$,~$n = 4$~atom/unit cell. This function for $\Delta P$ explains nearly all the unit cell volumes measured before, during, and after pulsed electrical heating of crystalline platinum at high temperature (Fig. \ref{fig:Big_XRD}). An intuitive understanding of Eq. \ref{eqn:deltaP} comes from comparing it to the isobar ($\Delta P = 0$), and isochore ($\Delta P = P_{th}(T)$). The permissible range of~$\Delta P $~spans from 25\% to 75\% of the isochore up to 2500 K, and is constant at~$ T \geq 2500$~K.

For the data at~$P_0 = 1$~bar and 99 GPa, we use the analogous function with a larger uncertainty that spans from the isobar to the isochore,~$\pm \frac{1}{2} P_{th} (T')$. The reason for added uncertainty at 99 GPa is that the anvils have beveled culets, whereas Eq. \ref{eqn:deltaP} is fit to data with 200 to 300~$\mu$m-diameter flat culets. At ambient pressure, the sample was not gasketed, the sample and thermal insulation were~$\sim 10$-times thicker, and the heating pulse was~$\sim 10$-times longer.

\subsection{Electrical resistance across melting}
Several of the melting experiments reveal changes in electrical resistance across the melting transition. Increases in total resistance during melting are documented from~3~to~10\%~for the samples melted at 7 to 86 GPa (Fig. \ref{fig:60GPa_electrical}, Table \ref{table:electrical}). These values represent lower bounds for the resistivity change across melting, because the resistance measurement includes cold regions of the sample that are not melted during the experiment. For comparison, at ambient pressure, platinum's resistivity increases from 0.6 to 0.9~$\mu \Omega$-m upon melting, an increase of 40\% \cite{Wilthan2004}.

\subsection{Latent heat of fusion}

Our measurements before, during, and after the plateau-like region provide information about the specific heat capacity of the high temperature solid, the latent heat of fusion, and the specific heat capacity of the liquid. However, all measurements are biased by conductive heat losses to the surrounding KCl medium and to relatively cold regions of platinum. We refer to the combination of both regions as the ``addenda''. Nevertheless, we can use our measurements plus a few simple assumptions to derive lower and upper bounds on the latent heat of fusion, $L$, and hence the entropy of fusion, $\Delta S_m = L/T_m$. Our final result is that the entropy of fusion at 86 GPa is between 0.5 and 22 kJ/mol, or between 0.05- and 2- times the ambient pressure entropy of fusion (Table \tableelectrical). While this is a wide range, it is the first experimental constraint for Pt at high pressure, and paves the way for narrower constraints in the future. 

Upper bounds on specific heat and latent heat are determined directly from the measurements of temperature and Joule-heating power. The total heat capacity of sample plus addenda is the temperature derivative of the time-integral of the Joule-heating power:  
\begin{equation}
C_\sam + C_\addenda = \frac{d E_J}{dT} = \frac{d \int{P_J dt}}{dT} = \frac{d\int{IV dt}}{dT}
\end{equation}
The time-evolution of temperature is determined by the two-step method described in the Discussion section of the main text: a Planck fit to the plateau-like region's thermal emissions spectra, followed by a pyrometry using the wavelength-integrated thermal emissions. The pyrometry step assumes emissivity is fixed to the value determined in the Planck fit.

\begin{figure}[tbhp]
	\centering
	\includegraphics[width=6.3in]{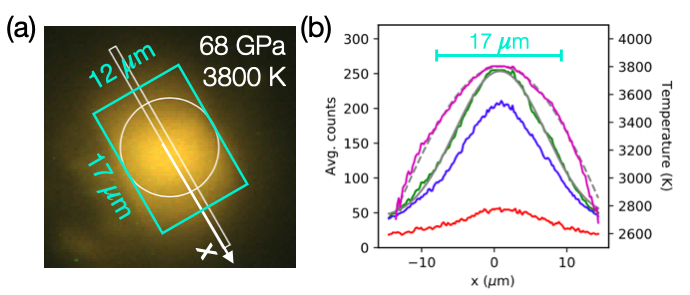}
	\caption{Spatial distribution of thermal emissions and of temperature of platinum sample \#6 heated from room temperature at~$60 \pm 3$~GPa during two heating runs to~$> 3000$~K,~$68 \pm 5$~GPa. (a) Image of thermal emissions on a color CCD camera during a heating pulse up to $\sim 3800$~K. White circle marks the 12~$\mu$m-diameter area from which thermal emissions enter the streak camera. White box and arrow marked ``x'' shows the area and x-axis used for (b). Cyan box marks the estimated area of sample that is melted during the plateau-like region of the highest temperature heating pulse (i.e., the green-shaded area in Fig. \ref{fig:60GPa_latent}b). (b, left axis) Red, green, and blue curves show average thermal emissions in the white box in (b), measured in red, green, and blue pixels of the color CCD. (c, right axis) Magenta shows the temperature profile inferred by fitting the green curve by the one-color method of Geballe et al. \cite{Geballe2020}. Grey lines show fits used in the process of determining temperature with the one-color method. The 17 $\mu$m scale bar shows the length of sample that melts during the plateau-like region of the highest temperature heating pulse.}
	\label{fig:60GPa_Tx}
\end{figure}

There are many ways to estimate an upper bound on the specific heat capacity, $c_\max$. Here we use a simple estimate: 
\begin{equation}
c_\max = \frac{C_\sam + C_\addenda}{N_\min}
\label{eq:cmax}
\end{equation}
where $N_\min$ is a lower estimate on the number moles of sample that are heated to within $(T_2-T_1)$ of the peak temperature, where $T_2$ and $T_1$ are the temperatures at the beginning and end of the plateau-like region (Fig. \ref{fig:60GPa_latent}). The number of moles is the product of length, width, thickness, and the molar density assuming the equation of state of platinum from Matsui et al. \cite{Matsui2009}: 
\begin{equation}
    N_\min = (l_\sam - dl_\sam)  (w_\sam-dw_\sam) (t_\sam - dt_\sam) \rho
\end{equation}
The thickness of the sample is measured by optical interferometry (Table \ref{table:electrical}). The width and length of the sample are measured by optical microscopy (Figs. \ref{fig:photos},\ref{fig:60GPa_Tx}). Specifically, the length of sample heated to within~$(T_2-T_1)$~of the peak temperature is estimated by the one-color method shown in Fig. \ref{fig:60GPa_Tx} and described in Supplementary Eqs. (S3-S4) of Geballe et al. \cite{Geballe2020}. Uncertainties in $l_\sam$ and $t_\sam$ are much larger than the uncertainty in $w_\sam$, so we set $dw_\sam = 0$.

Finally, we calculate an upper bound on latent heat by integrating the amount of excess Joule-heat required to overcome the plateau-like region. 
\begin{equation}
L_\max = \int_{T_1}^{T_2}{ c_\textrm{max}dT} - \frac{c_\max(T_2)+ c_\max(T_1)}{2}(T_2-T_1)
    \label{eq:Lmax}
\end{equation}

Estimates of lower bounds require two assumptions. First, we assume the specific heat of the solid at high temperature is the Dulong-Petit limit, $3R$ where $R$ is the gas constant. Second, we assume that $L/c \geq L_\max/c_\max$ for the upper bounds calculated above. This inequality is true for the simulation results shown in Figs. 3 and 6 of Geballe and Jeanloz \cite{Geballe2012}, where $L/c = 561$~K~$\geq \Delta T = L_\max/c_\max$. Combining these two assumptions and solving for $L$,
\begin{equation}
    L \geq \frac{L_\max}{c_\max} c =  \frac{L_\max}{c_\max} \times 3R
    \label{eq:Lmin}
\end{equation}.

All together, $ L_\max \geq L \geq L_\textrm{min}$ for $ L_\max$ and $L_\textrm{min}$ given by Eqs. \ref{eq:Lmax} and \ref{eq:Lmin}. The resulting upper and lower bounds derived from measurements of melting at 34 to 86 GPa are listed in Table \ref{table:electrical}, along with the bounds for entropy change across melting,~$\Delta S_m = L/T_m$.

In all cases, the ranges of estimates for~$L$~and~$\Delta S_m$~span the ambient pressure values,~$L = 22\pm2$~kJ/mol and~$\Delta S_m = 11\pm1$~J/mol/K \cite{Wilthan2004}. For example, at~86~GPa,~$2.0 \leq L \leq 89$~kJ/mol and~$0.5 \leq \Delta S_m \leq 22$~J/mol/K. The modest value of this upper bound,~$L_\max$, is evidence that latent heat is a plausible cause of the plateau-like region.

By contrast, in many previous diamond-cell heating experiments, unexpectedly large latent heats would be needed to explain the plateau-like regions; see Refs. \onlinecite{Geballe2012,GoncharovComment,Houtput2019} for analysis. One crucial difference here is the~$\mu$s-timescale of heating and temperature measurements, compared to second-timescale experiments in most of the works analyzed in Geballe and Jeanloz \cite{Geballe2012}. One important difference compared to pulsed-laser heating work (e.g. Deemyad and Silvera \cite{Deemyad2008} and Zaghoo et al. \cite{Zaghoo2016}) is that here the samples are heated more uniformly because of the internal resistive heating method.

A major improvement in the accuracy of latent heat and resistivity estimates would result from better control of the spatial extent of sample that melts. In the present experiments, 8 to 17 $\mu$m-long segments of the central platinum strips melt, leaving~$\sim 50$~to 80\% of the central strips un-melted. In cases where shear forces visibly narrow part of the central platinum strip, this is the region that melts (e.g. samples \#2, \#5). In other cases, the melting region is near the center of the platinum strip (e.g. samples \#3, \#4, \#6, \#7). It is possible that by suitable choice of gasket thickness and precise control of sample and insulator shapes, shear forces could be minimized, leading to larger spatial extents of melting in future experiments. On the other hand, the width of sample that melts is well-controlled, because the full width always appears to heat uniformly in these experiments (Fig. \ref{fig:photos}). 

\subsection{X-ray diffraction near melting}
\begin{figure}[tbhp]
	\centering
	\includegraphics[width=3in]{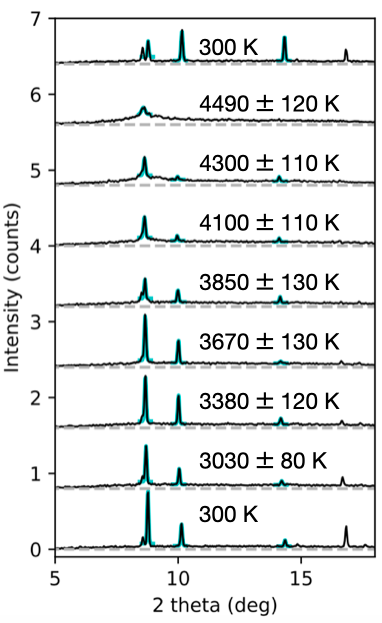}
	\caption{Stack of X-ray diffraction patterns from sample \#4 at 39 to 47 GPa during sets of 100 heating pulses (100 $\mu$s cumulative X-ray exposure time). Cyan curves mark Gaussian fits to three diffraction peaks of fcc platinum, (100), (200), and (220). Grey dashed lines mark the offsets used in the stack.}
	\label{fig:XRD_stack}
\end{figure}

\begin{figure}[tbhp]
\begin{subfigure}
	\centering
	\includegraphics[width=5in]{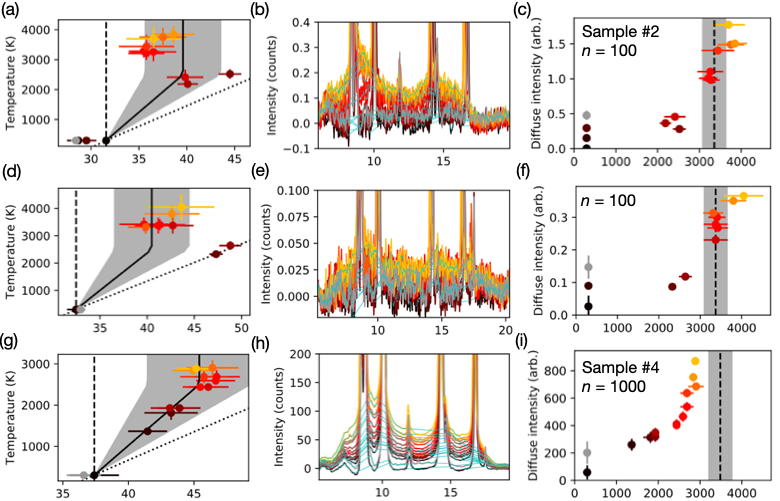}%5 in
\end{subfigure}
\begin{subfigure}
	\centering
	\includegraphics[width=5in]{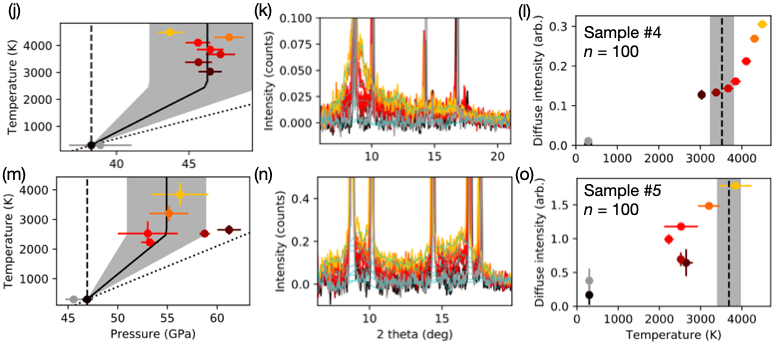}
\end{subfigure}
	\caption{		\linespread{1.0}\selectfont{}
	% \fontsize{11.5}\selecfont{}
X-ray diffraction from 30 to 60 GPa. Each row shows results from a sample heated $n$-times per diffraction pattern. $n$ is listed in the legend, along with the label ``Sample \#2'', \#4, or \#5, for the three samples that generated both streak camera and X-ray data. Each diffraction pattern is represented by one color (black, grey, or dark red through yellow). (Left Column) Temperature-Pressure conditions reached in the final 1 $\mu$s of the resistive heating pulse, when the X-ray data was collected. Circles show pressure inferred from the platinum equation of state \cite{Matsui2009}, and the average temperature of left-side and right-side temperature measurements. Errors in pressure are propagated from the standard deviation in the fcc lattice parameter determined from each of three lattice planes: (111), (200) and (220). Errors in temperature are the quadrature sum of the misfit error and the difference between left-side and right-side. The isobar and isochore are shown as dashed and dotted lines, respectively. The piecewise linear function we use to estimate heating-induced pressure is shown in the solid black line with grey error envelope. (Center Column)  Diffraction patterns (colors) and diffuse scattering (cyan), which is estimated by truncating sharp diffraction peaks. The X-ray energy is 37 keV. (Right column) Total diffuse intensity, calculated by the integral of cyan curves of the Center Column, as a function of temperature. Uncertainties in diffuse intensity are calculated by varying the limits of truncation of the sharp peaks. The black dashed line shows the melting temperature determined by fitting latent heat melting data to the Simon function; grey shading shows a $\pm 300$~K uncertainty envelope, matching Fig. \figsummary.}
	\label{fig:Big_XRD}
\end{figure}

\begin{figure}[tbhp]
\begin{subfigure}
	\centering
	\includegraphics[width=5in]{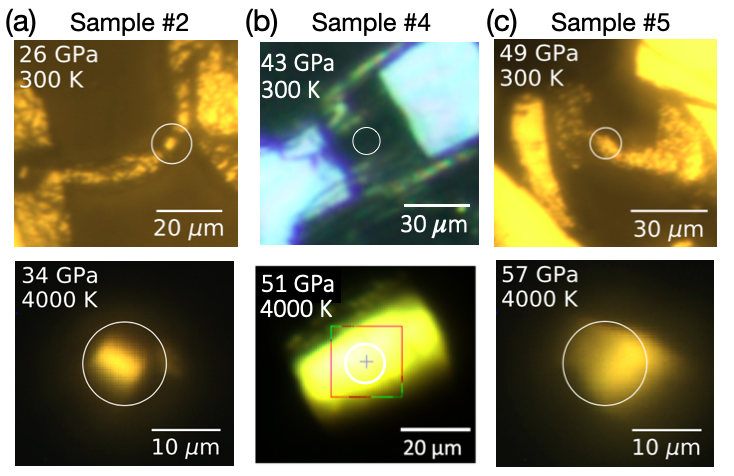}%5 in
\end{subfigure}
\begin{subfigure}
	\centering
	\includegraphics[width=5in]{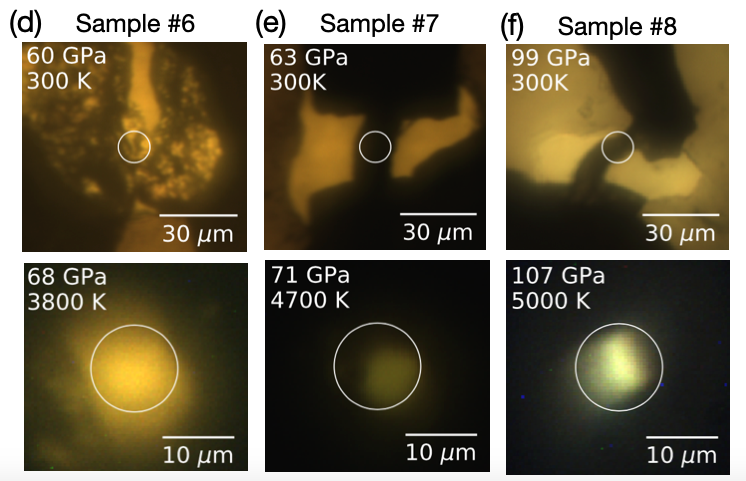}
\end{subfigure}
	\caption{		\linespread{1.0}\selectfont{}
    Photos of six samples at room temperature (top panels) and during heating to the temperature labeled (bottom panels). White circles show the 12~$\mu$m-diameter region of sample from which thermal emissions enter the streak camera. The red square marks a 20 x 20 $\mu$m region of the sample at GSECARS.}
	\label{fig:photos}
\end{figure}

\begin{figure}[tbhp]
	\centering
	\includegraphics[width=5in]{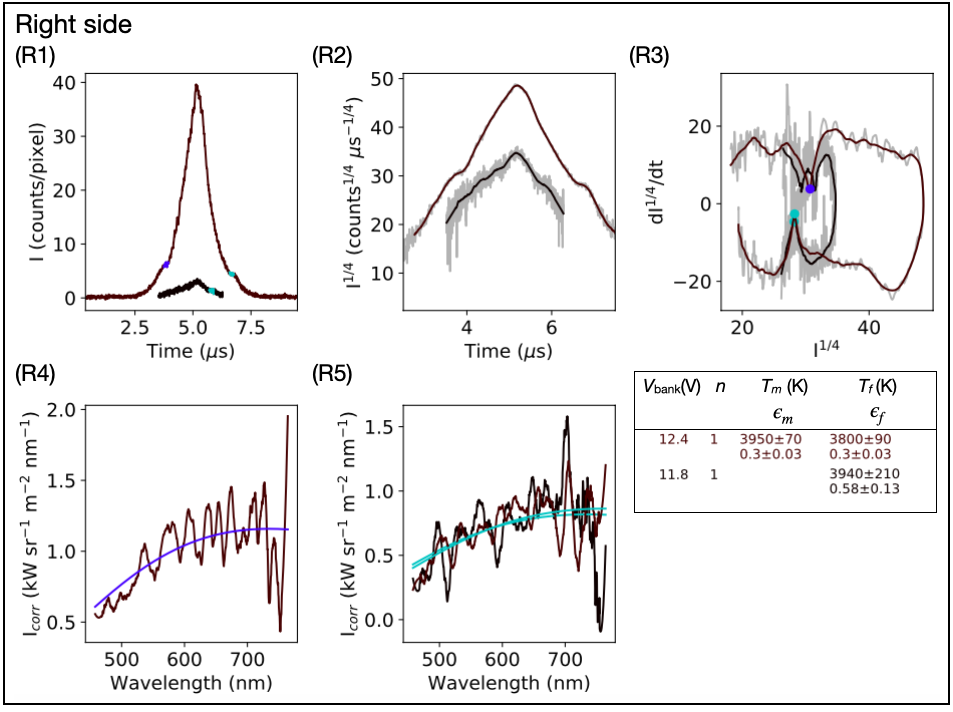}
	\caption{		\linespread{1.0}\selectfont{}
	% \fontsize{11.5}\selecfont{}
	Time-resolved thermal emissions of platinum sample \#6 heated from room temperature at~$60 \pm 3$~GPa to past its melting point at ~$4060 \pm 140$~K at~$68 \pm 5$~GPa. All measurements are performed on the right side of the sample, making this figure the complement of Fig. \figsixtyGPaIt. (R1) Average counts along the vertical dimension of the streak camera CCD. (R2) Fourth-root of average counts per microsecond, a proxy for temperature. Noisy grey curves show un-smoothed data, $I^{1/4}$, while colored curves show smoothed data, $I^{1/4}_s$. (R3) Time-derivatives, $\frac{dI^{1/4}_s}{dt}$ (grey), and smoothed time derivatives, $\frac{dI^{1/4}_s}{dt}_s$ (colors). The smoothing function is a second-order Savitzky-Golay filter with timescale~$\tau = 0.4$~$\mu$s for both $I^{1/4}_s$ and $\frac{dI^{1/4}_s}{dt}_s$. The minima during heating (blue circles) and maxima during cooling (cyan circles), are interpreted as melting and freezing. The corresponding times,~$t_{\melt}\pm \tau/2$~and~$t_\freeze \pm \tau/2$ are marked in blue and cyan in (R1), and used for the temperature fits in (R4) and (R5). (R4) Planck fits (Eq. \ref{eqn:greybody}; blue) to thermal emissions spectra during melting (warm colors). Fit parameters, $T_m$ and $\epsilon_m$, are listed in the legend. (R5) Planck fits (cyan) to thermal emissions spectra during freezing (warm colors).  Fit parameters, $T_f$ and $\epsilon_f$, are listed in the legend.}
	\label{fig:60GPa_SM}
\end{figure}

\begin{figure}[tbhp]
	\centering
	\includegraphics[width=5in]{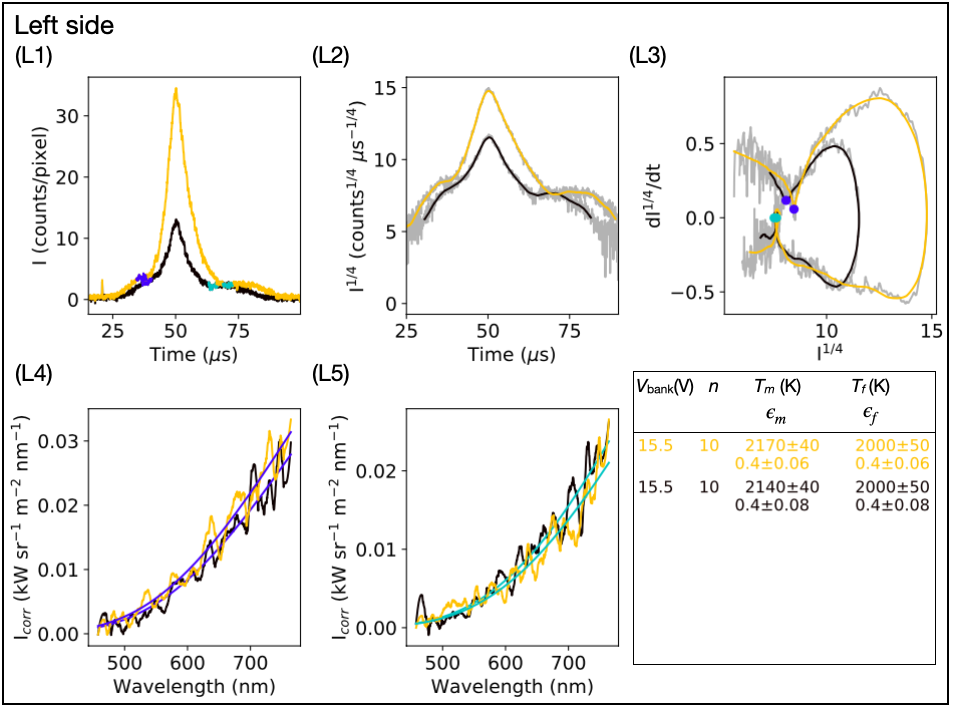}
	\caption{		\linespread{1.0}\selectfont{}
	% \fontsize{11.5}\selecfont{}
Time-resolved thermal emissions of platinum sample \#1 heated from room temperature at~$<0.1$~GPa to past its melting point at ~$2160 \pm 20$~K,~$6.8 \pm 6.8$~GPa. The filtering timescale used in (L2) and (L3) is ~$\tau=8$~$\mu$s. See caption of \ref{fig:60GPa_SM} for all other details.}
	\label{fig:0GPa_SM}
\end{figure}

\begin{figure}[tbhp]
\begin{subfigure}
	\centering
	\includegraphics[width=5in]{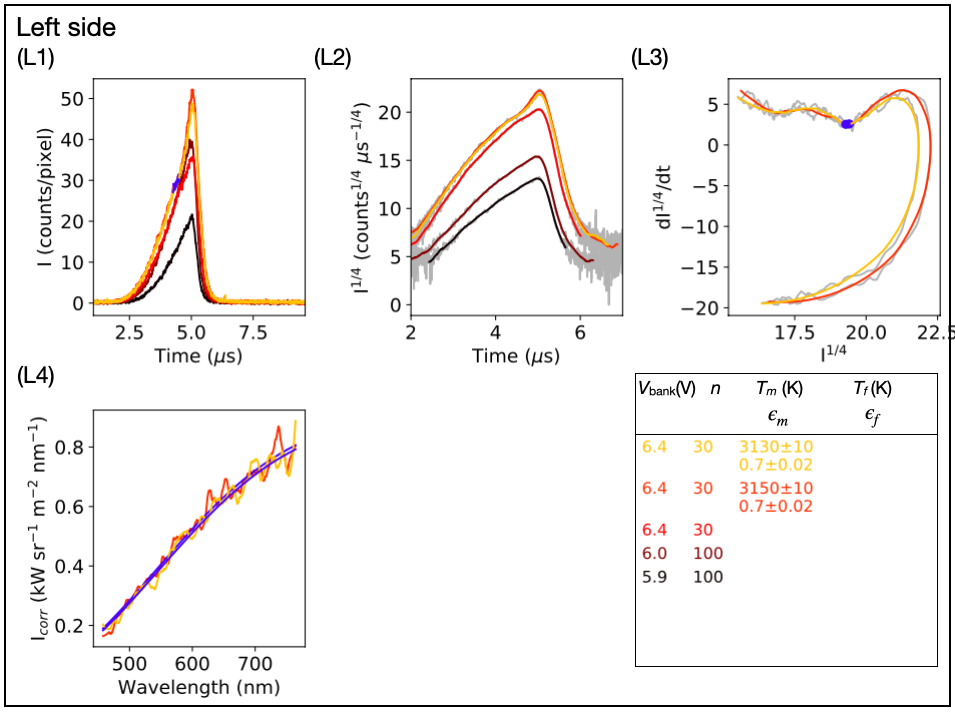}%5 in
\end{subfigure}
\begin{subfigure}
	\centering
	\includegraphics[width=5in]{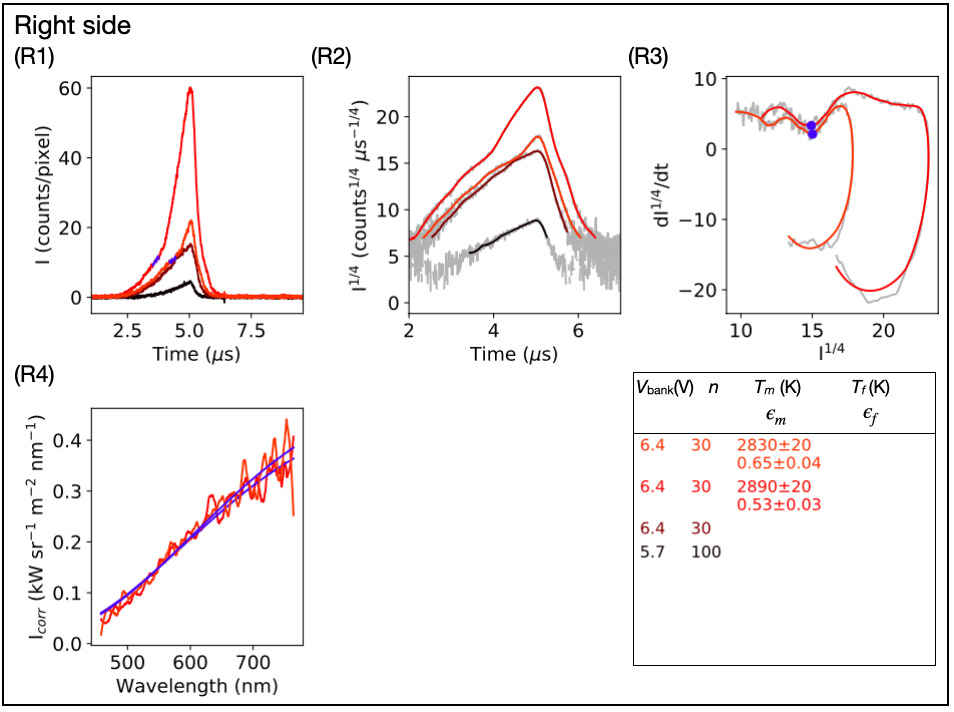}
\end{subfigure}
	\caption{		\linespread{1.0}\selectfont{}
	% \fontsize{11.5}\selecfont{}
Time-resolved thermal emissions of platinum sample \#2 heated from room temperature at~$26\pm 1.3$~GPa to past its melting point at ~$3000 \pm 140$~K, ~$34 \pm 4.2$~GPa. The filtering timescale used in (L2), (L3), (R2), and (R3) is ~$\tau=0.6$~$\mu$s. See caption of \ref{fig:60GPa_SM} for all other details.}
	\label{fig:26GPa_SM}
\end{figure}

\begin{figure}[tbhp]
\begin{subfigure}
	\centering
	\includegraphics[width=5in]{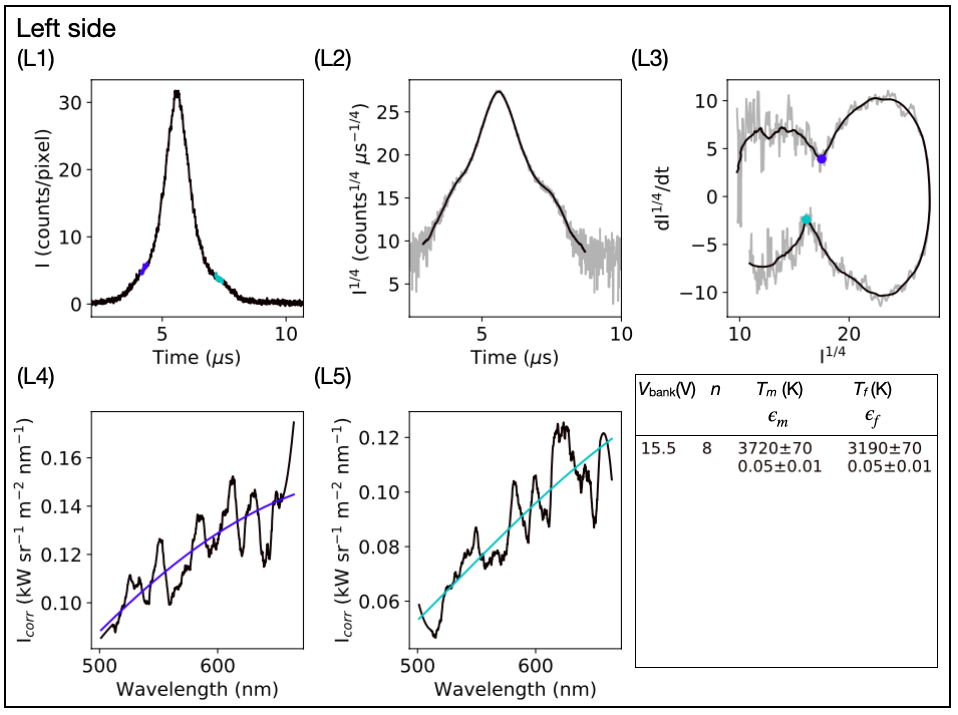}
\end{subfigure}
\begin{subfigure}
	\centering
	\includegraphics[width=5in]{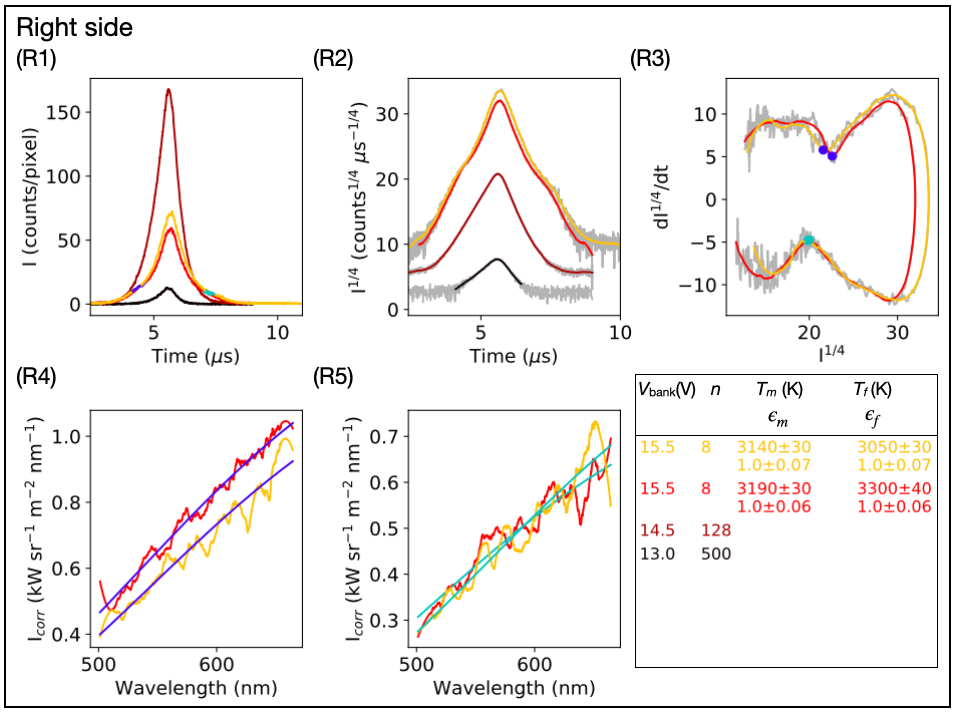}
\end{subfigure}
	\caption{		\linespread{1.0}\selectfont{}
	% \fontsize{11.5}\selecfont{}
Time-resolved thermal emissions of platinum sample \#3 heated from room temperature at~$31\pm 1.6$~GPa to past its melting point at ~$3430 \pm 250$~K,~$39 \pm 4.3$~GPa. The filtering timescale used in (L2), (L3), (R2), and (R3) is ~$\tau=0.8$~$\mu$s. See caption of \ref{fig:60GPa_SM} for all other details.}
	\label{fig:31GPa_SM}
\end{figure}

\begin{figure}[tbhp]
\begin{subfigure}
	\centering
	\includegraphics[width=5in]{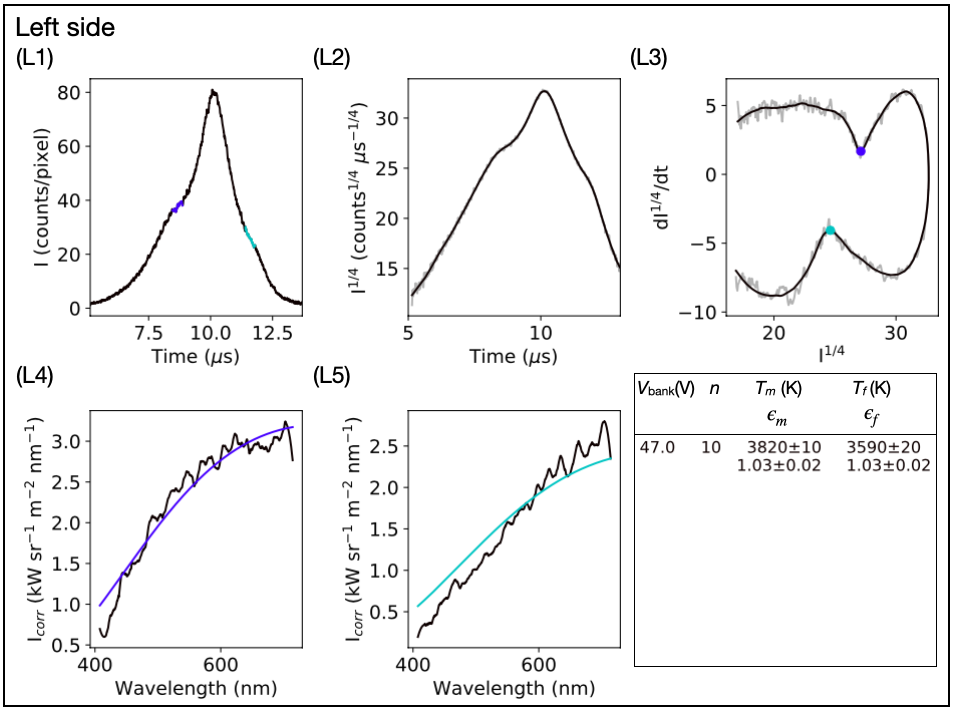}
\end{subfigure}
\begin{subfigure}
	\centering
	\includegraphics[width=5in]{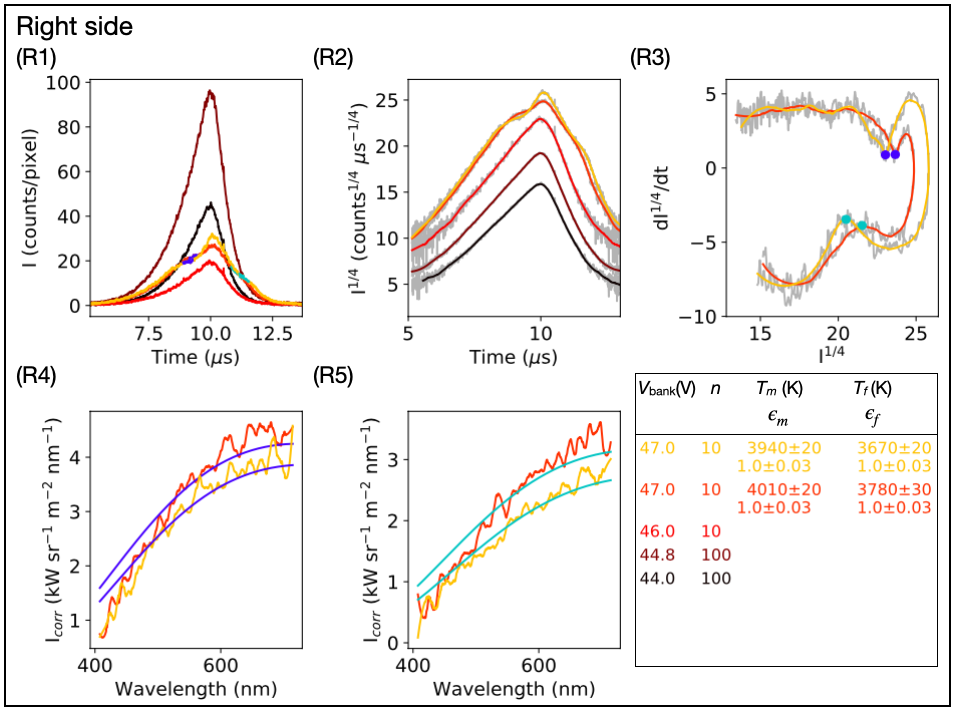}
\end{subfigure}
	\caption{		\linespread{1.0}\selectfont{}
	% \fontsize{11.5}\selecfont{}
Time-resolved thermal emissions of platinum sample \#4 heated from room temperature at~$43\pm 2.2$~GPa to past its melting point at ~$3890 \pm 70$~K, ~$51 \pm 4.5$~GPa. The filtering timescale used in (L2), (L3), (R2), and (R3) is ~$\tau=1.2$~$\mu$s. See caption of \ref{fig:60GPa_SM} for all other details.}
	\label{fig:43GPa_SM}
\end{figure}

\begin{figure}[tbhp]
\begin{subfigure}
	\centering
	\includegraphics[width=5in]{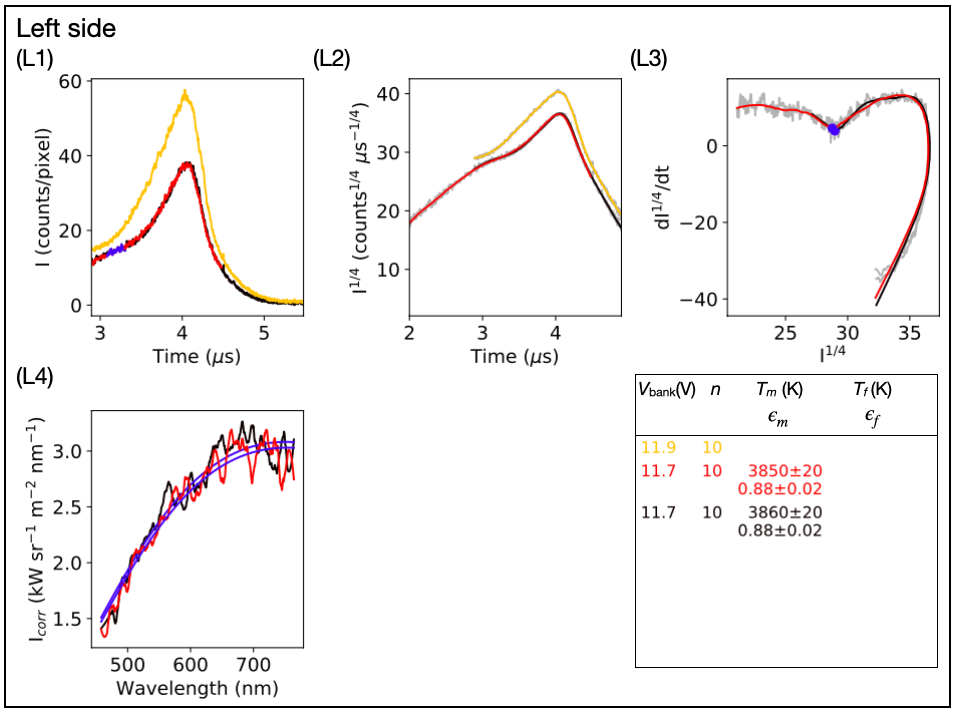}
\end{subfigure}
\begin{subfigure}
	\centering
	\includegraphics[width=5in]{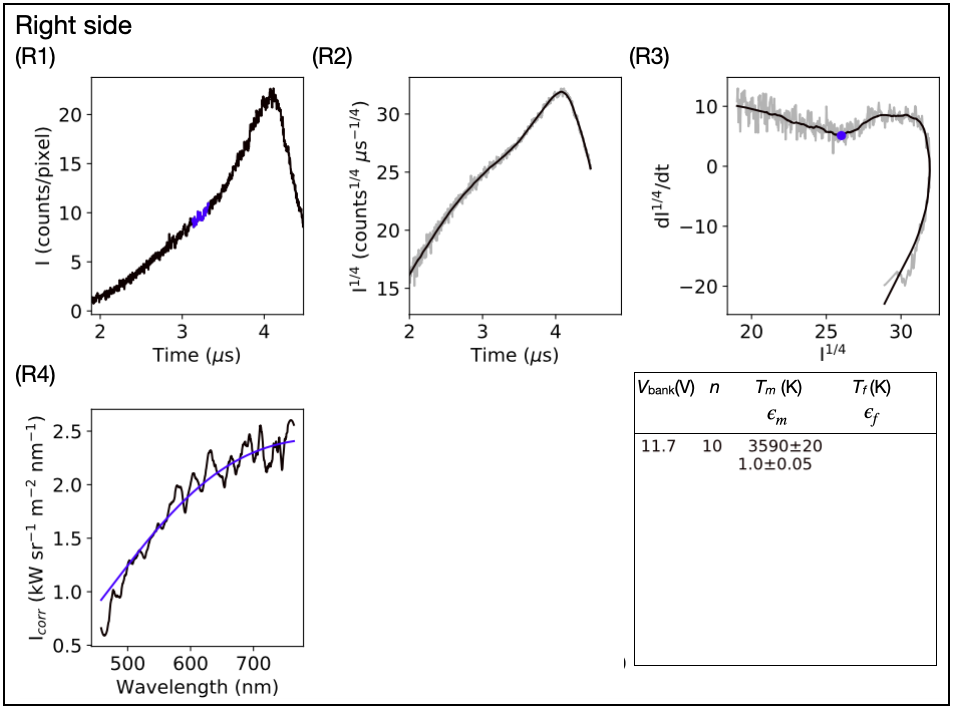}
\end{subfigure}
	\caption{		\linespread{1.0}\selectfont{}
	% \fontsize{11.5}\selecfont{}
Time-resolved thermal emissions of platinum sample \#5 heated from room temperature at~$49\pm 2.4$~GPa to past its melting point at~$3710 \pm 120$~K,~$57 \pm 4.7$~GPa. The filtering timescale used in (L2), (L3), (R2), and (R3) is~$\tau=0.6$~$\mu$s. See caption of \ref{fig:60GPa_SM} for all other details.}
	\label{fig:49GPa_SM}
\end{figure}

\begin{figure}[tbhp]
\begin{subfigure}
	\centering
	\includegraphics[width=5in]{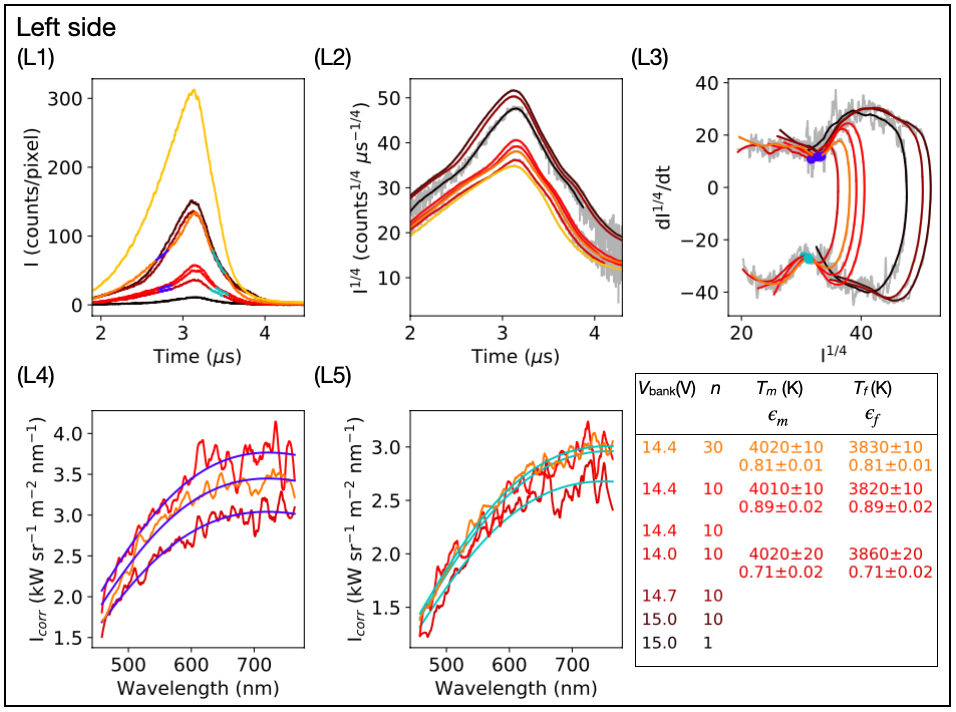}
\end{subfigure}
\begin{subfigure}
	\centering
	\includegraphics[width=5in]{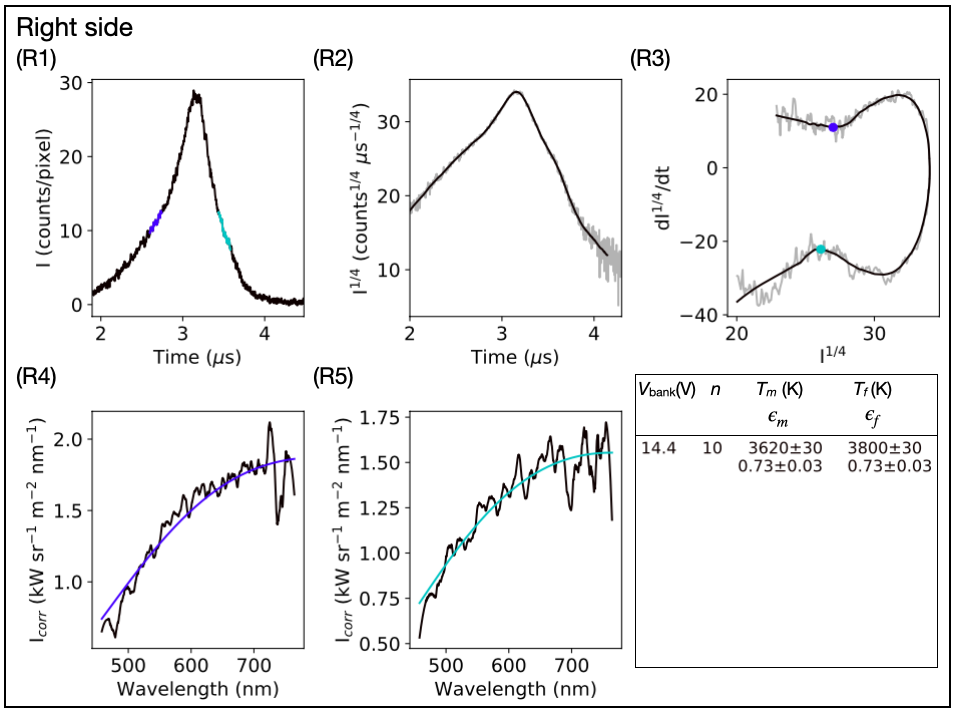}
\end{subfigure}
	\caption{		\linespread{1.0}\selectfont{}
	% \fontsize{11.5}\selecfont{}
Time-resolved thermal emissions of platinum sample \#7 heated from room temperature at~$63\pm 3.2$~GPa to past its melting point at~$3810 \pm 190$~K,~$ 71 \pm 5.1$~GPa. The filtering timescale used in (L2), (L3), (R2), and (R3) is~$\tau=0.4$~$\mu$s. See caption of \ref{fig:60GPa_SM} for all other details.}
	\label{fig:63GPa_SM}
\end{figure}

\begin{figure}[tbhp]
\begin{subfigure}
	\centering
	\includegraphics[width=5in]{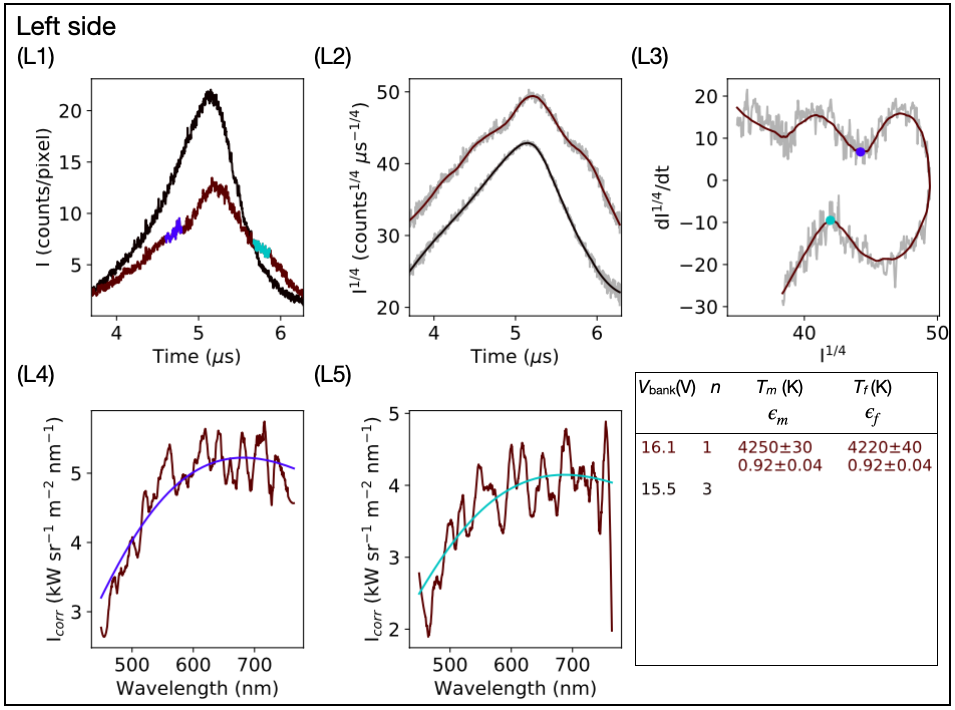}
\end{subfigure}
\begin{subfigure}
	\centering
	\includegraphics[width=5in]{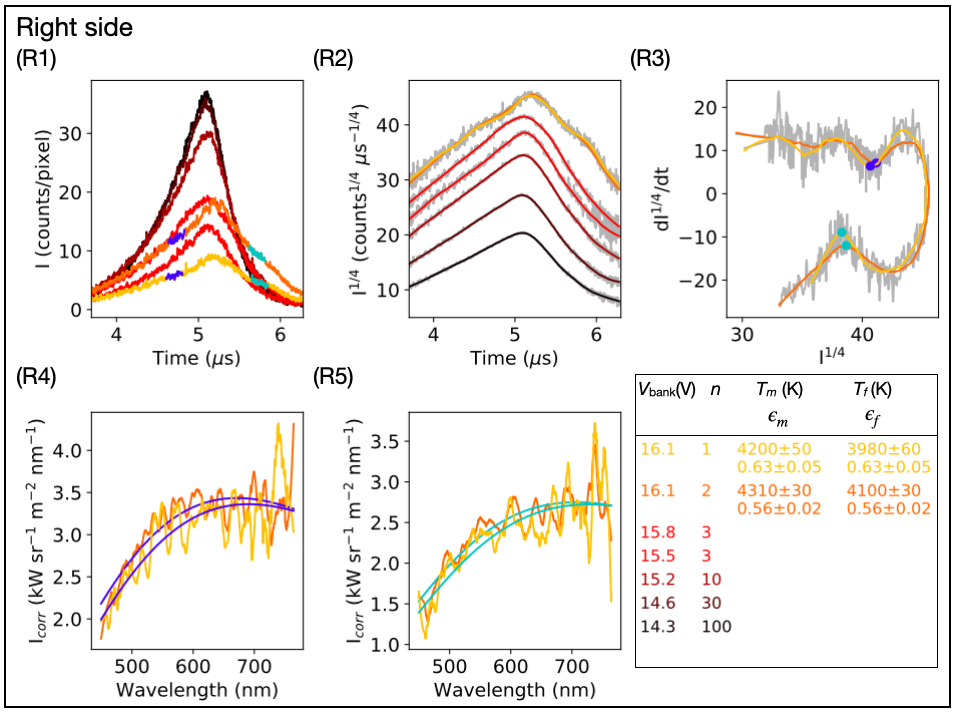}
\end{subfigure}
	\caption{		\linespread{1.0}\selectfont{}
	% \fontsize{11.5}\selecfont{}
Time-resolved thermal emissions of platinum sample \#6 heated from room temperature at~$78\pm 3.9$~GPa to past its melting point at~$4260 \pm 20$~K,~$ 85.9 \pm 5.6$~GPa. The filtering timescale used in (L2), (L3), (R2), and (R3) is~$\tau=0.6$~$\mu$s. See caption of \ref{fig:60GPa_SM} for all other details.}
	\label{fig:78GPa_SM}
\end{figure}

\begin{figure}[tbhp]
\begin{subfigure}
	\centering
	\includegraphics[width=5in]{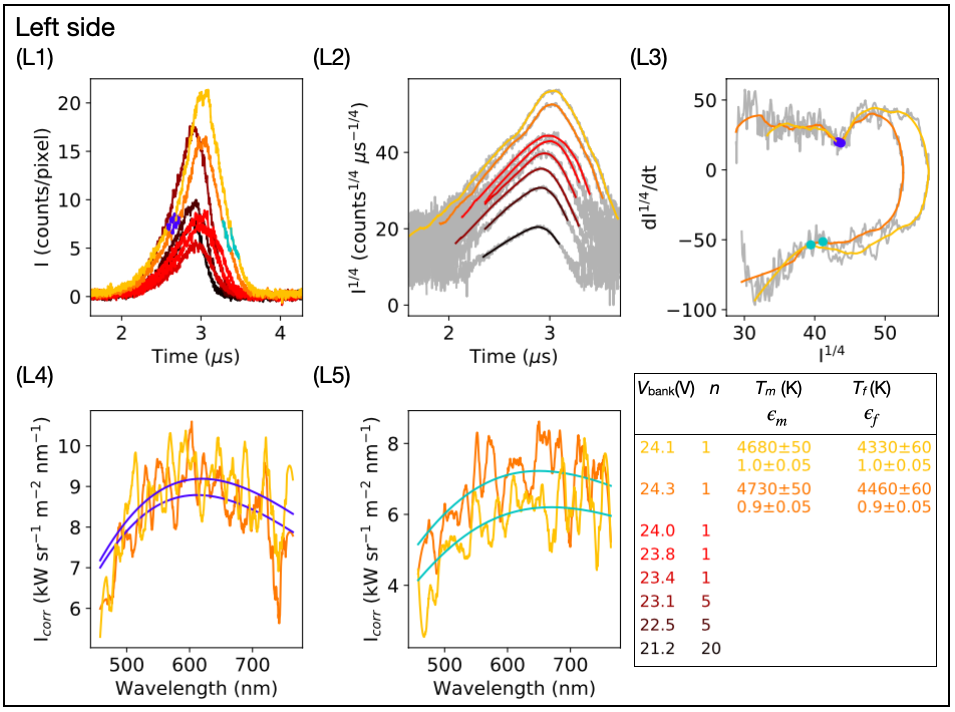}
\end{subfigure}
\begin{subfigure}
	\centering
	\includegraphics[width=5in]{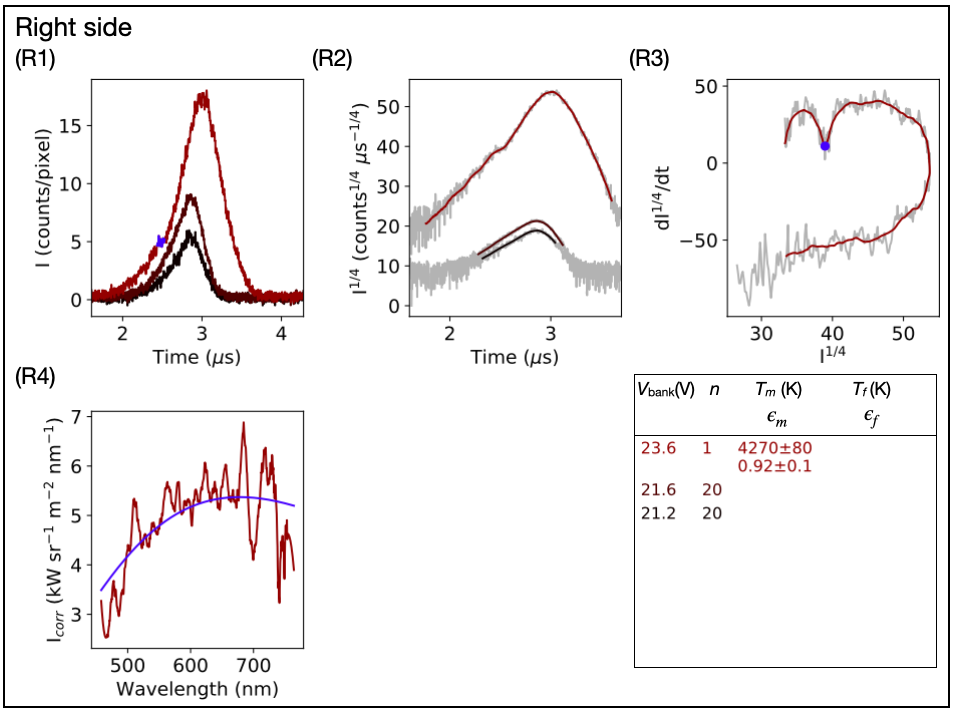}
\end{subfigure}
	\caption{		\linespread{1.0}\selectfont{}
	% \fontsize{11.5}\selecfont{}
Time-resolved thermal emissions of platinum sample \#8 heated from room temperature at~$99 \pm 5$~GPa to past its melting point at~$4480 \pm 170$~K,~$106.9 \pm 9.3$~GPa. The filtering timescale used in (L2), (L3), (R2), and (R3) is~$\tau=0.3$~$\mu$s. See caption of \ref{fig:60GPa_SM} for all other details. Freezing was not detected in this experiment.}
	\label{fig:99GPa_SM}
\end{figure}

X-ray diffraction on four platinum samples confirm an increase in Angstrom-scale disorder and the absence of new crystalline phases upon heating to 3000 to 4000 K at 35 to 55 GPa (Figs. \ref{fig:XRD_stack}, \ref{fig:Big_XRD}).

The intensity of diffuse scattering increases with temperature in all cases. For three of the four samples, it increases most rapidly within $\pm 600$ K of the melting curve determined by latent heat (black dashed lines in Fig. \ref{fig:Big_XRD}). For the fourth sample, Sample \#5, the increase is approximately linear with respect to temperature (Fig. \ref{fig:Big_XRD}o). Sample \#4 was heated to $>4000$ K with set of 100 heating pulses, and subsequently heated to~$\sim 3000$~K with sets of 1000 heating pulses. In the first case, the rapid increase in diffuse scattering occurred at a temperature above our melting curve, while in the second case it occurred at a lower temperature.  We conclude that although X-ray diffraction data set is consistent with the melting curve determined by latent heat, it does not improve the precision of our determination of melting temperature. 

The relative imprecision of X-ray diffraction melt-identification for our samples is likely caused by an instability in resistive heating to a molten state. During a single melt-freeze cycle, the platinum sample deforms slightly, causing a small change in electrical power deposition during the subsequent heating pulse. This effect causes variations in the peak temperature from pulse-to-pulse. Since X-ray measurements are accumulated during 100 to 1000 melting shots, slight variations from pulse-to-pulse can propagate to significant variations in the temperature of the sample from which X-rays are diffracted. These variations affect the temperature measurement in a different way than the X-ray measurement, causing scatter in the plots of diffuse X-ray intensity versus temperature. For streak camera measurements, by contrast, thermal emissions from no more than ten melting shots are accumulated, minimizing the effect of pulse-to-pulse variations.

Although large gradients in temperature often complicate interpretations of diffuse scattering in high temperature diamond-cell experiments, the spatial gradients are relatively small here. The full-width at half max of thermal emission intensity is 10 to 20~$\mu$m, even when the peak temperature is 4000 K, which propagates to a temperature differential of $\sim 400$~K over the 10 to 20~$\mu$m-long segment. For comparison, the X-ray beam is focused to a 3~$\mu$m~x~4~$\mu$m area.

\clearpage
% \begin{sidewaystable*}[h]
\begin{table*}[ht]
%  \scriptsize
\tiny
%Copy from output of make_Pt_data_table.py 
\begin{tabular}{ c c c c c c c c c c c |c c c c | c c c c }%| c c c c  }
% \hline 
\hline 
&&&&&&&&&& &&Left side  &&& &Right side && \\
 Sample \#&$P_\0$ & $dP_\0$ & $P_m$ & $dP_m$ & $V_m$ & $T_m$ & $dT_m$ & $T_f$ & $dT_f$ & $\tau$ & $T_{m,i}$ & $dT_{m,i}$ & $T_{f,i}$ & $dT_{f,i}$ & $T_{m,i}$ & $dT_{m,i}$ & $T_{f,i}$ & $dT_{f,i}$ \\
&(GPa) & (GPa) & (GPa) & (GPa) & (\AA$^3$/unit cell) & (K) & (K) & (K) & (K) & ($\mu$s) & (K) & (K) & (K) & (K) & (K) & (K) & (K) & (K) \\

\hline \#1& 0 & 0.0 & 6.8 & 6.8 & 62.04 & 2160 & 20 & 2000 & 0 & 8.0 & 2140 & 40 & 2000 & 50 &  &  &  &  \\
& & & & & & & & & & & 2170 & 40 & 2000 & 50 & & & & \\
\hline \#2& 26 & 1.3 & 34.0 & 4.2 & 57.65 & 3000 & 140 &  &  & 0.6 & 3130 & 10 &  &  & 2850 & 20 &  &  \\
& & & & & & & & & & & 3150 & 10 & & & 2870 & 20 & & \\
\hline \#3& 31 & 1.6 & 39.0 & 4.3 & 57.35 & 3430 & 250 & 3170 & 60 & 0.8 & 3680 & 70 & 3200 & 60 & 3200 & 20 & 3260 & 40 \\
& & & & & & & & & & & & & & & 3170 & 30 & 3050 & 30 \\
\hline \#4& 43 & 2.2 & 51.0 & 4.5 & 56.03 & 3890 & 70 & 3680 & 70 & 1.2 & 3830 & 10 & 3610 & 20 & 4010 & 20 & 3780 & 20 \\
& & & & & & & & & & & & & & & 3920 & 20 & 3700 & 20 \\
\hline \#5& 49 & 2.4 & 57.0 & 4.7 & 55.04 & 3710 & 120 &  &  & 0.6 & 3840 & 10 &  &  & 3590 & 20 &  &  \\
& & & & & & & & & & & 3830 & 20 & & & & & & \\
\hline \#6& 60 & 3.0 & 68.0 & 5.0 & 54.0 & 4060 & 140 & 3860 & 40 & 0.6 & 4120 & 40 & 3810 & 50 & 3920 & 60 & 3740 & 170 \\
& & & & & & & & & & & 4180 & 40 & 3920 & 50 & & & 3840 & 70 \\
& & & & & & & & & & & 4020 & 40 & 3890 & 30 & & & & \\
& & & & & & & & & & & 4180 & 30 & 3980 & 30 & & & & \\
& & & & & & & & & & & 4260 & 30 & 3840 & 30 & & & & \\
& & & & & & & & & & & 4290 & 30 & & & & & & \\
\hline \#7& 63 & 3.2 & 71.0 & 5.1 & 53.45 & 3810 & 190 & 3810 & 30 & 0.4 & 4020 & 10 & 3930 & 10 & 3620 & 20 & 3780 & 30 \\
& & & & & & & & & & & 4020 & 10 & 3850 & 10 & & & & \\
& & & & & & & & & & & 3940 & 10 & 3800 & 10 & & & & \\
& & & & & & & & & & & 3970 & 10 & 3780 & 10 & & & & \\
& & & & & & & & & & & & & 3840 & 10 & & & & \\
\hline \#8& 78 & 3.9 & 85.9 & 5.6 & 52.24 & 4260 & 20 & 4110 & 70 & 0.6 & 4270 & 30 & 4180 & 30 & 4280 & 30 & 4070 & 30 \\
& & & & & & & & & & & & & & & 4170 & 40 & 3980 & 50 \\
\hline \#9& 99 & 5.0 & 106.9 & 9.3 & 50.51 & 4480 & 170 & 4800 & 70 & 0.3 & 4670 & 40 & 4730 & 40 & 4310 & 70 &  &  \\
& & & & & & & & & & & 4630 & 40 & 4870 & 40 & & & & \\
\hline \hline
\end{tabular}

\caption{Pressure and temperature of melting and freezing for each individual melting run, identified by plateaus in $I^{1/4}$. In order, the columns are pressure before and after heating ($P_0 \pm dP_0$), pressure at the melting point ($P_m \pm dP_m$), weighted average of measured melting temperature ($T_m \pm dT_m$) and of measured freezing temperature ($T_f \pm dT_f$), and individual measurements of melting and freezing temperatures from the right side and left side of the sample. All individual temperature measurements are shown in Figs. \figsixtyGPaIt, \ref{fig:60GPa_SM}-\ref{fig:99GPa_SM}.
}
\label{table:errors}
\end{table*}
% \end{sidewaystable*}

\begin{table*} %From Tables4Pt_melting_paper google spreadsheet in LaTex folder
%\normalsize
%\small
%\footnotesize
\scriptsize
%\tiny
	\begin{tabular}{ 	c		c		c		c		c		c		c		c		c		c		c		c		c		c	}
\hline \hline	$P_m$	&	$dP_m$	&	$l_\sam$ 	&	$dl_\sam$ 	&	$w_\sam$ 	&	$t_\sam$	&	$dt_\sam$	&	$r_\max^{Tm,\sol}$	&	$\frac{\Delta R_m}{R_{Tm}^\sol}$	&	$\frac{L_\max}{c_\max}$	&	$L_\max$	&	$L_\min$	&	$\Delta S_\max$	&	$\Delta S_\min$	\\
	(GPa)	&	(GPa)	&	($\mu$m)	&	($\mu$m)	&	($\mu$m)	&	($\mu$m)	&	($\mu$m)	&	($\mu \Omega$~m)	&		&	(K)	&	(kJ/mol)	&	(kJ/mol)	&	(J/mol K)	&	(J/mol K)	\\
\hline	6.8	&	6.8	&	120	&	20	&	22	&	22	&	4	&	1.3	&	4\%	&	95	&	200	&	1.6	&	91	&	0.8	\\
	34	&	4.2	&	8	&	3	&	8	&	5.3	&	0.6	&	2.0	&	-	&	48	&	101	&	0.9	&	32	&	0.3	\\
	39	&	4.3	&	13	&	4	&	13	&	3	&	1	&	0.6	&	10\%	&	111	&	260	&	1.6	&	80	&	0.5	\\
	51	&	4.5	&	15	&	4	&	44	&	6.4	&	0.4	&	2.0	&	5\%	&	168	&	1431	&	3.9	&	374	&	1.0	\\
	57	&	4.7	&	10	&	4	&	12	&	5	&	1	&	4.0	&	4\%	&	121	&	239	&	2.4	&	60	&	0.6	\\
	68	&	5	&	17	&	3	&	12	&	5	&	0.8	&	1.9	&	4\%	&	327	&	161	&	5.3	&	37	&	1.2	\\
	71	&	5.1	&	8	&	3	&	5	&	2.3	&	1	&	1.0	&	-	&	104	&	1170	&	1.1	&	292	&	0.3	\\
	85.9	&	5.6	&	12	&	3	&	12	&	4.9	&	0.4	&	1.8	&	3\%	&	173	&	89	&	2.0	&	22	&	0.5	\\
	106.9	&	9.3	&	8	&	3	&	8	&	0.4	&	1	&	-	&	-	&	-	&	-	&	-	&	-	&	-	\\
\hline	Literature	&		&	50,000	&	-	&	500	&	(diameter)	&	-	&	0.6	&	40\%	&	-	&	24	&	20	&	12	&	10	\\
	(ambient 	&	pressure)	&		&		&		&		&		&		&		&		&		&		&		&		\\
\hline \hline
\end{tabular}
\caption{Sample dimensions, resistance measurements, and calorimetric properties. The data collected here is compared to the literature values at ambient pressure of Wilthan et al. \cite{Wilthan2004}.
}
\label{table:electrical}
\end{table*}

\clearpage

% Sample prep schematic in keynote, or pictures? 

% \pagebreak
\clearpage
\nocite{*}
\bibliography{Pt_melting_arXiv_Feb2021}% Produces the bibliography via BibTeX.

\begin{thebibliography}{64}%
\makeatletter
\providecommand \@ifxundefined [1]{%
 \@ifx{#1\undefined}
}%
\providecommand \@ifnum [1]{%
 \ifnum #1\expandafter \@firstoftwo
 \else \expandafter \@secondoftwo
 \fi
}%
\providecommand \@ifx [1]{%
 \ifx #1\expandafter \@firstoftwo
 \else \expandafter \@secondoftwo
 \fi
}%
\providecommand \natexlab [1]{#1}%
\providecommand \enquote  [1]{``#1''}%
\providecommand \bibnamefont  [1]{#1}%
\providecommand \bibfnamefont [1]{#1}%
\providecommand \citenamefont [1]{#1}%
\providecommand \href@noop [0]{\@secondoftwo}%
\providecommand \href [0]{\begingroup \@sanitize@url \@href}%
\providecommand \@href[1]{\@@startlink{#1}\@@href}%
\providecommand \@@href[1]{\endgroup#1\@@endlink}%
\providecommand \@sanitize@url [0]{\catcode `\\12\catcode `\$12\catcode
  `\&12\catcode `\#12\catcode `\^12\catcode `\_12\catcode `\%12\relax}%
\providecommand \@@startlink[1]{}%
\providecommand \@@endlink[0]{}%
\providecommand \url  [0]{\begingroup\@sanitize@url \@url }%
\providecommand \@url [1]{\endgroup\@href {#1}{\urlprefix }}%
\providecommand \urlprefix  [0]{URL }%
\providecommand \Eprint [0]{\href }%
\providecommand \doibase [0]{http://dx.doi.org/}%
\providecommand \selectlanguage [0]{\@gobble}%
\providecommand \bibinfo  [0]{\@secondoftwo}%
\providecommand \bibfield  [0]{\@secondoftwo}%
\providecommand \translation [1]{[#1]}%
\providecommand \BibitemOpen [0]{}%
\providecommand \bibitemStop [0]{}%
\providecommand \bibitemNoStop [0]{.\EOS\space}%
\providecommand \EOS [0]{\spacefactor3000\relax}%
\providecommand \BibitemShut  [1]{\csname bibitem#1\endcsname}%
\let\auto@bib@innerbib\@empty
%</preamble>
\bibitem [{\citenamefont {{Kraut}}\ and\ \citenamefont
  {{Kennedy}}(1966{\natexlab{a}})}]{Kraut1966_PRL}%
  \BibitemOpen
  \bibfield  {author} {\bibinfo {author} {\bibfnamefont {E.~A.}\ \bibnamefont
  {{Kraut}}}\ and\ \bibinfo {author} {\bibfnamefont {G.~C.}\ \bibnamefont
  {{Kennedy}}},\ }\href {\doibase 10.1103/PhysRevLett.16.608} {\bibfield
  {journal} {\bibinfo  {journal} {\prl}\ }\textbf {\bibinfo {volume} {16}},\
  \bibinfo {pages} {608} (\bibinfo {year} {1966}{\natexlab{a}})}\BibitemShut
  {NoStop}%
\bibitem [{\citenamefont {{Gilvarry}}(1956)}]{Gilvarry1956}%
  \BibitemOpen
  \bibfield  {author} {\bibinfo {author} {\bibfnamefont {J.~J.}\ \bibnamefont
  {{Gilvarry}}},\ }\href {\doibase 10.1103/PhysRev.102.308} {\bibfield
  {journal} {\bibinfo  {journal} {Phys. Rev.}\ }\textbf {\bibinfo {volume}
  {102}},\ \bibinfo {pages} {308} (\bibinfo {year} {1956})}\BibitemShut
  {NoStop}%
\bibitem [{\citenamefont {{Belonoshko}}\ and\ \citenamefont
  {{Rosengren}}(2012)}]{Belonoshko2012}%
  \BibitemOpen
  \bibfield  {author} {\bibinfo {author} {\bibfnamefont {A.~B.}\ \bibnamefont
  {{Belonoshko}}}\ and\ \bibinfo {author} {\bibfnamefont {A.}~\bibnamefont
  {{Rosengren}}},\ }\href {\doibase 10.1103/PhysRevB.85.174104} {\bibfield
  {journal} {\bibinfo  {journal} {\prb}\ }\textbf {\bibinfo {volume} {85}},\
  \bibinfo {eid} {174104} (\bibinfo {year} {2012})}\BibitemShut {NoStop}%
\bibitem [{\citenamefont {{Hirose}}\ \emph {et~al.}(2013)\citenamefont
  {{Hirose}}, \citenamefont {{Labrosse}},\ and\ \citenamefont
  {{Hernlund}}}]{Hirose2013}%
  \BibitemOpen
  \bibfield  {author} {\bibinfo {author} {\bibfnamefont {K.}~\bibnamefont
  {{Hirose}}}, \bibinfo {author} {\bibfnamefont {S.}~\bibnamefont
  {{Labrosse}}}, \ and\ \bibinfo {author} {\bibfnamefont {J.}~\bibnamefont
  {{Hernlund}}},\ }\href {\doibase 10.1146/annurev-earth-050212-124007}
  {\bibfield  {journal} {\bibinfo  {journal} {Ann. Rev. Earth Planet. Sci.}\
  }\textbf {\bibinfo {volume} {41}},\ \bibinfo {pages} {657} (\bibinfo {year}
  {2013})}\BibitemShut {NoStop}%
\bibitem [{\citenamefont {{Kraut}}\ and\ \citenamefont
  {{Kennedy}}(1966{\natexlab{b}})}]{Kraut1966_PR}%
  \BibitemOpen
  \bibfield  {author} {\bibinfo {author} {\bibfnamefont {E.~A.}\ \bibnamefont
  {{Kraut}}}\ and\ \bibinfo {author} {\bibfnamefont {G.~C.}\ \bibnamefont
  {{Kennedy}}},\ }\href {\doibase 10.1103/PhysRev.151.668} {\bibfield
  {journal} {\bibinfo  {journal} {Phys. Rev.}\ }\textbf {\bibinfo {volume}
  {151}},\ \bibinfo {pages} {668} (\bibinfo {year}
  {1966}{\natexlab{b}})}\BibitemShut {NoStop}%
\bibitem [{\citenamefont {{Errandonea}}\ \emph {et~al.}(2019)\citenamefont
  {{Errandonea}}, \citenamefont {{MacLeod}}, \citenamefont {{Burakovsky}},
  \citenamefont {{Santamaria-Perez}}, \citenamefont {{Proctor}}, \citenamefont
  {{Cynn}},\ and\ \citenamefont {{Mezouar}}}]{Errandonea2019}%
  \BibitemOpen
  \bibfield  {author} {\bibinfo {author} {\bibfnamefont {D.}~\bibnamefont
  {{Errandonea}}}, \bibinfo {author} {\bibfnamefont {S.~G.}\ \bibnamefont
  {{MacLeod}}}, \bibinfo {author} {\bibfnamefont {L.}~\bibnamefont
  {{Burakovsky}}}, \bibinfo {author} {\bibfnamefont {D.}~\bibnamefont
  {{Santamaria-Perez}}}, \bibinfo {author} {\bibfnamefont {J.~E.}\ \bibnamefont
  {{Proctor}}}, \bibinfo {author} {\bibfnamefont {H.}~\bibnamefont {{Cynn}}}, \
  and\ \bibinfo {author} {\bibfnamefont {M.}~\bibnamefont {{Mezouar}}},\ }\href
  {\doibase 10.1103/PhysRevB.100.094111} {\bibfield  {journal} {\bibinfo
  {journal} {\prb}\ }\textbf {\bibinfo {volume} {100}},\ \bibinfo {eid}
  {094111} (\bibinfo {year} {2019})}\BibitemShut {NoStop}%
\bibitem [{\citenamefont {Errandonea}\ \emph {et~al.}(2020)\citenamefont
  {Errandonea}, \citenamefont {Burakovsky}, \citenamefont {Preston},
  \citenamefont {MacLeod}, \citenamefont {Santamaría-Perez}, \citenamefont
  {Chen}, \citenamefont {Cynn}, \citenamefont {Simak}, \citenamefont {McMahon},
  \citenamefont {Proctor},\ and\ \citenamefont {Mezouar}}]{Errandonea2020}%
  \BibitemOpen
  \bibfield  {author} {\bibinfo {author} {\bibfnamefont {D.}~\bibnamefont
  {Errandonea}}, \bibinfo {author} {\bibfnamefont {L.}~\bibnamefont
  {Burakovsky}}, \bibinfo {author} {\bibfnamefont {D.~L.}\ \bibnamefont
  {Preston}}, \bibinfo {author} {\bibfnamefont {S.~G.}\ \bibnamefont
  {MacLeod}}, \bibinfo {author} {\bibfnamefont {D.}~\bibnamefont
  {Santamaría-Perez}}, \bibinfo {author} {\bibfnamefont {S.}~\bibnamefont
  {Chen}}, \bibinfo {author} {\bibfnamefont {H.}~\bibnamefont {Cynn}}, \bibinfo
  {author} {\bibfnamefont {S.~I.}\ \bibnamefont {Simak}}, \bibinfo {author}
  {\bibfnamefont {M.~I.}\ \bibnamefont {McMahon}}, \bibinfo {author}
  {\bibfnamefont {J.~E.}\ \bibnamefont {Proctor}}, \ and\ \bibinfo {author}
  {\bibfnamefont {M.}~\bibnamefont {Mezouar}},\ }\href {\doibase
  10.1038/s43246-020-00058-2} {\bibfield  {journal} {\bibinfo  {journal}
  {Communications Materials}\ }\textbf {\bibinfo {volume} {1}},\ \bibinfo
  {pages} {60} (\bibinfo {year} {2020})}\BibitemShut {NoStop}%
\bibitem [{\citenamefont {{Sinmyo}}\ \emph {et~al.}(2019)\citenamefont
  {{Sinmyo}}, \citenamefont {{Hirose}},\ and\ \citenamefont
  {{Ohishi}}}]{Sinmyo2019}%
  \BibitemOpen
  \bibfield  {author} {\bibinfo {author} {\bibfnamefont {R.}~\bibnamefont
  {{Sinmyo}}}, \bibinfo {author} {\bibfnamefont {K.}~\bibnamefont {{Hirose}}},
  \ and\ \bibinfo {author} {\bibfnamefont {Y.}~\bibnamefont {{Ohishi}}},\
  }\href {\doibase 10.1016/j.epsl.2019.01.006} {\bibfield  {journal} {\bibinfo
  {journal} {Earth Planet. Sci. Lett.}\ }\textbf {\bibinfo {volume} {510}},\
  \bibinfo {pages} {45} (\bibinfo {year} {2019})}\BibitemShut {NoStop}%
\bibitem [{\citenamefont {{Hrubiak}}\ \emph {et~al.}(2017)\citenamefont
  {{Hrubiak}}, \citenamefont {{Meng}},\ and\ \citenamefont
  {{Shen}}}]{Hrubiak2017}%
  \BibitemOpen
  \bibfield  {author} {\bibinfo {author} {\bibfnamefont {R.}~\bibnamefont
  {{Hrubiak}}}, \bibinfo {author} {\bibfnamefont {Y.}~\bibnamefont {{Meng}}}, \
  and\ \bibinfo {author} {\bibfnamefont {G.}~\bibnamefont {{Shen}}},\ }\href
  {\doibase 10.1038/ncomms14562} {\bibfield  {journal} {\bibinfo  {journal}
  {Nature Comm.}\ }\textbf {\bibinfo {volume} {8}},\ \bibinfo {eid} {14562}
  (\bibinfo {year} {2017})}\BibitemShut {NoStop}%
\bibitem [{\citenamefont {{Stutzmann}}\ \emph {et~al.}(2015)\citenamefont
  {{Stutzmann}}, \citenamefont {{Dewaele}}, \citenamefont {{Bouchet}},
  \citenamefont {{Bottin}},\ and\ \citenamefont {{Mezouar}}}]{Stutzmann2015}%
  \BibitemOpen
  \bibfield  {author} {\bibinfo {author} {\bibfnamefont {V.}~\bibnamefont
  {{Stutzmann}}}, \bibinfo {author} {\bibfnamefont {A.}~\bibnamefont
  {{Dewaele}}}, \bibinfo {author} {\bibfnamefont {J.}~\bibnamefont
  {{Bouchet}}}, \bibinfo {author} {\bibfnamefont {F.}~\bibnamefont {{Bottin}}},
  \ and\ \bibinfo {author} {\bibfnamefont {M.}~\bibnamefont {{Mezouar}}},\
  }\href {\doibase 10.1103/PhysRevB.92.224110} {\bibfield  {journal} {\bibinfo
  {journal} {\prb}\ }\textbf {\bibinfo {volume} {92}},\ \bibinfo {eid} {224110}
  (\bibinfo {year} {2015})}\BibitemShut {NoStop}%
\bibitem [{\citenamefont {{Parisiades}}\ \emph {et~al.}(2019)\citenamefont
  {{Parisiades}}, \citenamefont {{Cova}},\ and\ \citenamefont
  {{Garbarino}}}]{Parisiades2019}%
  \BibitemOpen
  \bibfield  {author} {\bibinfo {author} {\bibfnamefont {P.}~\bibnamefont
  {{Parisiades}}}, \bibinfo {author} {\bibfnamefont {F.}~\bibnamefont
  {{Cova}}}, \ and\ \bibinfo {author} {\bibfnamefont {G.}~\bibnamefont
  {{Garbarino}}},\ }\href {\doibase 10.1103/PhysRevB.100.054102} {\bibfield
  {journal} {\bibinfo  {journal} {\prb}\ }\textbf {\bibinfo {volume} {100}},\
  \bibinfo {eid} {054102} (\bibinfo {year} {2019})}\BibitemShut {NoStop}%
\bibitem [{\citenamefont {{Anzellini}}\ \emph {et~al.}(2019)\citenamefont
  {{Anzellini}}, \citenamefont {{Monteseguro}}, \citenamefont {{Bandiello}},
  \citenamefont {{Dewaele}}, \citenamefont {{Burakovsky}},\ and\ \citenamefont
  {{Errandonea}}}]{Anzellini2019}%
  \BibitemOpen
  \bibfield  {author} {\bibinfo {author} {\bibfnamefont {S.}~\bibnamefont
  {{Anzellini}}}, \bibinfo {author} {\bibfnamefont {V.}~\bibnamefont
  {{Monteseguro}}}, \bibinfo {author} {\bibfnamefont {E.}~\bibnamefont
  {{Bandiello}}}, \bibinfo {author} {\bibfnamefont {A.}~\bibnamefont
  {{Dewaele}}}, \bibinfo {author} {\bibfnamefont {L.}~\bibnamefont
  {{Burakovsky}}}, \ and\ \bibinfo {author} {\bibfnamefont {D.}~\bibnamefont
  {{Errandonea}}},\ }\href {\doibase 10.1038/s41598-019-49676-y} {\bibfield
  {journal} {\bibinfo  {journal} {Sci. Rep.}\ }\textbf {\bibinfo {volume}
  {9}},\ \bibinfo {eid} {13034} (\bibinfo {year} {2019})}\BibitemShut {NoStop}%
\bibitem [{\citenamefont {{Karandikar}}\ and\ \citenamefont
  {{Boehler}}(2016)}]{Karandikar2016}%
  \BibitemOpen
  \bibfield  {author} {\bibinfo {author} {\bibfnamefont {A.}~\bibnamefont
  {{Karandikar}}}\ and\ \bibinfo {author} {\bibfnamefont {R.}~\bibnamefont
  {{Boehler}}},\ }\href {\doibase 10.1103/PhysRevB.93.054107} {\bibfield
  {journal} {\bibinfo  {journal} {\prb}\ }\textbf {\bibinfo {volume} {93}},\
  \bibinfo {eid} {054107} (\bibinfo {year} {2016})}\BibitemShut {NoStop}%
\bibitem [{\citenamefont {{Errandonea}}(2013)}]{Errandonea2013}%
  \BibitemOpen
  \bibfield  {author} {\bibinfo {author} {\bibfnamefont {D.}~\bibnamefont
  {{Errandonea}}},\ }\href {\doibase 10.1103/PhysRevB.87.054108} {\bibfield
  {journal} {\bibinfo  {journal} {\prb}\ }\textbf {\bibinfo {volume} {87}},\
  \bibinfo {eid} {054108} (\bibinfo {year} {2013})}\BibitemShut {NoStop}%
\bibitem [{\citenamefont {{Kavner}}\ and\ \citenamefont
  {{Jeanloz}}(1998)}]{Kavner1998}%
  \BibitemOpen
  \bibfield  {author} {\bibinfo {author} {\bibfnamefont {A.}~\bibnamefont
  {{Kavner}}}\ and\ \bibinfo {author} {\bibfnamefont {R.}~\bibnamefont
  {{Jeanloz}}},\ }\href {\doibase 10.1063/1.367520} {\bibfield  {journal}
  {\bibinfo  {journal} {J. Appl. Phys.}\ }\textbf {\bibinfo {volume} {83}},\
  \bibinfo {pages} {7553} (\bibinfo {year} {1998})}\BibitemShut {NoStop}%
\bibitem [{\citenamefont {{Patel}}\ and\ \citenamefont
  {{Sunder}}(2018)}]{Patel2018}%
  \BibitemOpen
  \bibfield  {author} {\bibinfo {author} {\bibfnamefont {N.~N.}\ \bibnamefont
  {{Patel}}}\ and\ \bibinfo {author} {\bibfnamefont {M.}~\bibnamefont
  {{Sunder}}},\ }in\ \href {\doibase 10.1063/1.5028588} {\emph {\bibinfo
  {booktitle} {American Institute of Physics Conference Series}}},\ \bibinfo
  {series} {American Institute of Physics Conference Series}, Vol.\ \bibinfo
  {volume} {1942}\ (\bibinfo {year} {2018})\ p.\ \bibinfo {pages}
  {030007}\BibitemShut {NoStop}%
\bibitem [{\citenamefont {{Lo Nigro}}(2011)}]{LoNigro2011}%
  \BibitemOpen
  \bibfield  {author} {\bibinfo {author} {\bibfnamefont {G.}~\bibnamefont {{Lo
  Nigro}}},\ }\emph {\bibinfo {title} {Experimental investigation of the deep
  mantle melting properties}},\ \href@noop {} {Ph.D. thesis},\ \bibinfo
  {school} {Universite Blaise Pascal-Clermont-Ferrand II} (\bibinfo {year}
  {2011})\BibitemShut {NoStop}%
\bibitem [{\citenamefont {{Lazicki}}\ \emph {et~al.}(2010)\citenamefont
  {{Lazicki}}, \citenamefont {{Fei}},\ and\ \citenamefont
  {{Hemley}}}]{Lazicki2010}%
  \BibitemOpen
  \bibfield  {author} {\bibinfo {author} {\bibfnamefont {A.}~\bibnamefont
  {{Lazicki}}}, \bibinfo {author} {\bibfnamefont {Y.}~\bibnamefont {{Fei}}}, \
  and\ \bibinfo {author} {\bibfnamefont {R.~J.}\ \bibnamefont {{Hemley}}},\
  }\href {\doibase 10.1016/j.ssc.2009.12.029} {\bibfield  {journal} {\bibinfo
  {journal} {Solid State Comm.}\ }\textbf {\bibinfo {volume} {150}},\ \bibinfo
  {pages} {625} (\bibinfo {year} {2010})}\BibitemShut {NoStop}%
\bibitem [{\citenamefont {{Luedemann}}\ and\ \citenamefont
  {{Kennedy}}(1968)}]{Luedemann1968}%
  \BibitemOpen
  \bibfield  {author} {\bibinfo {author} {\bibfnamefont {H.~D.}\ \bibnamefont
  {{Luedemann}}}\ and\ \bibinfo {author} {\bibfnamefont {G.~C.}\ \bibnamefont
  {{Kennedy}}},\ }\href {\doibase 10.1029/JB073i008p02795} {\bibfield
  {journal} {\bibinfo  {journal} {J. Geophys. Res.}\ }\textbf {\bibinfo
  {volume} {73}},\ \bibinfo {pages} {2795} (\bibinfo {year}
  {1968})}\BibitemShut {NoStop}%
\bibitem [{\citenamefont {{Wolf}}\ and\ \citenamefont
  {{Jeanloz}}(1984)}]{Wolf1984}%
  \BibitemOpen
  \bibfield  {author} {\bibinfo {author} {\bibfnamefont {G.~H.}\ \bibnamefont
  {{Wolf}}}\ and\ \bibinfo {author} {\bibfnamefont {R.}~\bibnamefont
  {{Jeanloz}}},\ }\href {\doibase 10.1029/JB089iB09p07821} {\bibfield
  {journal} {\bibinfo  {journal} {J. Geophys. Res.}\ }\textbf {\bibinfo
  {volume} {89}},\ \bibinfo {pages} {7821} (\bibinfo {year}
  {1984})}\BibitemShut {NoStop}%
\bibitem [{\citenamefont {{Stacey}}\ \emph {et~al.}(1989)\citenamefont
  {{Stacey}}, \citenamefont {{Spiliopoulos}},\ and\ \citenamefont
  {{Barton}}}]{Stacey1989}%
  \BibitemOpen
  \bibfield  {author} {\bibinfo {author} {\bibfnamefont {F.~D.}\ \bibnamefont
  {{Stacey}}}, \bibinfo {author} {\bibfnamefont {S.~S.}\ \bibnamefont
  {{Spiliopoulos}}}, \ and\ \bibinfo {author} {\bibfnamefont {M.~A.}\
  \bibnamefont {{Barton}}},\ }\href {\doibase 10.1016/0031-9201(89)90068-X}
  {\bibfield  {journal} {\bibinfo  {journal} {Physics of the Earth and
  Planetary Interiors}\ }\textbf {\bibinfo {volume} {55}},\ \bibinfo {pages}
  {201} (\bibinfo {year} {1989})}\BibitemShut {NoStop}%
\bibitem [{\citenamefont {{Alf{\`e}}}\ \emph {et~al.}(2011)\citenamefont
  {{Alf{\`e}}}, \citenamefont {{Cazorla}},\ and\ \citenamefont
  {{Gillan}}}]{Alfe2011}%
  \BibitemOpen
  \bibfield  {author} {\bibinfo {author} {\bibfnamefont {D.}~\bibnamefont
  {{Alf{\`e}}}}, \bibinfo {author} {\bibfnamefont {C.}~\bibnamefont
  {{Cazorla}}}, \ and\ \bibinfo {author} {\bibfnamefont {M.~J.}\ \bibnamefont
  {{Gillan}}},\ }\href {\doibase 10.1063/1.3605601} {\bibfield  {journal}
  {\bibinfo  {journal} {\jcp}\ }\textbf {\bibinfo {volume} {135}},\ \bibinfo
  {pages} {024102} (\bibinfo {year} {2011})}\BibitemShut {NoStop}%
\bibitem [{\citenamefont {{Braithwaite}}\ and\ \citenamefont
  {{Stixrude}}(2019)}]{Braithwaite2019}%
  \BibitemOpen
  \bibfield  {author} {\bibinfo {author} {\bibfnamefont {J.}~\bibnamefont
  {{Braithwaite}}}\ and\ \bibinfo {author} {\bibfnamefont {L.}~\bibnamefont
  {{Stixrude}}},\ }\href {\doibase 10.1029/2018GL081805} {\bibfield  {journal}
  {\bibinfo  {journal} {Geophys. Res. Lett.}\ }\textbf {\bibinfo {volume}
  {46}},\ \bibinfo {pages} {2037} (\bibinfo {year} {2019})}\BibitemShut
  {NoStop}%
\bibitem [{\citenamefont {{Geballe}}\ and\ \citenamefont
  {{Jeanloz}}(2012)}]{Geballe2012}%
  \BibitemOpen
  \bibfield  {author} {\bibinfo {author} {\bibfnamefont {Z.~M.}\ \bibnamefont
  {{Geballe}}}\ and\ \bibinfo {author} {\bibfnamefont {R.}~\bibnamefont
  {{Jeanloz}}},\ }\href {\doibase 10.1063/1.4729905} {\bibfield  {journal}
  {\bibinfo  {journal} {J. Appl. Phys.}\ }\textbf {\bibinfo {volume} {111}},\
  \bibinfo {eid} {123518-123518-15} (\bibinfo {year} {2012})}\BibitemShut
  {NoStop}%
\bibitem [{\citenamefont {{Karandikar}}(2006)}]{KarandikarThesis}%
  \BibitemOpen
  \bibfield  {author} {\bibinfo {author} {\bibfnamefont {A.}~\bibnamefont
  {{Karandikar}}},\ }\emph {\bibinfo {title} {Development of the Flash-heating
  Method for Measuring Melting Temperatures in the Diamond Anvil Cell}},\
  \href@noop {} {Ph.D. thesis},\ \bibinfo  {school} {Goethe-Universitat,
  Frankfurt am Main} (\bibinfo {year} {2006})\BibitemShut {NoStop}%
\bibitem [{\citenamefont {{Hultgren}}\ \emph {et~al.}(1963)\citenamefont
  {{Hultgren}}, \citenamefont {{Orr}}, \citenamefont {{Anderson}},\ and\
  \citenamefont {{Kelley}}}]{Hultgren1963}%
  \BibitemOpen
  \bibfield  {author} {\bibinfo {author} {\bibfnamefont {R.}~\bibnamefont
  {{Hultgren}}}, \bibinfo {author} {\bibfnamefont {L.}~\bibnamefont {{Orr}}},
  \bibinfo {author} {\bibfnamefont {P.~D.}\ \bibnamefont {{Anderson}}}, \ and\
  \bibinfo {author} {\bibfnamefont {K.~K.}\ \bibnamefont {{Kelley}}},\
  }\href@noop {} {\emph {\bibinfo {title} {Selected Values of Thermodynamic
  Properties of Metals and Alloys}}}\ (\bibinfo  {publisher} {John Wiley and
  Sons},\ \bibinfo {address} {New York},\ \bibinfo {year} {1963})\BibitemShut
  {NoStop}%
\bibitem [{\citenamefont {Cagran}\ and\ \citenamefont
  {Pottlacher}(2008)}]{Cagran2008}%
  \BibitemOpen
  \bibfield  {author} {\bibinfo {author} {\bibfnamefont {C.}~\bibnamefont
  {Cagran}}\ and\ \bibinfo {author} {\bibfnamefont {G.}~\bibnamefont
  {Pottlacher}},\ }in\ \href@noop {} {\emph {\bibinfo {booktitle} {Handbook of
  Thermal Analysis and Calorimetry - Recent Advances, Techniques, and
  Applications}}},\ Vol.~\bibinfo {volume} {5},\ \bibinfo {editor} {edited by\
  \bibinfo {editor} {\bibfnamefont {M.~E.}\ \bibnamefont {Brown}}\ and\
  \bibinfo {editor} {\bibfnamefont {P.~K.}\ \bibnamefont {Gallagher}}}\
  (\bibinfo  {publisher} {Elsevier},\ \bibinfo {year} {2008})\ \bibinfo
  {edition} {5th}\ ed.,\ Chap.~\bibinfo {chapter} {9}, pp.\ \bibinfo {pages}
  {299--320}\BibitemShut {NoStop}%
\bibitem [{\citenamefont {{Deemyad}}\ and\ \citenamefont
  {{Silvera}}(2008)}]{Deemyad2008}%
  \BibitemOpen
  \bibfield  {author} {\bibinfo {author} {\bibfnamefont {S.}~\bibnamefont
  {{Deemyad}}}\ and\ \bibinfo {author} {\bibfnamefont {I.~F.}\ \bibnamefont
  {{Silvera}}},\ }\href {\doibase 10.1103/PhysRevLett.100.155701} {\bibfield
  {journal} {\bibinfo  {journal} {\prl}\ }\textbf {\bibinfo {volume} {100}},\
  \bibinfo {eid} {155701} (\bibinfo {year} {2008})}\BibitemShut {NoStop}%
\bibitem [{\citenamefont {{Zaghoo}}\ \emph {et~al.}(2016)\citenamefont
  {{Zaghoo}}, \citenamefont {{Salamat}},\ and\ \citenamefont
  {{Silvera}}}]{Zaghoo2016}%
  \BibitemOpen
  \bibfield  {author} {\bibinfo {author} {\bibfnamefont {M.}~\bibnamefont
  {{Zaghoo}}}, \bibinfo {author} {\bibfnamefont {A.}~\bibnamefont {{Salamat}}},
  \ and\ \bibinfo {author} {\bibfnamefont {I.~F.}\ \bibnamefont {{Silvera}}},\
  }\href {\doibase 10.1103/PhysRevB.93.155128} {\bibfield  {journal} {\bibinfo
  {journal} {\prb}\ }\textbf {\bibinfo {volume} {93}},\ \bibinfo {eid} {155128}
  (\bibinfo {year} {2016})}\BibitemShut {NoStop}%
\bibitem [{\citenamefont {{Houtput}}\ \emph {et~al.}(2019)\citenamefont
  {{Houtput}}, \citenamefont {{Tempere}},\ and\ \citenamefont
  {{Silvera}}}]{Houtput2019}%
  \BibitemOpen
  \bibfield  {author} {\bibinfo {author} {\bibfnamefont {M.}~\bibnamefont
  {{Houtput}}}, \bibinfo {author} {\bibfnamefont {J.}~\bibnamefont
  {{Tempere}}}, \ and\ \bibinfo {author} {\bibfnamefont {I.~F.}\ \bibnamefont
  {{Silvera}}},\ }\href {\doibase 10.1103/PhysRevB.100.134106} {\bibfield
  {journal} {\bibinfo  {journal} {\prb}\ }\textbf {\bibinfo {volume} {100}},\
  \bibinfo {eid} {134106} (\bibinfo {year} {2019})}\BibitemShut {NoStop}%
\bibitem [{\citenamefont {{Montoya}}\ and\ \citenamefont
  {{Goncharov}}(2012)}]{Montoya2012}%
  \BibitemOpen
  \bibfield  {author} {\bibinfo {author} {\bibfnamefont {J.~A.}\ \bibnamefont
  {{Montoya}}}\ and\ \bibinfo {author} {\bibfnamefont {A.~F.}\ \bibnamefont
  {{Goncharov}}},\ }\href {\doibase 10.1063/1.4726231} {\bibfield  {journal}
  {\bibinfo  {journal} {J. Appl. Phys.}\ }\textbf {\bibinfo {volume} {111}},\
  \bibinfo {eid} {112617-112617-9} (\bibinfo {year} {2012})}\BibitemShut
  {NoStop}%
\bibitem [{\citenamefont {{Goncharov}}\ and\ \citenamefont
  {{Geballe}}(2017)}]{GoncharovComment}%
  \BibitemOpen
  \bibfield  {author} {\bibinfo {author} {\bibfnamefont {A.~F.}\ \bibnamefont
  {{Goncharov}}}\ and\ \bibinfo {author} {\bibfnamefont {Z.~M.}\ \bibnamefont
  {{Geballe}}},\ }\href {\doibase 10.1103/PhysRevB.96.157101} {\bibfield
  {journal} {\bibinfo  {journal} {\prb}\ }\textbf {\bibinfo {volume} {96}},\
  \bibinfo {eid} {157101} (\bibinfo {year} {2017})}\BibitemShut {NoStop}%
\bibitem [{\citenamefont {{Zha}}\ \emph {et~al.}(2008)\citenamefont {{Zha}},
  \citenamefont {{Mibe}}, \citenamefont {{Bassett}}, \citenamefont
  {{Tschauner}}, \citenamefont {{Mao}},\ and\ \citenamefont
  {{Hemley}}}]{Zha2008}%
  \BibitemOpen
  \bibfield  {author} {\bibinfo {author} {\bibfnamefont {C.-S.}\ \bibnamefont
  {{Zha}}}, \bibinfo {author} {\bibfnamefont {K.}~\bibnamefont {{Mibe}}},
  \bibinfo {author} {\bibfnamefont {W.~A.}\ \bibnamefont {{Bassett}}}, \bibinfo
  {author} {\bibfnamefont {O.}~\bibnamefont {{Tschauner}}}, \bibinfo {author}
  {\bibfnamefont {H.-K.}\ \bibnamefont {{Mao}}}, \ and\ \bibinfo {author}
  {\bibfnamefont {R.~J.}\ \bibnamefont {{Hemley}}},\ }\href {\doibase
  10.1063/1.2844358} {\bibfield  {journal} {\bibinfo  {journal} {J. Appl.
  Phys.}\ }\textbf {\bibinfo {volume} {103}},\ \bibinfo {eid}
  {054908-054908-10} (\bibinfo {year} {2008})}\BibitemShut {NoStop}%
\bibitem [{\citenamefont {Akahama}\ and\ \citenamefont
  {Kawamura}(2006)}]{Akahama2006}%
  \BibitemOpen
  \bibfield  {author} {\bibinfo {author} {\bibfnamefont {Y.}~\bibnamefont
  {Akahama}}\ and\ \bibinfo {author} {\bibfnamefont {H.}~\bibnamefont
  {Kawamura}},\ }\href {\doibase 10.1063/1.2335683} {\bibfield  {journal}
  {\bibinfo  {journal} {J. Appl. Phys.}\ }\textbf {\bibinfo {volume} {100}},\
  \bibinfo {pages} {043516} (\bibinfo {year} {2006})}\BibitemShut {NoStop}%
\bibitem [{\citenamefont {{Matsui}}\ \emph {et~al.}(2009)\citenamefont
  {{Matsui}}, \citenamefont {{Ito}}, \citenamefont {{Katsura}}, \citenamefont
  {{Yamazaki}}, \citenamefont {{Yoshino}}, \citenamefont {{Yokoyama}},\ and\
  \citenamefont {{Funakoshi}}}]{Matsui2009}%
  \BibitemOpen
  \bibfield  {author} {\bibinfo {author} {\bibfnamefont {M.}~\bibnamefont
  {{Matsui}}}, \bibinfo {author} {\bibfnamefont {E.}~\bibnamefont {{Ito}}},
  \bibinfo {author} {\bibfnamefont {T.}~\bibnamefont {{Katsura}}}, \bibinfo
  {author} {\bibfnamefont {D.}~\bibnamefont {{Yamazaki}}}, \bibinfo {author}
  {\bibfnamefont {T.}~\bibnamefont {{Yoshino}}}, \bibinfo {author}
  {\bibfnamefont {A.}~\bibnamefont {{Yokoyama}}}, \ and\ \bibinfo {author}
  {\bibfnamefont {K.-i.}\ \bibnamefont {{Funakoshi}}},\ }\href {\doibase
  10.1063/1.3054331} {\bibfield  {journal} {\bibinfo  {journal} {J. Appl.
  Phys.}\ }\textbf {\bibinfo {volume} {105}},\ \bibinfo {eid} {013505-013505-7}
  (\bibinfo {year} {2009})}\BibitemShut {NoStop}%
\bibitem [{\citenamefont {{McWilliams}}\ \emph {et~al.}(2015)\citenamefont
  {{McWilliams}}, \citenamefont {{Kon{\^o}pkov{\'a}}},\ and\ \citenamefont
  {{Goncharov}}}]{McWilliams2015}%
  \BibitemOpen
  \bibfield  {author} {\bibinfo {author} {\bibfnamefont {R.~S.}\ \bibnamefont
  {{McWilliams}}}, \bibinfo {author} {\bibfnamefont {Z.}~\bibnamefont
  {{Kon{\^o}pkov{\'a}}}}, \ and\ \bibinfo {author} {\bibfnamefont {A.~F.}\
  \bibnamefont {{Goncharov}}},\ }\href {\doibase 10.1016/j.pepi.2015.06.002}
  {\bibfield  {journal} {\bibinfo  {journal} {Phys. Earth Planet. Inter.}\
  }\textbf {\bibinfo {volume} {247}},\ \bibinfo {pages} {17} (\bibinfo {year}
  {2015})}\BibitemShut {NoStop}%
\bibitem [{\citenamefont {{Prakapenka}}\ \emph {et~al.}(2008)\citenamefont
  {{Prakapenka}}, \citenamefont {{Kubo}}, \citenamefont {{Kuznetsov}},
  \citenamefont {{Laskin}}, \citenamefont {{Shkurikhin}}, \citenamefont
  {{Dera}}, \citenamefont {{Rivers}},\ and\ \citenamefont
  {{Sutton}}}]{Prakapenka2008}%
  \BibitemOpen
  \bibfield  {author} {\bibinfo {author} {\bibfnamefont {V.~B.}\ \bibnamefont
  {{Prakapenka}}}, \bibinfo {author} {\bibfnamefont {A.}~\bibnamefont
  {{Kubo}}}, \bibinfo {author} {\bibfnamefont {A.}~\bibnamefont {{Kuznetsov}}},
  \bibinfo {author} {\bibfnamefont {A.}~\bibnamefont {{Laskin}}}, \bibinfo
  {author} {\bibfnamefont {O.}~\bibnamefont {{Shkurikhin}}}, \bibinfo {author}
  {\bibfnamefont {P.}~\bibnamefont {{Dera}}}, \bibinfo {author} {\bibfnamefont
  {M.~L.}\ \bibnamefont {{Rivers}}}, \ and\ \bibinfo {author} {\bibfnamefont
  {S.~R.}\ \bibnamefont {{Sutton}}},\ }\href {\doibase
  10.1080/08957950802050718} {\bibfield  {journal} {\bibinfo  {journal} {High
  Pressure Res.}\ }\textbf {\bibinfo {volume} {28}},\ \bibinfo {pages} {225}
  (\bibinfo {year} {2008})}\BibitemShut {NoStop}%
\bibitem [{\citenamefont {{Benedetti}}\ and\ \citenamefont
  {{Loubeyre}}(2004)}]{Benedetti2004}%
  \BibitemOpen
  \bibfield  {author} {\bibinfo {author} {\bibfnamefont {L.~R.}\ \bibnamefont
  {{Benedetti}}}\ and\ \bibinfo {author} {\bibfnamefont {P.}~\bibnamefont
  {{Loubeyre}}},\ }\href {\doibase 10.1080/08957950412331331718} {\bibfield
  {journal} {\bibinfo  {journal} {High Pressure Res.}\ }\textbf {\bibinfo
  {volume} {24}},\ \bibinfo {pages} {423} (\bibinfo {year} {2004})}\BibitemShut
  {NoStop}%
\bibitem [{\citenamefont {{Prescher}}\ and\ \citenamefont
  {{Prakapenka}}(2015)}]{Prescher2015}%
  \BibitemOpen
  \bibfield  {author} {\bibinfo {author} {\bibfnamefont {C.}~\bibnamefont
  {{Prescher}}}\ and\ \bibinfo {author} {\bibfnamefont {V.~B.}\ \bibnamefont
  {{Prakapenka}}},\ }\href {\doibase 10.1080/08957959.2015.1059835} {\bibfield
  {journal} {\bibinfo  {journal} {High Pressure Res.}\ }\textbf {\bibinfo
  {volume} {35}},\ \bibinfo {pages} {223} (\bibinfo {year} {2015})}\BibitemShut
  {NoStop}%
\bibitem [{\citenamefont {{Mitra}}\ \emph {et~al.}(1967)\citenamefont
  {{Mitra}}, \citenamefont {{Decker}},\ and\ \citenamefont
  {{Vanfleet}}}]{Mitra1967}%
  \BibitemOpen
  \bibfield  {author} {\bibinfo {author} {\bibfnamefont {N.~R.}\ \bibnamefont
  {{Mitra}}}, \bibinfo {author} {\bibfnamefont {D.~L.}\ \bibnamefont
  {{Decker}}}, \ and\ \bibinfo {author} {\bibfnamefont {H.~B.}\ \bibnamefont
  {{Vanfleet}}},\ }\href {\doibase 10.1103/PhysRev.161.613} {\bibfield
  {journal} {\bibinfo  {journal} {Phys. Rev.}\ }\textbf {\bibinfo {volume}
  {161}},\ \bibinfo {pages} {613} (\bibinfo {year} {1967})}\BibitemShut
  {NoStop}%
\bibitem [{\citenamefont {{Anderson}}\ and\ \citenamefont
  {{Isaak}}(2000)}]{Anderson2000}%
  \BibitemOpen
  \bibfield  {author} {\bibinfo {author} {\bibfnamefont {O.~L.}\ \bibnamefont
  {{Anderson}}}\ and\ \bibinfo {author} {\bibfnamefont {D.~G.}\ \bibnamefont
  {{Isaak}}},\ }\href {\doibase 10.2138/am-2000-2-317} {\bibfield  {journal}
  {\bibinfo  {journal} {American Mineralogist}\ }\textbf {\bibinfo {volume}
  {85}},\ \bibinfo {pages} {376} (\bibinfo {year} {2000})}\BibitemShut
  {NoStop}%
\bibitem [{\citenamefont {Fei}\ \emph {et~al.}(2007)\citenamefont {Fei},
  \citenamefont {Ricolleau}, \citenamefont {Frank}, \citenamefont {Mibe},
  \citenamefont {Shen},\ and\ \citenamefont {Prakapenka}}]{Fei2007}%
  \BibitemOpen
  \bibfield  {author} {\bibinfo {author} {\bibfnamefont {Y.}~\bibnamefont
  {Fei}}, \bibinfo {author} {\bibfnamefont {A.}~\bibnamefont {Ricolleau}},
  \bibinfo {author} {\bibfnamefont {M.}~\bibnamefont {Frank}}, \bibinfo
  {author} {\bibfnamefont {K.}~\bibnamefont {Mibe}}, \bibinfo {author}
  {\bibfnamefont {G.}~\bibnamefont {Shen}}, \ and\ \bibinfo {author}
  {\bibfnamefont {V.}~\bibnamefont {Prakapenka}},\ }\href {\doibase
  10.1073/pnas.0609013104} {\bibfield  {journal} {\bibinfo  {journal} {Proc.
  Nat. Acad. Sci. USA}\ }\textbf {\bibinfo {volume} {104}},\ \bibinfo {pages}
  {9182} (\bibinfo {year} {2007})}\BibitemShut {NoStop}%
\bibitem [{\citenamefont {{Arblaster}}(2017)}]{Arblaster2017}%
  \BibitemOpen
  \bibfield  {author} {\bibinfo {author} {\bibfnamefont {J.~W.}\ \bibnamefont
  {{Arblaster}}},\ }\href {\doibase 10.1595/205651317X694461} {\bibfield
  {journal} {\bibinfo  {journal} {Johnson Matthey Technology Review}\ }\textbf
  {\bibinfo {volume} {62}},\ \bibinfo {pages} {80} (\bibinfo {year}
  {2017})}\BibitemShut {NoStop}%
\bibitem [{\citenamefont {{Shen}}\ and\ \citenamefont
  {{Lazor}}(1995)}]{Shen1995}%
  \BibitemOpen
  \bibfield  {author} {\bibinfo {author} {\bibfnamefont {G.}~\bibnamefont
  {{Shen}}}\ and\ \bibinfo {author} {\bibfnamefont {P.}~\bibnamefont
  {{Lazor}}},\ }\href {\doibase 10.1029/95JB01864} {\bibfield  {journal}
  {\bibinfo  {journal} {J. Geophys. Res.}\ }\textbf {\bibinfo {volume} {100}},\
  \bibinfo {pages} {17,699} (\bibinfo {year} {1995})}\BibitemShut {NoStop}%
\bibitem [{\citenamefont {{Lord}}\ \emph {et~al.}(2009)\citenamefont {{Lord}},
  \citenamefont {{Walter}}, \citenamefont {{Dasgupta}}, \citenamefont
  {{Walker}},\ and\ \citenamefont {{Clark}}}]{Lord2009}%
  \BibitemOpen
  \bibfield  {author} {\bibinfo {author} {\bibfnamefont {O.~T.}\ \bibnamefont
  {{Lord}}}, \bibinfo {author} {\bibfnamefont {M.~J.}\ \bibnamefont
  {{Walter}}}, \bibinfo {author} {\bibfnamefont {R.}~\bibnamefont
  {{Dasgupta}}}, \bibinfo {author} {\bibfnamefont {D.}~\bibnamefont
  {{Walker}}}, \ and\ \bibinfo {author} {\bibfnamefont {S.~M.}\ \bibnamefont
  {{Clark}}},\ }\href {\doibase 10.1016/j.epsl.2009.04.017} {\bibfield
  {journal} {\bibinfo  {journal} {Earth and Planetary Science Letters}\
  }\textbf {\bibinfo {volume} {284}},\ \bibinfo {pages} {157} (\bibinfo {year}
  {2009})}\BibitemShut {NoStop}%
\bibitem [{\citenamefont {{Lord}}\ \emph {et~al.}(2010)\citenamefont {{Lord}},
  \citenamefont {{Walter}}, \citenamefont {{Dobson}}, \citenamefont
  {{Armstrong}}, \citenamefont {{Clark}},\ and\ \citenamefont
  {{Kleppe}}}]{Lord2010}%
  \BibitemOpen
  \bibfield  {author} {\bibinfo {author} {\bibfnamefont {O.~T.}\ \bibnamefont
  {{Lord}}}, \bibinfo {author} {\bibfnamefont {M.~J.}\ \bibnamefont
  {{Walter}}}, \bibinfo {author} {\bibfnamefont {D.~P.}\ \bibnamefont
  {{Dobson}}}, \bibinfo {author} {\bibfnamefont {L.}~\bibnamefont
  {{Armstrong}}}, \bibinfo {author} {\bibfnamefont {S.~M.}\ \bibnamefont
  {{Clark}}}, \ and\ \bibinfo {author} {\bibfnamefont {A.}~\bibnamefont
  {{Kleppe}}},\ }\href {\doibase 10.1029/2009JB006528} {\bibfield  {journal}
  {\bibinfo  {journal} {Journal of Geophysical Research (Solid Earth)}\
  }\textbf {\bibinfo {volume} {115}},\ \bibinfo {eid} {B06208} (\bibinfo {year}
  {2010})}\BibitemShut {NoStop}%
\bibitem [{\citenamefont {{Lord}}\ \emph {et~al.}(2014)\citenamefont {{Lord}},
  \citenamefont {{Wood}}, \citenamefont {{Dobson}}, \citenamefont
  {{Vo{\v{c}}adlo}}, \citenamefont {{Wang}}, \citenamefont {{Thomson}},
  \citenamefont {{Wann}}, \citenamefont {{Morard}}, \citenamefont {{Mezouar}},\
  and\ \citenamefont {{Walter}}}]{Lord2014}%
  \BibitemOpen
  \bibfield  {author} {\bibinfo {author} {\bibfnamefont {O.~T.}\ \bibnamefont
  {{Lord}}}, \bibinfo {author} {\bibfnamefont {I.~G.}\ \bibnamefont {{Wood}}},
  \bibinfo {author} {\bibfnamefont {D.~P.}\ \bibnamefont {{Dobson}}}, \bibinfo
  {author} {\bibfnamefont {L.}~\bibnamefont {{Vo{\v{c}}adlo}}}, \bibinfo
  {author} {\bibfnamefont {W.}~\bibnamefont {{Wang}}}, \bibinfo {author}
  {\bibfnamefont {A.~R.}\ \bibnamefont {{Thomson}}}, \bibinfo {author}
  {\bibfnamefont {E.~T.~H.}\ \bibnamefont {{Wann}}}, \bibinfo {author}
  {\bibfnamefont {G.}~\bibnamefont {{Morard}}}, \bibinfo {author}
  {\bibfnamefont {M.}~\bibnamefont {{Mezouar}}}, \ and\ \bibinfo {author}
  {\bibfnamefont {M.~J.}\ \bibnamefont {{Walter}}},\ }\href {\doibase
  10.1016/j.epsl.2014.09.046} {\bibfield  {journal} {\bibinfo  {journal} {Earth
  and Planetary Science Letters}\ }\textbf {\bibinfo {volume} {408}},\ \bibinfo
  {pages} {226} (\bibinfo {year} {2014})}\BibitemShut {NoStop}%
\bibitem [{\citenamefont {{Dewaele}}\ \emph {et~al.}(2007)\citenamefont
  {{Dewaele}}, \citenamefont {{Mezouar}}, \citenamefont {{Guignot}},\ and\
  \citenamefont {{Loubeyre}}}]{Dewale2007}%
  \BibitemOpen
  \bibfield  {author} {\bibinfo {author} {\bibfnamefont {A.}~\bibnamefont
  {{Dewaele}}}, \bibinfo {author} {\bibfnamefont {M.}~\bibnamefont
  {{Mezouar}}}, \bibinfo {author} {\bibfnamefont {N.}~\bibnamefont
  {{Guignot}}}, \ and\ \bibinfo {author} {\bibfnamefont {P.}~\bibnamefont
  {{Loubeyre}}},\ }\href {\doibase 10.1103/PhysRevB.76.144106} {\bibfield
  {journal} {\bibinfo  {journal} {\prb}\ }\textbf {\bibinfo {volume} {76}},\
  \bibinfo {eid} {144106} (\bibinfo {year} {2007})}\BibitemShut {NoStop}%
\bibitem [{\citenamefont {{Dewaele}}\ \emph {et~al.}(2010)\citenamefont
  {{Dewaele}}, \citenamefont {{Mezouar}}, \citenamefont {{Guignot}},\ and\
  \citenamefont {{Loubeyre}}}]{Dewaele2010}%
  \BibitemOpen
  \bibfield  {author} {\bibinfo {author} {\bibfnamefont {A.}~\bibnamefont
  {{Dewaele}}}, \bibinfo {author} {\bibfnamefont {M.}~\bibnamefont
  {{Mezouar}}}, \bibinfo {author} {\bibfnamefont {N.}~\bibnamefont
  {{Guignot}}}, \ and\ \bibinfo {author} {\bibfnamefont {P.}~\bibnamefont
  {{Loubeyre}}},\ }\href {\doibase 10.1103/PhysRevLett.104.255701} {\bibfield
  {journal} {\bibinfo  {journal} {\prl}\ }\textbf {\bibinfo {volume} {104}},\
  \bibinfo {eid} {255701} (\bibinfo {year} {2010})}\BibitemShut {NoStop}%
\bibitem [{\citenamefont {{Kimura}}\ \emph {et~al.}(2017)\citenamefont
  {{Kimura}}, \citenamefont {{Ohfuji}}, \citenamefont {{Nishi}},\ and\
  \citenamefont {{Irifune}}}]{Kimura2017}%
  \BibitemOpen
  \bibfield  {author} {\bibinfo {author} {\bibfnamefont {T.}~\bibnamefont
  {{Kimura}}}, \bibinfo {author} {\bibfnamefont {H.}~\bibnamefont {{Ohfuji}}},
  \bibinfo {author} {\bibfnamefont {M.}~\bibnamefont {{Nishi}}}, \ and\
  \bibinfo {author} {\bibfnamefont {T.}~\bibnamefont {{Irifune}}},\ }\href
  {\doibase 10.1038/ncomms15735} {\bibfield  {journal} {\bibinfo  {journal}
  {Nature Communications}\ }\textbf {\bibinfo {volume} {8}},\ \bibinfo {eid}
  {15735} (\bibinfo {year} {2017})}\BibitemShut {NoStop}%
\bibitem [{\citenamefont {{Anzellini}}\ \emph {et~al.}(2013)\citenamefont
  {{Anzellini}}, \citenamefont {{Dewaele}}, \citenamefont {{Mezouar}},
  \citenamefont {{Loubeyre}},\ and\ \citenamefont {{Morard}}}]{Anzellini2013}%
  \BibitemOpen
  \bibfield  {author} {\bibinfo {author} {\bibfnamefont {S.}~\bibnamefont
  {{Anzellini}}}, \bibinfo {author} {\bibfnamefont {A.}~\bibnamefont
  {{Dewaele}}}, \bibinfo {author} {\bibfnamefont {M.}~\bibnamefont
  {{Mezouar}}}, \bibinfo {author} {\bibfnamefont {P.}~\bibnamefont
  {{Loubeyre}}}, \ and\ \bibinfo {author} {\bibfnamefont {G.}~\bibnamefont
  {{Morard}}},\ }\href {\doibase 10.1126/science.1233514} {\bibfield  {journal}
  {\bibinfo  {journal} {Science}\ }\textbf {\bibinfo {volume} {340}},\ \bibinfo
  {pages} {464} (\bibinfo {year} {2013})}\BibitemShut {NoStop}%
\bibitem [{\citenamefont {{Jeong}}\ and\ \citenamefont
  {{Chang}}(1999)}]{Jeong1999}%
  \BibitemOpen
  \bibfield  {author} {\bibinfo {author} {\bibfnamefont {J.-W.}\ \bibnamefont
  {{Jeong}}}\ and\ \bibinfo {author} {\bibfnamefont {K.~J.}\ \bibnamefont
  {{Chang}}},\ }\href {\doibase 10.1088/0953-8984/11/19/302} {\bibfield
  {journal} {\bibinfo  {journal} {J. Phys. Condens. Matt.}\ }\textbf {\bibinfo
  {volume} {11}},\ \bibinfo {pages} {3799} (\bibinfo {year}
  {1999})}\BibitemShut {NoStop}%
\bibitem [{\citenamefont {{Liu}}\ \emph {et~al.}(2010)\citenamefont {{Liu}},
  \citenamefont {{Yang}}, \citenamefont {{Zhao}}, \citenamefont {{Cai}},\ and\
  \citenamefont {{Jing}}}]{Liu2010}%
  \BibitemOpen
  \bibfield  {author} {\bibinfo {author} {\bibfnamefont {Z.-L.}\ \bibnamefont
  {{Liu}}}, \bibinfo {author} {\bibfnamefont {J.-H.}\ \bibnamefont {{Yang}}},
  \bibinfo {author} {\bibfnamefont {Z.-G.}\ \bibnamefont {{Zhao}}}, \bibinfo
  {author} {\bibfnamefont {L.-C.}\ \bibnamefont {{Cai}}}, \ and\ \bibinfo
  {author} {\bibfnamefont {F.-Q.}\ \bibnamefont {{Jing}}},\ }\href {\doibase
  10.1016/j.physleta.2010.01.063} {\bibfield  {journal} {\bibinfo  {journal}
  {Phys. Lett. A}\ }\textbf {\bibinfo {volume} {374}},\ \bibinfo {pages} {1579}
  (\bibinfo {year} {2010})}\BibitemShut {NoStop}%
\bibitem [{\citenamefont {Zha}\ \emph {et~al.}(2017)\citenamefont {Zha},
  \citenamefont {Liu}, \citenamefont {Tse},\ and\ \citenamefont
  {Hemley}}]{Zha2017}%
  \BibitemOpen
  \bibfield  {author} {\bibinfo {author} {\bibfnamefont {C.-S.}\ \bibnamefont
  {Zha}}, \bibinfo {author} {\bibfnamefont {H.}~\bibnamefont {Liu}}, \bibinfo
  {author} {\bibfnamefont {J.~S.}\ \bibnamefont {Tse}}, \ and\ \bibinfo
  {author} {\bibfnamefont {R.~J.}\ \bibnamefont {Hemley}},\ }\href {\doibase
  10.1103/PhysRevLett.119.075302} {\bibfield  {journal} {\bibinfo  {journal}
  {Phys. Rev. Lett.}\ }\textbf {\bibinfo {volume} {119}},\ \bibinfo {pages}
  {075302} (\bibinfo {year} {2017})}\BibitemShut {NoStop}%
\bibitem [{\citenamefont {{Giesselmann}}\ \emph {et~al.}(2005)\citenamefont
  {{Giesselmann}}, \citenamefont {{Palmer}}, \citenamefont {{Neuber}},\ and\
  \citenamefont {{Donlon}}}]{Giesselmann2005}%
  \BibitemOpen
  \bibfield  {author} {\bibinfo {author} {\bibfnamefont {M.}~\bibnamefont
  {{Giesselmann}}}, \bibinfo {author} {\bibfnamefont {B.}~\bibnamefont
  {{Palmer}}}, \bibinfo {author} {\bibfnamefont {A.}~\bibnamefont {{Neuber}}},
  \ and\ \bibinfo {author} {\bibfnamefont {J.}~\bibnamefont {{Donlon}}},\ }in\
  \href@noop {} {\emph {\bibinfo {booktitle} {2005 IEEE Pulsed Power
  Conference}}}\ (\bibinfo {year} {2005})\ pp.\ \bibinfo {pages}
  {763--766}\BibitemShut {NoStop}%
\bibitem [{\citenamefont {{Gallob}}\ \emph {et~al.}(1986)\citenamefont
  {{Gallob}}, \citenamefont {{J{\"a}ger}},\ and\ \citenamefont
  {{Pottlacher}}}]{Gallob1986}%
  \BibitemOpen
  \bibfield  {author} {\bibinfo {author} {\bibfnamefont {R.}~\bibnamefont
  {{Gallob}}}, \bibinfo {author} {\bibfnamefont {H.}~\bibnamefont
  {{J{\"a}ger}}}, \ and\ \bibinfo {author} {\bibfnamefont {G.}~\bibnamefont
  {{Pottlacher}}},\ }\href {\doibase 10.1007/BF00503805} {\bibfield  {journal}
  {\bibinfo  {journal} {Int. J. Thermophys.}\ }\textbf {\bibinfo {volume}
  {7}},\ \bibinfo {pages} {139} (\bibinfo {year} {1986})}\BibitemShut {NoStop}%
\bibitem [{\citenamefont {{Cezairliyan}}\ and\ \citenamefont
  {{McClure}}(1987)}]{Cezairliyan1987}%
  \BibitemOpen
  \bibfield  {author} {\bibinfo {author} {\bibfnamefont {A.}~\bibnamefont
  {{Cezairliyan}}}\ and\ \bibinfo {author} {\bibfnamefont {J.~L.}\ \bibnamefont
  {{McClure}}},\ }\href {\doibase 10.1007/BF00503644} {\bibfield  {journal}
  {\bibinfo  {journal} {Int. J. Thermophys.}\ }\textbf {\bibinfo {volume}
  {8}},\ \bibinfo {pages} {577} (\bibinfo {year} {1987})}\BibitemShut {NoStop}%
\bibitem [{\citenamefont {Giampaoli}\ \emph {et~al.}(2018)\citenamefont
  {Giampaoli}, \citenamefont {Kantor}, \citenamefont {Mezouar}, \citenamefont
  {Boccato}, \citenamefont {Rosa}, \citenamefont {Torchio}, \citenamefont
  {Garbarino}, \citenamefont {Mathon},\ and\ \citenamefont
  {Pascarelli}}]{Giampaoli2018}%
  \BibitemOpen
  \bibfield  {author} {\bibinfo {author} {\bibfnamefont {R.}~\bibnamefont
  {Giampaoli}}, \bibinfo {author} {\bibfnamefont {I.}~\bibnamefont {Kantor}},
  \bibinfo {author} {\bibfnamefont {M.}~\bibnamefont {Mezouar}}, \bibinfo
  {author} {\bibfnamefont {S.}~\bibnamefont {Boccato}}, \bibinfo {author}
  {\bibfnamefont {A.~D.}\ \bibnamefont {Rosa}}, \bibinfo {author}
  {\bibfnamefont {R.}~\bibnamefont {Torchio}}, \bibinfo {author} {\bibfnamefont
  {G.}~\bibnamefont {Garbarino}}, \bibinfo {author} {\bibfnamefont
  {O.}~\bibnamefont {Mathon}}, \ and\ \bibinfo {author} {\bibfnamefont
  {S.}~\bibnamefont {Pascarelli}},\ }\href {\doibase
  10.1080/08957959.2018.1480017} {\bibfield  {journal} {\bibinfo  {journal}
  {High Pressure Research}\ }\textbf {\bibinfo {volume} {38}},\ \bibinfo
  {pages} {250} (\bibinfo {year} {2018})}\BibitemShut {NoStop}%
\bibitem [{\citenamefont {Ubbelohde}(1978)}]{Ubbelohde1978}%
  \BibitemOpen
  \bibfield  {author} {\bibinfo {author} {\bibfnamefont {A.~R.}\ \bibnamefont
  {Ubbelohde}},\ }\href@noop {} {\emph {\bibinfo {title} {{The Molten State of
  Matter}}}}\ (\bibinfo  {publisher} {John Wiley},\ \bibinfo {address}
  {Chichester},\ \bibinfo {year} {1978})\ p.\ \bibinfo {pages}
  {454}\BibitemShut {NoStop}%
\bibitem [{\citenamefont {{Geballe}}\ \emph {et~al.}(2020)\citenamefont
  {{Geballe}}, \citenamefont {{Sime}}, \citenamefont {{Badro}}, \citenamefont
  {{van Keken}},\ and\ \citenamefont {{Goncharov}}}]{Geballe2020}%
  \BibitemOpen
  \bibfield  {author} {\bibinfo {author} {\bibfnamefont {Z.~M.}\ \bibnamefont
  {{Geballe}}}, \bibinfo {author} {\bibfnamefont {N.}~\bibnamefont {{Sime}}},
  \bibinfo {author} {\bibfnamefont {J.}~\bibnamefont {{Badro}}}, \bibinfo
  {author} {\bibfnamefont {P.~E.}\ \bibnamefont {{van Keken}}}, \ and\ \bibinfo
  {author} {\bibfnamefont {A.~F.}\ \bibnamefont {{Goncharov}}},\ }\href
  {\doibase 10.1016/j.epsl.2020.116161} {\bibfield  {journal} {\bibinfo
  {journal} {Earth Planet. Sci. Lett.}\ }\textbf {\bibinfo {volume} {536}},\
  \bibinfo {eid} {116161} (\bibinfo {year} {2020})}\BibitemShut {NoStop}%
\bibitem [{\citenamefont {{Arveson}}\ \emph {et~al.}(2018)\citenamefont
  {{Arveson}}, \citenamefont {{Kiefer}}, \citenamefont {{Deng}}, \citenamefont
  {{Liu}},\ and\ \citenamefont {{Lee}}}]{Arveson2018}%
  \BibitemOpen
  \bibfield  {author} {\bibinfo {author} {\bibfnamefont {S.~M.}\ \bibnamefont
  {{Arveson}}}, \bibinfo {author} {\bibfnamefont {B.}~\bibnamefont {{Kiefer}}},
  \bibinfo {author} {\bibfnamefont {J.}~\bibnamefont {{Deng}}}, \bibinfo
  {author} {\bibfnamefont {Z.}~\bibnamefont {{Liu}}}, \ and\ \bibinfo {author}
  {\bibfnamefont {K.~K.~M.}\ \bibnamefont {{Lee}}},\ }\href {\doibase
  10.1103/PhysRevB.97.094103} {\bibfield  {journal} {\bibinfo  {journal}
  {\prb}\ }\textbf {\bibinfo {volume} {97}},\ \bibinfo {eid} {094103} (\bibinfo
  {year} {2018})}\BibitemShut {NoStop}%
\bibitem [{\citenamefont {{Wilthan}}\ \emph {et~al.}(2004)\citenamefont
  {{Wilthan}}, \citenamefont {{Cagran}},\ and\ \citenamefont
  {{Pottlacher}}}]{Wilthan2004}%
  \BibitemOpen
  \bibfield  {author} {\bibinfo {author} {\bibfnamefont {B.}~\bibnamefont
  {{Wilthan}}}, \bibinfo {author} {\bibfnamefont {C.}~\bibnamefont {{Cagran}}},
  \ and\ \bibinfo {author} {\bibfnamefont {G.}~\bibnamefont {{Pottlacher}}},\
  }\href {\doibase 10.1007/s10765-004-5756-7} {\bibfield  {journal} {\bibinfo
  {journal} {Int. J. Thermophys.}\ }\textbf {\bibinfo {volume} {25}},\ \bibinfo
  {pages} {1519} (\bibinfo {year} {2004})}\BibitemShut {NoStop}%
\bibitem [{\citenamefont {Jiang}\ \emph {et~al.}(2020)\citenamefont {Jiang},
  \citenamefont {Holtgrewe}, \citenamefont {Geballe}, \citenamefont {Lobanov},
  \citenamefont {Mahmood}, \citenamefont {McWilliams},\ and\ \citenamefont
  {Goncharov}}]{Jiang2020}%
  \BibitemOpen
  \bibfield  {author} {\bibinfo {author} {\bibfnamefont {S.}~\bibnamefont
  {Jiang}}, \bibinfo {author} {\bibfnamefont {N.}~\bibnamefont {Holtgrewe}},
  \bibinfo {author} {\bibfnamefont {Z.~M.}\ \bibnamefont {Geballe}}, \bibinfo
  {author} {\bibfnamefont {S.~S.}\ \bibnamefont {Lobanov}}, \bibinfo {author}
  {\bibfnamefont {M.~F.}\ \bibnamefont {Mahmood}}, \bibinfo {author}
  {\bibfnamefont {R.~S.}\ \bibnamefont {McWilliams}}, \ and\ \bibinfo {author}
  {\bibfnamefont {A.~F.}\ \bibnamefont {Goncharov}},\ }\href {\doibase
  10.1002/advs.201901668} {\bibfield  {journal} {\bibinfo  {journal} {Advanced
  Science}\ }\textbf {\bibinfo {volume} {7}},\ \bibinfo {pages} {1901668}
  (\bibinfo {year} {2020})}\BibitemShut {NoStop}%
\bibitem [{\citenamefont {{Vopson}}\ \emph {et~al.}(2020)\citenamefont
  {{Vopson}}, \citenamefont {{Rogers}},\ and\ \citenamefont
  {{Hepburn}}}]{Vopson2020}%
  \BibitemOpen
  \bibfield  {author} {\bibinfo {author} {\bibfnamefont {M.~M.}\ \bibnamefont
  {{Vopson}}}, \bibinfo {author} {\bibfnamefont {N.}~\bibnamefont {{Rogers}}},
  \ and\ \bibinfo {author} {\bibfnamefont {I.}~\bibnamefont {{Hepburn}}},\
  }\href {\doibase 10.1016/j.ssc.2020.113977} {\bibfield  {journal} {\bibinfo
  {journal} {Solid State Communications}\ }\textbf {\bibinfo {volume} {318}},\
  \bibinfo {eid} {113977} (\bibinfo {year} {2020})}\BibitemShut {NoStop}%
\end{thebibliography}%

\end{document}